\documentclass[12pt]{article}
\usepackage[T1]{fontenc}
\usepackage[a4paper]{geometry}
\usepackage{mathtools}
\usepackage{amsmath}
\usepackage{amsthm}
\usepackage{amssymb}
\usepackage{cases}
 \usepackage{soul}
\usepackage{tikz,pgfplots}
\usepackage{color}
\usepackage[round]{natbib}
\usepackage{xr-hyper}
\usepackage{subfigure}
\usepackage{comment}
\makeatletter
\newcommand*{\addFileDependency}[1]{
  \typeout{(#1)}
  \@addtofilelist{#1}
  \IfFileExists{#1}{}{\typeout{No file #1.}}
}
\makeatother

\usetikzlibrary{arrows}
\usetikzlibrary{decorations.markings,arrows.meta,automata,positioning,quotes}
\tikzset{
arrowmark/.style 2 args={decoration={markings,mark=at     position #1 with \arrow{Stealth[color=red]}}}}
\tikzset{
revarrowmark/.style 2 args={decoration={markings,mark=at     position #1 with \arrow{Stealth[color=red,reversed]}}}}
\usetikzlibrary{arrows,arrows.meta,decorations,decorations.pathreplacing,calc,matrix}

\usepackage{xcolor}
\usepackage[colorlinks,citecolor=northwestern-purple,urlcolor=chicago-maroon,linkcolor=chicago-maroon,filecolor=chicago-maroon,backref=page]{hyperref}

\usepackage[nameinlink,noabbrev,capitalize]{cleveref}

\usepackage{enumitem}

\crefname{equation}{}{}

\theoremstyle{plain}
\newtheorem{lem}{\protect\lemmaname}
\newtheorem{cor}{Corollary}
\theoremstyle{plain}
\newtheorem{claim}{\protect\claimname}
\theoremstyle{plain}
\newtheorem{prop}{\protect\propositionname}
\theoremstyle{plain}

\theoremstyle{plain}
\newtheorem{thm}{\protect\theoremname}
\theoremstyle{remark}
\newtheorem{rem}{\protect\remarkname}

\newenvironment{proprestated}[1]
{\begin{trivlist}
\item[\hskip\labelsep\bfseries Proposition #1.]\itshape}
{\end{trivlist}}

\definecolor{darkgreen}{rgb}{0.0, 0.5, 0.0}
\definecolor{brightpink}{rgb}{1.0, 0.0, 0.5}
\definecolor{brightgreen}{rgb}{0.4, 1.0, 0.0}
\definecolor{shockingpink}{rgb}{0.99, 0.06, 0.75}
\definecolor{persiangreen}{rgb}{0.0, 0.65, 0.58}
\definecolor{clemson-orange}{RGB}{234,106,32}
\definecolor{chicago-maroon}{RGB}{128,0,0}
\definecolor{northwestern-purple}{RGB}{82,0,99}
\definecolor{cornell-red}{RGB}{179,27,27}
\definecolor{sauder-green}{RGB}{171,180,0}
\definecolor{gray}{RGB}{192,192,192}
\definecolor{lawngreen}{RGB}{0,250,154}
\definecolor{clemson-orange}{RGB}{234,106,32}
\definecolor{chicago-maroon}{RGB}{128,0,0}
\definecolor{northwestern-purple}{RGB}{82,0,99}
\definecolor{cornell-red}{RGB}{179,27,27}
\definecolor{sauder-green}{RGB}{171,180,0}
\definecolor{gray}{RGB}{192,192,192}
\definecolor{lawngreen}{RGB}{0,250,154}

\colorlet{DarkBlue}{blue!80!black!200}
\colorlet{Darkblue}{blue!70!black!200}
\colorlet{DarkRed}{red!80!black!200}

\newcommand{\ykk}[1]{\textcolor{cornell-red}{ #1}}

\allowdisplaybreaks
\makeatother

\providecommand{\assumptionname}{Condition}
\providecommand{\claimname}{Claim}
\providecommand{\lemmaname}{Lemma}
\providecommand{\propositionname}{Proposition}
\providecommand{\remarkname}{Remark}
\providecommand{\theoremname}{Theorem}

\global\long\def\p{\pi}

\global\long\def\g{x}
\global\long\def\l{\lambda}
\global\long\def\ll{\underline{\l}}%
\global\long\def\lh{\overline{\l}}

\def\hp{\pm}
\def\ynp{y_{\text{\tiny SP}}}
\def\ypm{y_{\text{\tiny MP}}}
\def\ymp{y_{\text{\tiny MP}}}
\def\gnp{x_{\text{\tiny SP}}}

 \def\gsp{x_{\text{\tiny MP}}}
 \def\ysp{y_{\text{\tiny MP}}}
  \def\xop{x_{\text{\tiny OP}}}
 \def\yop{y_{\text{\tiny OP}}}
  \def\pop{\hat p_{\text{\tiny OP}}}
\def\pm{\hat p_{\text{\tiny M}}}
\usepackage{setspace}
\title{\bf Data-Driven Monitoring and Deterrence in a Changing Environment\thanks{We are grateful to the participants of numerous seminars and conferences for valuable comments and suggestions.    The authors gratefully acknowledge Mingeon Kim and Tialing Luo for their excellent research assistance.   This work was supported by the Ministry of Education of the Republic of Korea and the National Research Foundation of Korea (NRF-2024S1A5A2A03038509).  Our dear friend and coauthor, Konrad Mierendorff, passed away in August of 2021. All the ideas and results in this paper are the collaborative work of all three authors, but Che and Kim are responsible for any remaining errors.}}
\author{\textsc{Yeon-Koo Che},\thinspace\thinspace\ \textsc{Jinwoo Kim},\thinspace\thinspace\ \textsc{Konrad Mierendorff}\thanks{Che: Department of Economics, Columbia University (email: yeonkooche@gmail.com); Kim: Department of Economics, Hong Kong University of Science and Technology (email: jikim72@gmail.com); Mierendorff: Department of Economics, University College London.}}
\date{\today}
 
\begin{document}
\maketitle

\begin{abstract}
We study a dynamic model in which a principal monitors agents based on historical data of infractions. This data informs when and at what intensity to monitor; the monitoring decision, in turn, selects the collected data, shaping the principal's future learning. We analyze this feedback loop using a bandit model in which the underlying monitoring environment evolves according to a hidden Markov process. Because data collection is endogenous, how the principal uses this information is critical: surprisingly, a myopic approach renders historical data completely valueless. By endogenizing the agent's incentives, we demonstrate that the principal's purely informational motive to explore serves as an endogenous commitment device. This inherent drive to gather data compels persistent vigilance, strictly lowering the equilibrium infraction rate and restoring the power of deterrence.

\vspace{0.2in}

 \end{abstract}

\section{Introduction}

The proliferation of big data and algorithmic prediction has fundamentally transformed how organizations monitor compliance, allocate inspection resources, and deter infractions. A prominent example is New York City's inspection of ``illegal conversions''---the illicit division of dwellings into smaller units, which poses severe fire hazards. By linking various municipal data sources to predict likely violations, the City's analytics team achieved a five-fold improvement in inspection effectiveness.\footnote{``Before the big-data analysis, inspectors followed up the complaints they deemed most dire, but only in 13 percent of cases did they find conditions severe enough to warrant a vacate order. Now they were issuing vacate orders on more than 70 percent of the buildings they inspected. By indicating which buildings most needed their attention, big data improved their efficiency five-fold. ... Flowers and his kids looked like wizards with a crystal ball that let them see into the future and predict the most risky places. They took massive quantities of data lying around for years, largely unused after it was collected, and harnessed it in a novel way to extract real value.'' See \cite{Mayer-schonberger:14}.
} Similar data-driven monitoring is now ubiquitous across diverse domains: digital platforms utilize machine learning to moderate content and flag compromised accounts distributing malware, environmental protection agencies use data-driven models to inspect facilities likely to bypass emissions controls, and financial regulators deploy anomaly detection algorithms to identify divisions at risk of accounting fraud. In all these settings, the principal's objective is twofold: to mitigate the immediate harm of infractions through active intervention, and to deter strategic agents from committing infractions in the first place.

While data-driven monitoring holds the promise of unprecedented efficiency, it introduces a profound dynamic challenge: the inherent endogeneity of the data generation process. Data is rarely collected randomly; it is selectively generated only where the principal actively looks. Consequently, monitoring serves a dual purpose. It not only halts current infractions but also generates the very information required to predict future ones. This creates a complex feedback loop. A principal who myopically trusts a predictive algorithm may cease monitoring an entity that currently appears safe. However, without ongoing monitoring, the principal creates an inferential blind spot, failing to recognize when the underlying conditions inevitably deteriorate. The fundamental challenge is understanding how to optimally trade off the short-term exploitation of current data against the long-term informational value of continued exploration.

The purpose of this paper is to formally investigate this tension. First, we develop a tractable, continuous-time framework for dynamic monitoring and deterrence that explicitly incorporates the feedback loop between monitoring intensity and endogenous datafication. Second, we extend the literature on optimal experimentation---specifically, the seminal bandit framework of \cite{keller:05}---in two critical dimensions to capture the realities of dynamic enforcement. First, we embed the learning problem in a changing world. In standard bandit models, the unobservable state of the arm is fixed; once the principal learns the true state, the need for experimentation permanently resolves. By modeling the environment as a hidden Markov state, we ensure that the need for learning and vigilance persists indefinitely. Second, we endogenize the value of the arm. While the standard bandit literature treats the payoff of the arm as exogenous, in regulatory and monitoring contexts, the arm represents a strategic agent. We therefore model the agent's infraction rate as a dynamic best response to the principal's anticipated monitoring policy.

We formulate this problem as a continuous-time dynamic inspection game. At each time $t \ge 0$, the environment is in an unobservable state that is either ``high risk'' ($H$) or ``low risk'' ($L$), transitioning according to a continuous-time Markov chain  with stationary belief $\pi_0$. Accordingly, in the absence of monitoring, the principal’s belief converges to this natural benchmark.   Opportunities for infractions arise at a Poisson rate only in the high-risk state. The principal chooses a monitoring intensity $y_t \in [0,1]$ at a flow cost $c$, while agents, upon receiving an opportunity, choose to commit an infraction with probability $x_t \in [0,1]$. Crucially, a detection perfectly reveals that the state is $H$. However, if no infraction is detected, the principal updates her posterior belief $p_t$ of state $H$ downward. This learning is strictly endogenous: the speed of the downward drift depends entirely on the principal's chosen monitoring intensity $y_t$ and the agent's infraction rate $x_t$. 


To isolate the value of data and the necessity of exploration, we evaluate  three distinct monitoring policy regimes: a static policy (SP)  that ignores data and relies solely on the stationary prior; a myopic policy (MP) that updates beliefs using data but acts myopically, ignoring the informational option value of monitoring; and the optimal  policy (OP), which balances exploitation with the need for persistent learning.

Our analysis yields four principal insights that challenge conventional intuition regarding data-driven monitoring.

First, a striking payoff equivalence emerges: the predictive value of data is entirely neutralized when used myopically. While it is natural to expect that a standard, data-driven algorithm should intuitively outperform a policy that ignores historical data entirely, we show that this is not the case: the myopic use of data yields no long-run performance gain over a static baseline. This equivalence is driven fundamentally by the endogeneity of data—the reality that new data is generated only through active monitoring. Because of this endogeneity, a myopic principal ceases enforcement the moment an entity currently appears safe. This creates a critical inferential blind spot, preventing her from recognizing when the underlying conditions inevitably deteriorate. Ultimately, this equivalence provides a crucial cautionary lesson: the value of predictive data is not automatic, and simply adopting a data-driven approach does not guarantee an enforcement gain.

Second, unlocking the true value of data requires a dynamically optimal policy that explicitly values exploration.  In a changing-world setup, data-driven policies can feature belief freezing at an interior cutoff, where partial monitoring exactly offsets the natural drift of the environment. The optimal policy uses this feature as an optimal ``hedging'' device. When data are valuable, the principal cannot optimally abandon monitoring, since the environment can revert to a high-risk state at any time.  Thus, when the belief reaches the optimal monitoring threshold, the principal does not stop entirely; instead, she chooses an interior monitoring intensity that freezes the belief at the cutoff, trapping the system in a partially absorbing state of persistent vigilance.

Third, we demonstrate that this exploratory motive acts as an endogenous commitment device for deterrence. When we endogenize the agents' behavior, the problem transforms into a dynamic inspection game where the principal lacks commitment power. Under myopic regimes, the principal cannot credibly commit to high monitoring rates, and agents simply adjust their infraction rate upward to keep the principal exactly indifferent.  However, under the optimal policy, the principal's inherent informational desire to explore organically pushes her monitoring intensity above what is myopically justified. Knowing the principal is eager to gather data, agents are forced to strictly lower their equilibrium infraction rate to maintain the deterrence condition.

Finally, we establish that the deterrence benefits of data-driven monitoring rely entirely on the principal maintaining an informational advantage. We analyze a scenario in which agents are ``data-aware,'' having access to the same historical detection data, and perfectly tracking the principal's posterior belief. Under symmetric information, agents perfectly best-respond at every instant, which entirely neutralizes the principal's exploratory benefit. Without the advantage of private learning, the endogenous commitment mechanism unravels, and the expected losses across all three policy regimes collapse to the exact same level.

\subsection*{Related Literature}

This paper contributes primarily to two strands of economic theory: optimal experimentation in non-stationary environments and dynamic monitoring or inspection. It also relates to the broader literature on dynamic agency with endogenous learning.

Our model builds on the continuous-time experimentation framework pioneered by \cite{keller:05}. In their canonical setting, the unobservable state of the arm is fixed; the principal learns through Poisson arrivals, and experimentation eventually ceases once uncertainty is sufficiently resolved. We extend this framework by embedding the learning problem in a changing world governed by a hidden Markov chain. Because the underlying state can continually transition, the need for learning never permanently disappears, fundamentally altering the optimal policy.

This changing-world feature naturally connects our work to the restless bandit literature \citep{whittle1988restless, nino2001restless}, recently applied in economics by \cite{fryer2018two}  and \cite{urgun2021restless}. Whereas Fryer and Harms feature a fixed hidden type and Urgun focuses on indexability in contracting, our emphasis lies on the endogenous generation of data.  We highlight a distinctive form of belief ``freezing''---an interior monitoring intensity that exactly offsets the natural drift of beliefs. Under the optimal policy, this belief freezing has the nature of hedging, as the principal must continually balance exploitation against the option value of vigilance in a changing environment.


We also depart from the standard bandit paradigm by treating the arm not as a passive source of rewards, but as an endogenous object whose future evolution depends on current actions. From this perspective, our paper is related to \cite{marinovic2022monitoring}, which also studies a dynamic monitoring problem with an endogenous arm. However, their monitoring quality depends on the regulator's effort incentives, whereas our monitoring effectiveness is driven by the agents' infraction incentives. More broadly, our model connects to a recent literature exploring the interplay between endogenous data collection and agent incentives (e.g., \cite{che2018recommender}, \cite{che2020statistical}, and \cite{madsen2024collective}). While this literature often focuses on designing mechanisms to incentivize agents to actively provide or reveal data, data in our dynamic inspection game is generated purely as a byproduct of the principal's enforcement and the agents' non-compliance. In our setting, the principal's informational motive to explore serves as an endogenous commitment device for deterrence.

Finally, we contribute to the literature on dynamic inspection games and the economics of deterrence. Traditional enforcement models (e.g., \cite{becker1968crime}) typically assume perfect commitment. In settings without commitment, principals are often assumed to optimize myopically, yielding mixed-strategy equilibria where agents merely keep the principal indifferent \citep{khalil1997auditing}. While several recent papers explore the optimal timing or frequency of inspections in dynamic settings  \citep{dilme2019residual, varas2020random, wagner2021relational, ball2023should}, we depart from this literature by emphasizing how monitoring endogenously generates data. Incorporating this informational option value into a dynamic inspection game provides a rational foundation for why a principal lacking formal commitment power may nonetheless credibly sustain high levels of monitoring.


 \section{The Model}
\label{sec:model}
\subsection{The Environment and the Hidden State}
We consider a continuous-time dynamic inspection game between a principal and a continuum of agents. Time is continuous and indexed by $t \ge 0$. The environment is characterized by an unobservable, time-varying state $\omega_t \in \{H, L\}$. The state evolves according to a continuous-time Markov chain, transitioning from $L$ to $H$ at a Poisson rate $\rho_L > 0$, and from $H$ to $L$ at a Poisson rate $\rho_H > 0$. 

\begin{center}
\begin{tikzpicture}[
    node distance = 79mm and 47mm,
every edge/.style = {draw, -{Stealth[scale=1.2]}, bend left=15},
every edge quotes/.append style = {auto, inner sep=2pt, font=\footnotesize}]
\node (n1)  [state] {$H$};
\node (n2)  [state,right=of n1]   {$L$};
\path   (n1)    edge ["$\rho_H$"] (n2)
        (n2)    edge ["$\rho_L$"] (n1);
\end{tikzpicture}
\end{center}

The unobservable state $\omega_t$ represents a shifting, systemic vulnerability that periodically generates opportunities for infractions. In \textit{financial auditing}, this could be a newly invented tax-evasion scheme or accounting loophole: the high-risk state $H$ indicates the scheme is prevalent and available to agents, while the low-risk state $L$ implies it has been regulated away. Alternatively, in \textit{cybersecurity}, $\omega_t$ might represent a zero-day exploit or ransomware strain, where $H$ implies the exploit is actively circulating and creating opportunities for network compromise, while $L$ indicates the threat has been patched or has naturally died out. In both contexts, the hidden state captures an invisible, fluctuating environment that dictates when strategic agents possess the opportunity to offend.

\subsection{Infraction Opportunities and Agent Behavior}
Opportunities to exploit this vulnerability arise according to a Poisson process with rate $\lambda > 0$ only when the state is $H$; in state $L$, the arrival rate is zero. Crucially, we assume these opportunities are drawn from a large population of agents, such that each opportunity is received by a different, randomly selected agent. 

When an opportunity arises at time $t$, the selected agent must decide whether to commit an infraction. We denote by $x_t \in [0,1]$ the probability that the agent commits the infraction. Because the population is large and any given agent receives an opportunity only rarely, the acting agent does not internalize how their individual choice affects the principal's future beliefs or long-run monitoring policy. Consequently, the agent acts myopically, optimizing their decision strictly against the principal's instantaneous monitoring intensity at time $t$. For the initial analysis in Section 3, we will treat the infraction rate $x$ as an exogenous parameter to isolate the principal's learning problem. In Section 4, we will fully endogenize $x_t$ as a strategic best response.

\subsection{The Principal's Monitoring Policy and Payoffs}
At each instant $t$, the principal chooses a monitoring intensity $y_t \in [0,1]$. Maintaining this monitoring intensity imposes a flow cost of $c \cdot y_t$ on the principal, where $c > 0$ represents the marginal cost of deploying audit resources. If an opportunity arises, the agent commits an infraction, and the principal is actively monitoring, the infraction is detected and halted with probability $y_t$. The principal incurs a fixed loss $\ell > 0$ for every infraction that goes undetected, which we normalize to 1 without loss of generality.

Because the true state $\omega_t$ is hidden, the principal's policy must be measurable with respect to her information. Let $\mathcal{H}_t$ denote the history of the game up to time $t$, which consists of the realized arrival times of all detected infractions prior to $t$.  A monitoring policy is a process $y = \{y_t\}_{t \ge 0}$ that is adapted to the filtration generated by $\mathcal{H}_t$ and admits well-defined belief dynamics.\footnote{As will be seen, the belief dynamics are characterized by an ordinary differential equation. The admissibility requirement simply ensures that the solution to this equation is well-defined. In our context, this requirement has an effect in only one circumstance: if a belief $p$ attracts drift from both above and below, $y(p)$ must be chosen at an interior level to ensure that the drift at $p$ vanishes. This prevents the belief process from oscillating infinitely often around $p$. See further discussion in the remark at the end of this subsection.} The principal's objective, under a prior belief $p_0 = p$ that $\omega_0 = H$, is to choose an admissible monitoring policy to minimize the expected discounted loss:

\begin{equation} \label{eq:loss}
L(p): = \min_{\{y_t\}_{t\ge 0}}  \mathbb{E}  \left[ \int_0^\infty e^{-rt} \left( c y_t + \lambda x_t (1 - y_t) \mathbf{1}_{\{\omega_t = H\}} \right) dt    \bigg|  p_0 = p   \right],
\end{equation} where $r>0$ is the discount rate, and the expectation is taken over the stochastic evolution of the hidden state $\omega_t$ and the history of detected infractions that determines $\{y_t\}_{t\ge 0}$.

\subsection{Endogenous Datafication and the Fixed-State Benchmark}
The central friction in this environment is that learning is strictly endogenous: the principal only generates data when she actively monitors and an infraction actually occurs. If the principal exerts monitoring intensity $y_t$ and the agent offends with probability $x_t$, a detection occurs at a Poisson arrival rate of $\lambda x_t y_t$ conditional on the state being $H$. Because opportunities only arise in state $H$, a detection is perfectly revealing. If an infraction is detected at time $t$, the principal's posterior belief that the state is $H$, denoted by $p_t$, immediately jumps to 1.

In the absence of a detection, the principal continues to update her belief based on no news. Although no news pushes the belief downward, the overall drift may still be upward when the current belief is sufficiently low, due to the underlying Markov state transitions. By standard Bayesian updating, the evolution of the belief $p_t$ between detections is governed by the deterministic differential equation $\dot{p}_t = f(p_t, y_t)$, where the drift function is defined as:\footnote{To derive this drift, consider a small time interval $dt$. Suppose no detection occurs in $[t, t+dt)$. The probability of this event is $1 - \lambda x_t y_t dt$ if the state is $H$, and $1$ if the state is $L$. By Bayes' rule, the posterior belief at $t+dt$ is 
$p_{t+dt} = \frac{p_t (1 - \rho_H dt)(1 - \lambda x_t y_t dt) + (1 - p_t) \rho_L dt}{p_t (1 - \lambda x_t y_t dt) + 1 - p_t} + o(dt)$. 
Rearranging terms, dividing by $dt$, and taking the limit as $dt \to 0$ yields the differential equation $\dot{p}_t = \rho_L(1 - p_t) - \rho_H p_t - p_t(1 - p_t)\lambda x_t y_t$.}
\begin{equation} \label{eq:drift}
f(p, y) = \rho_L(1 - p) - \rho_H p - p(1 - p)\lambda x y.
\end{equation}

To appreciate the role of the changing environment, it is instructive to define two natural bounds on the belief. Let $\p_0= \rho_L / (\rho_L + \rho_H)$ denote the stationary probability of the high-risk state. Equivalently, $\pi_0$ is the unique belief satisfying $f(\pi_0,0)=0$. Thus, if the principal stops monitoring entirely ($y_t = 0$), the learning term vanishes, and the belief simply follows the natural rate of the Markov chain, moving upward when $p_t<\pi_0$ and downward when $p_t>\pi_0$. Conversely, let $\pi_1$ be the belief where $f(\pi_1, 1) = 0$. If the principal always chooses full enforcement ($y_t = 1$), her belief drifts down if it is above $\pi_1$ and drifts up if it is below $\pi_1$, as long as no infraction is detected. Naturally, $\pi_1$ is strictly lower than the stationary probability $\pi_0$. 

\begin{figure}[h]
    \begin{center}
    \begin{tikzpicture}[baseline=-2pt,x=2cm,y=2cm]
    \draw[postaction={decorate},
    arrowmark={0.13}{},
    arrowmark={0.22}{},
    arrowmark={0.28}{}, 	arrowmark={0.33}{}, revarrowmark={0.695}{}
    ,revarrowmark={0.73}{},
    revarrowmark={0.8}{},
    revarrowmark={0.9}{}] 
    (3,0)  node [below]   {$0$}  --  (4, 0)     node [above]   {\footnotesize \ykk{$ f(p,y)>0$}}
    -- (5,0)     node [below]   {{$\p_1$}}  -- (6,0)  node [above] {\footnotesize     \ykk{$f(p,y)\lesseqgtr 0$} }  (6,0)  node [below]   {\footnotesize \ykk{depends on $y$}}    --  (7,0) node [below]   {$\p_0$}  --  (8, 0)     node [above]   {\footnotesize \ykk{$ f(p,y)<0$}} 
    -- (9,0)  node [below]  {1} ; \fill  (3,0)  circle (1.5pt);
    \fill   (5,0)  circle (1.8pt);  \fill   (7,0)  circle (1.8pt);
    \fill   (9,0)  circle (1.5pt);
    \end{tikzpicture}
    \end{center}
    \caption{Principal's belief updating given no detection.\label{fig:belief-updating}}
\end{figure}

Figure \ref{fig:belief-updating} illustrates these belief dynamics. Without detection, the belief drifts up if $p_t < \pi_1$ and down if $p_t > \pi_0$. The belief's evolution within the interval $(\pi_1, \pi_0)$ depends heavily on the principal's active monitoring choice $y_t$. 

This changing-state framework represents a sharp departure from the classic fixed-state bandit model of \cite{keller:05}. If the underlying state were permanently fixed ($\rho_L = \rho_H = 0$), the drift equation would reduce to $-p(1-p)\lambda x y$. In such a world, a sufficiently long period without detections causes the belief $p_t$ to decline monotonically toward zero. Eventually, as long as the monitoring continues, the principal becomes confident the state is safe, ceases monitoring entirely, and uncertainty is permanently resolved. 
   In our model, however, the state transitions autonomously. The principal can never permanently rule out the high-risk state because the belief is bounded below by $\pi_1$. This interplay between a constantly shifting underlying reality and the endogenous generation of data forms the core of the dynamic monitoring challenge.

\paragraph{Admissibility.} We focus on Markovian strategies where the principal's action depends solely on her current belief $p$. This is without loss of generality for exogenous infractions and natural for endogenous infractions, allowing us to suppress time dependence. However, the belief dynamics in equation \eqref{eq:drift} must be well-defined, imposing an admissibility restriction on the strategy. Essentially, if a belief $p$ attracts drifts from both above and below, it must be a stationary point, implying $y(p) = z(p)$, where $z(p)$ satisfies $f(p, z(p)) = 0$. This ensures the belief process doesn't oscillate arbitrarily frequently around $p$. While admissibility may seem technical, it captures a substantive property: in a discrete-time approximation, $z(p)$ represents the fraction of time the principal chooses full monitoring (as opposed to no monitoring) in the ``neighborhood'' of $p$.  The interior solution $z(\hat{p})$ in the continuous-time model corresponds to this relative fraction of time in the limit as the time length approaches zero.

\subsection{Alternative Monitoring Regimes}

To evaluate the value of endogenous datafication, we define three alternative regimes regarding the principal's use of data for monitoring decisions. These regimes serve as benchmarks to isolate the role of dynamically optimal learning.

\paragraph{Static Policy (SP):}  The principal does not update her belief based on historical detection data, resulting in a constant monitoring policy dependent solely on the stationary prior, $\pi_0$. We assume this prior matches the ground truth. The optimal static policy solves $\min_{y\in [0,1]} \pi_0 \lambda x(1-y) + cy$, which yields:
\begin{equation} \label{eq:npe}
 \ynp(p) = \left\{\begin{array}{ll}
1 & \text{for } \p_0> \pm\\
0 & \text{for }  \p_0< \pm,\\
\end{array}\right.
\end{equation}
where $\hat{p}_M := c/(\lambda x)$ is the myopic cutoff. That is, the principal enforces fully if and only if the expected long-run harm from an infraction exceeds the monitoring cost. 

\paragraph{Myopic Policy (MP):} In this benchmark, the principal updates her posterior belief $p_t$ accurately based on the history of detections according to \eqref{eq:drift}, but acts myopically, disregarding the informational option value of monitoring. The myopic policy solves $\min_{y\in [0,1]} p \lambda x(1-y) + cy$ for the updated posterior $p$, yielding:
\begin{equation} \label{eq:pme}
 \ypm(p) = \left\{\begin{array}{ll}
1 & \text{for } p> \pm\\
0 & \text{for } p < \pm.\\
\end{array}\right.
\end{equation}
While MP utilizes the predictive power of data, it fundamentally neglects the impact of current monitoring decisions on future information generation.

\paragraph{Optimal Policy (OP):} Our central focus is the dynamically optimal policy where the principal not only updates her belief $p_t$ accurately but also explicitly internalizes the informational value of monitoring into her decision-making to achieve long-term optimality. The formal derivation of this policy is presented in \Cref{sec:exog}.

\section{Analysis of Exogenous Infractions} \label{sec:exog}

This section examines the scenario of exogenous infractions, where the infraction rate $x$ is fixed within the interval $(0, 1]$. We begin by characterizing the optimal policy (OP) and then compare its outcomes with those of the static policy (SP) and the myopic policy (MP) benchmarks introduced in \Cref{sec:model}.

\subsection{The Optimal Policy}

This section focuses on the optimal policy (OP) scenario, where the principal dynamically optimizes monitoring decisions, considering both current predictions of the state and the long-term informational option value of monitoring. The problem is formulated as a Markov Decision Problem (MDP) with the belief $p$ as the state variable. The optimal policy, denoted by $p \rightarrow y(p)$, is time-independent due to the Markovian nature of the problem.

The analysis employs dynamic programming, focusing on the value function that maximizes the net present value of mitigated infractions minus monitoring costs.\footnote{\label{foot:equivalence} This value function, which represents the maximized net present value of mitigated infractions minus monitoring costs, is defined as: 
$$ V (p) :=  \max_{\{y_t\}_{t\ge 0}}  \mathbb{E} \left[  \int_0^\infty(p_t \lambda x -c) y_t  e^{-rt} dt \bigg| p_0 =p   \right].$$ 
\Cref{lem:equivalence} in \Cref{sec:omitted proofs} of the Online Appendix shows that any policy $y_t$ solves this problem if and only if it solves the minimization problem in \cref{eq:loss}.}    
The value function satisfies the Hamilton-Jacobi-Bellman (HJB) equation:
\begin{align}
  rV(p)  = & \max_{y}   \left(  p\lambda  x -c  \right) y   +p\lambda  x y [V(1) - V(p)] +f (p,y) V'(p).\label{HJB}
\end{align}
The HJB equation balances the cost of discounting future value (the LHS) with the immediate benefits of monitoring $(p\lambda  x -c)y$ and the value appreciation from Bayesian belief updates. The condition resembles standard bandit frameworks, but the changing state implies that learning remains essential even in the long run. In particular, the value function is not pinned down even for $p=0$ or $p=1$, since reaching these extreme states does not mean a permanent resolution of uncertainty. Instead, the value function at any belief depends on the principal's actions at other beliefs, leading to a fixed-point characterization. This presents a new analytical challenge absent in bandit models with a non-changing state.

The optimal policy, derived from the HJB equation, takes a cutoff form: for some $\hat p\in [0,1]$,
\begin{equation} \label{eq:cutoff}
 y(p) = \left\{\begin{array}{ll}
1, & \text{for } p> \hat p\\
0, & \text{for }  p< \hat p.\\
\end{array}\right.
\end{equation} 
For this structure to hold, the principal must strictly prefer to raise enforcement $y$ as the belief $p$ rises---a standard single-crossing property. However, our model's changing state introduces a novel methodological twist. In standard models (e.g., KRC), this single-crossing condition is slack at the cutoff because abandonment is permanent. Here, because the optimal policy must continually hedge against the risk of the state transitioning back to high risk, the single-crossing condition binds strictly at $\hat{p}$. We utilize this uniquely binding condition as the key analytical tool to pin down the exact value of the cutoff, which plays a central role in characterizing the partial monitoring---i.e., $y(\hat{p}) \in (0,1)$---seen in \ref{item:case-3} of \Cref{thm:main}.

The following theorem characterizes OP. 
\begin{thm} \label{thm:main}  For each $c>0$, there exist $ \lh>\frac{c}{\pi_0}>\ll>c$  such that the optimal monitoring policy denoted $y^*(p)$ is of the form in \eqref{eq:cutoff} where the cutoff is
\begin{enumerate}[label=\textsf{Case \arabic*:},ref=\textsf{Case \arabic*},itemindent=0.5cm]
	\item  $\p_0 \le   \hat p  < \pm$ if $\l \g \le \ll $ (``low'' infraction rate);
	\label{item:case-1}
	
	\item $\hat p  = \pm \le \p_1$ if  $ \l \g \ge \lh$ (``high'' infraction rate);
	\label{item:case-2}
	
	\item  $\hat p\in (\p_1,\p_0)$ and $\hat p < \pm$ if $\l \g \in (\ll, \lh )$ (``intermediate'' infraction rate).
	\label{item:case-3}
\end{enumerate}
The policy at the cutoff belief, $y^* (\hat p)$, is arbitrary in \ref{item:case-1} and \ref{item:case-2},    but in \ref{item:case-3}, $y^* (\hat p)=z(\hat p)  \in (0,1)$, where $z(p)$ satisfies $f(p, z(p))=0$.\footnote{In the knife-edge cases where $\hat p=\pi_0$ or $\hat p=\pi_1$, admissibility pins down the cutoff action: $y^*(\hat p)=0$ when $\hat p=\pi_0$, and $y^*(\hat p)=1$ when $\hat p=\pi_1$.} Moreover, $\ll $ and $\lh$ are increasing in $c$.  Finally,  $\hat p$ is continuously increasing in $c$ and decreasing in $\lambda x$, with $\hat p = \p_0$ if $\l x = \ll$ and $\hat p =\p_1$ if $\l x = \lh$.
\end{thm}
 \begin{proof}
     See \Cref{sec:proof of main}.
 \end{proof}

As stated in \Cref{thm:main}, there are three cases, depending on where the optimal cutoff $\hat p$ lies relative to $\p_1$ and $\p_0$, as depicted previously in \Cref{fig:belief-updating}.

\paragraph{\ref{item:case-1}: Low infraction rates.}  
In this scenario, the infraction rate $\lambda x$ is so low that the optimal cutoff $\hat{p}$ exceeds $\pi_0$. The cutoff is set lower than the myopic level $\pm:=c/(\g\l)$ to maintain an optimal tradeoff between exploitation and exploration. See \Cref{fig:low-infraction}. While it would be optimal to monitor fully if and only if $p>\pm$ from a pure exploitation perspective, there is a long-term benefit in exploring even when $p$ is slightly below the myopic cutoff. Hence, if the initial prior is above $\hat{p}$, the principal starts by monitoring fully. If an infraction is detected, the belief jumps to 1; otherwise, it drifts down. The prospect of irreversible abandonment causes the principal to explore below the myopic cutoff, but the changing world dictates that the principal will eventually cease monitoring as the belief stabilizes at $\p_0$.

\begin{figure}[h]
    \centering
  \begin{tikzpicture}
\begin{axis} 
[hide axis,
width=12.5cm,
height=6.5cm,
ymin =-0.25,
ymax =0.25,
xmax=1.1,
xmin=-0.1,
xlabel style={color=black},
ylabel style={align=center,rotate=-90,color=blue!50!cyan},
x tick label style={
	/pgf/number format/assume math mode, font=\sf\scriptsize},
y tick label style={
	/pgf/number format/assume math mode, font=\sf\scriptsize},
]
\draw [line width=0.4pt]  (axis cs:0,-0.14)  --  (axis cs:0.782,-0.14) node [pos=0, left] {0};
\draw [dotted, line width=0.8pt]  (axis cs:0,0.15)  --  (axis cs:0.782,0.15) node [pos=0, left] {1};
\draw [line width=1pt]  (axis cs:0.782,0.15)  --  (axis cs:1,0.15) node [right] {$y (p)$};
\draw[dotted, line width=0.8]  (axis cs: 0.782, -0.14) -- (axis cs: 0.782, 0.145);
\draw [-latex]  (axis cs:0,-0.14)  --  (axis cs:0,0.24); 
 \node at (axis cs: -0.04,0.22)  {$y$};
\draw[-latex,postaction={decorate},
arrowmark={0.23}{},
arrowmark={0.4}{},
arrowmark={0.52}{},
arrowmark={0.565}{},    revarrowmark={0.603}{},
revarrowmark={0.69}{}, revarrowmark={0.81}{}
] (axis cs: 0,-0.14)-- (axis cs:  1.05,-0.14) node [right] {\normalsize $p$} ;
\path[-{>[scale=1.3]}, >=Stealth, color=red]
(axis cs:  0.785,-0.135) edge[bend left=30] (axis cs:  1,-0.135);
\path[->, >=Stealth, color=red]
(axis cs:  0.88,-0.135) edge[bend left=30] (axis cs:  1,-0.135);
\addplot [draw=none,fill=blue, fill opacity =0.2]coordinates {    (0.782,-0.14)    (0.782,0.15)  (1,0.15)  (1,-0.14) };
\fill   (axis cs: 0.35,-0.14)  circle (1.5pt)  node [below=0.1cm] {$\p_1$};
\fill[color=DarkRed]   (axis cs: 0.6,-0.14)  circle (2pt) node [below=0.1cm] {$\p_0$};
\fill (axis cs: 0.9,-0.14)  circle (1.5pt) node [below=0.1cm] {$\hp$};
\fill   (axis cs: 0.782,-0.14)   circle (1.5pt) node [below=0.1cm] {\ykk{$\hat p$}};
\fill   (axis cs: 1,-0.14)  circle (1.5pt) node [below=0.1cm] { 1};
\end{axis}
\end{tikzpicture}
    \caption{Low Infraction Rates Case}
    \label{fig:low-infraction}
\end{figure}
 
\paragraph{\ref{item:case-2}: High infraction rates.}

Here, the infraction rate $\lambda x$ is high enough that the optimal cutoff $\hat{p}$ falls below $\pi_1$, the lowest possible stationary belief. Hence, the principal monitors fully for any long-run belief above $\pi_1$. The belief eventually cycles within the region $[\pi_1, 1]$, jumping to 1 upon detection and drifting down to $\pi_1$ otherwise (see \Cref{fig:high-infraction}). Monitoring is always full and never lets up, eliminating the need to trade off exploitation against exploration, resulting in $\hat{p} = \hat{p}_M$. This differs from the fixed-state model, where irreversible abandonment of exploration might occur. The changing world motivates persistent vigilance, ensuring monitoring even in the prolonged absence of infractions.

\begin{figure}[h]
    \centering
   \begin{tikzpicture}
 \begin{axis} [hide axis,
 width=12.5cm,
 height=6.5cm,
 ymin =-0.25,
 ymax =0.25,
 xmax=1.1,
 xmin=-0.1,
 xlabel={$p$},
 xlabel style={color=black},
 ylabel style={align=center,rotate=-90,color=blue!50!cyan},
 x tick label style={
 	/pgf/number format/assume math mode, font=\sf\scriptsize},
 y tick label style={
 	/pgf/number format/assume math mode, font=\sf\scriptsize},
 ]
 \draw[dotted, line width=0.8]  (axis cs: 0.3, -0.15) -- (axis cs: 0.3, 0.15);
 \node at (axis cs: 0.15,0.45)  {\textcolor{northwestern-purple}{$\lambda \ge  \overline{\lambda}$}};
\draw[-latex,postaction={decorate},
 arrowmark={0.18}{},
 arrowmark={0.32}{},
 arrowmark={0.42}{},
 arrowmark={0.47}{},
 revarrowmark={0.507}{},    revarrowmark={0.57}{},
 revarrowmark={0.67}{}, revarrowmark={0.8}{}
 ] (axis cs: 0,-0.15)-- (axis cs:  1.05,-0.15)  node [right] { $p$} ;
\path[-{>[scale=1.3]}, >=Stealth, color=red]
 (axis cs:  0.3, -0.14) edge[bend left=20] (axis cs:  1,-0.14);
\path[->, >=Stealth,color=red]
 (axis cs:  0.55,-0.14) edge[bend left=20] (axis cs:  1,-0.14);
\path[->, >=Stealth, color=red]
 (axis cs:  0.72,-0.14) edge[bend left=20] (axis cs:  1,-0.14);
\path[->, >=Stealth, color=red]
 (axis cs:  0.87,-0.14) edge[bend left=20] (axis cs:  1,-0.14);
\draw [-latex]  (axis cs:0,-0.15)  --  (axis cs:0,0.24);
\addplot [draw=none,fill=blue, fill opacity =0.2]coordinates {    (0.3,-0.15)    (0.3,0.15)  (1,0.15)  (1,-0.15) };
\draw [line width=0.4pt]  (axis cs:0,-0.15)  --  (axis cs:0.3,-0.15) node [pos=0, left] {0};
\draw [dotted, line width=0.8pt]  (axis cs:0,0.15)  --  (axis cs:0.3,0.15) node [pos=0, left] {1};
\draw [line width=1pt]  (axis cs:0.3,0.15)  --  (axis cs:1,0.15) node [right] {$y (p)$};
\node at (axis cs: -0.04,0.22)  {$y$};
 \fill   (axis cs: 0.3,-0.15)  circle (1.5pt)  node [below=0.15cm] {\ykk{$\hat p=\hp$}};
\fill[color=DarkRed] (axis cs: 0.5,-0.15)  circle (2pt) node [below =0.15cm] { $\p_1$};
\fill   (axis cs: 0.75,-0.15)   circle (1.5pt) node [below=0.15cm] {$\p_0$}  ;
\fill   (axis cs: 1,-0.15)  circle (1.5pt) node [below=0.15cm] {1};
\end{axis}
 \end{tikzpicture}
    \caption{High Infraction Rates Case}
    \label{fig:high-infraction}
\end{figure}
 
\paragraph{\ref{item:case-3}: Intermediate infraction rates.}  

In this case, the monitoring cutoff lies between $\pi_1$ and $\pi_0$. For beliefs above $\hat{p}$, the principal monitors fully. If no infraction is detected, the belief drifts down, potentially reaching $\hat{p}$. At $\hat{p}$, partial monitoring at an interior level $y = z(\hat{p}) \in (0,1)$ is employed, freezing the belief at $\hat{p}$ until a detection triggers a jump to 1. See \Cref{fig:intermediate-infraction}.

\begin{figure}[htb]
    \centering
    \begin{tikzpicture}
\begin{axis} [hide axis,
width=12.5cm,
height=6cm,
ymin =-0.25,
ymax =0.25,
xmax=1.1,
xmin=-0.1,
xlabel={$p$},
xlabel style={color=black},
ylabel style={align=center,rotate=-90,color=blue!50!cyan},
x tick label style={
	/pgf/number format/assume math mode, font=\sf\scriptsize},
y tick label style={
	/pgf/number format/assume math mode, font=\sf\scriptsize},
]
\addplot [draw=none,fill=blue, fill opacity =0.2]coordinates {    (0.611,-0.14)    (0.611,0.15)  (1,0.15)  (1,-0.14) };
\node at (axis cs: 0.15,0.32)  {\textcolor{northwestern-purple}{$\underline{\lambda} < \lambda <  \overline{\lambda}$}};
\draw[-latex, postaction={decorate},
arrowmark={0.2}{},
arrowmark={0.37}{},
arrowmark={0.5}{},
arrowmark={0.575}{},
revarrowmark={0.613}{},    revarrowmark={0.665}{},
revarrowmark={0.76}{}, revarrowmark={0.9}{}
] (axis cs: 0,-0.14)-- (axis cs:  1.05,-0.14)   node [right] {\normalsize $p$};
\path[-{>[scale=1.3]}, >=Stealth, color=red]
(axis cs:  0.611,-0.135) edge[bend left=30] (axis cs:  1,-0.135);
\path[->, >=Stealth, color=red]
(axis cs:  0.72,-0.135) edge[bend left=30] (axis cs:  1,-0.135);
\path[->, >=Stealth, color=red]
(axis cs:  0.82,-0.135) edge[bend left=30] (axis cs:  1,-0.135);
\draw [line width=0.4pt]  (axis cs:0,-0.14)  --  (axis cs:0.621,-0.14) node [pos=0, left] {0};
\draw [dotted, line width=0.8pt]  (axis cs:0,0.15)  --  (axis cs:0.605,0.15) node [pos=0, left] {1};
\draw [line width=1pt,{Circle[open]}-]  (axis cs:0.6,0.15)  --  (axis cs:1,0.15) node [right] {$y (p)$};
\draw[dotted, line width=0.8]  (axis cs: 0.611, -0.14) -- (axis cs: 0.611, 0.145);
\draw[dotted, line width=0.8]  (axis cs: 0, 0.06) -- (axis cs: 0.605, 0.06) node [pos=0, left] {$y (\hat p)$};
\draw [-latex]  (axis cs:0,-0.14)  --  (axis cs:0,0.24);
\fill   (axis cs: 0.35,-0.14)  circle (1.5pt) node [below=0.15cm] {$\p_1$};
\fill[color=DarkRed]   (axis cs: 0.611,-0.14)  circle (2pt) node [below=0.1cm, xshift=0cm, text width=0.25cm] {$\hat p$};
\fill   (axis cs: 0.83,-0.14)   circle (1.5pt) node [below=0.15cm] {$\p_0$};
\fill   (axis cs: 1,-0.14)  circle (1.5pt) node [below=0.15cm] {1};
\fill   (axis cs: 0.611,0.06)  circle (2pt);
\end{axis}
\end{tikzpicture}
    \caption{Intermediate Infraction Rates Case}
    \label{fig:intermediate-infraction}
\end{figure}

The system spends a significant amount of time at $\hat{p}$, and monitoring alternates between full and partial levels. This pattern deviates from the standard bandit model, where action eventually becomes either full or nonexistent.  The changing world leads to this belief-freezing behavior. The optimal policy calibrates this feature to optimally ``hedge”---to balance the increasing likelihood of the safe state against the possibility that the environment can become high-risk again. 
Because partial monitoring suppresses exploration, the standard exploitation-exploration logic dictates pushing the cutoff below the myopic level ($\hat p<\pm$). 

\begin{rem}[{\bf Long-run distribution}] 
One can characterize the stationary distribution of the principal's belief and the monitoring level under OP and MP. If the cutoff lies within $(\p_1, \p_0)$, monitoring has a two-point distribution with support $\{z(\hat p), 1\}$ under either policy. The long-run proportion of time the monitoring takes each level can be computed precisely; we do this in \Cref{sec:criminals belief formation}. Comparing the long-run monitoring levels in states $H$ and $L$ reveals the potential benefit of data-driven monitoring.   Under SP, the principal chooses the same monitoring level in both states, but under either MP or OP, the principal chooses, on average, a strictly higher monitoring level in $\omega=H$ than in $\omega=L$.
\end{rem}

\subsection{Comparison of Regimes}

This section compares the optimal policy (OP) with the static (SP) and myopic (MP) benchmarks, focusing on scenarios where OP strictly outperforms them. We are specifically interested in their long-run performance—evaluating the principal's expected discounted loss once the system reaches a steady state and the posterior belief $p$ enters its long-run support.

Because the policies induce varying degrees of exploration and learning, their long-run supports differ. However, they are structurally nested: the support under OP contains the support under MP, which in turn contains the degenerate support under SP, $\{\pi_0\}$.\footnote{In \ref{item:case-2} and \ref{item:case-3}, the support under OP is $[\hat p, 1]$ and the support under MP is $[\hat p_M, 1]$, where $\hat p\le \hat p_M.$ In \ref{item:case-1}, all three regimes degenerate to the exact same singleton support, $\{\pi_0\}$.} Because expected discounted payoffs depend heavily on the starting belief, meaningful welfare comparisons require evaluating losses from a common baseline: $p_0 = \pi_0$. This represents the unique intersection of the long-run supports across all three regimes. Anchoring the evaluation at $\pi_0$ does not render other belief states irrelevant; rather, the divergence in long-run performance is driven precisely by the possibility of  the long-run belief visiting these other states across the different regimes.


We initially examine cases where data-driven monitoring, even when optimized, offers no long-run advantage.

\begin{prop}\label{prop:comparison-exog0} In \ref{item:case-1} and \ref{item:case-2},  the monitoring policy is identical in the long run across all three regimes.
\end{prop}

In scenarios with low or high infraction rates, the OP, SP, and MP converge to the same actions in the long run because the prediction problem is relatively simple, and past data does not significantly improve decision-making. The more interesting scenario is \ref{item:case-3}, where prediction is valuable. There are two possibilities:

In \ref{item:case-3}(a), $\hat{p}_M\ge \pi_0$. Under MP, the belief eventually converges to $\pi_0$, resulting in the cessation of monitoring and exploration, mirroring the outcome under SP. Consequently, both SP and MP under-monitor relative to OP, which prescribes positive monitoring even in the long run.

In \ref{item:case-3}(b), $\hat{p}_M<\pi_0$. In this case, SP prescribes full monitoring, whereas MP alternates between full monitoring (when $p>\hat{p}_M$) and partial monitoring (when $p=\hat{p}_M$). The comparison reveals that MP under-monitors relative to OP, as the optimal cutoff $\hat{p}$ is set below $\hat{p}_M$ to encourage more exploration (see \Cref{fig:MP}). In contrast, SP results in over-monitoring compared to OP.

\begin{figure}[htb]
    \centering
    \begin{tikzpicture}
\begin{axis} [hide axis,
width=17cm,
height=6cm,
ymin =-0.25,
ymax =0.25,
xmax=1.1,
xmin=-0.1,
xlabel={$p$},
xlabel style={color=black},
ylabel style={align=center,rotate=-90,color=blue!50!cyan},
x tick label style={
	/pgf/number format/assume math mode, font=\sf\scriptsize},
y tick label style={
	/pgf/number format/assume math mode, font=\sf\scriptsize},
]
\addplot [draw=none,fill=blue, fill opacity =0.2]coordinates {    (0.611,-0.14)    (0.611,0.15)  (1,0.15)  (1,-0.14) };
\node at (axis cs: 0.15,0.32)  {\textcolor{northwestern-purple}{$\underline{\lambda} < \lambda <  \overline{\lambda}$}};
\draw[-latex, postaction={decorate},
arrowmark={0.2}{},
arrowmark={0.37}{},
arrowmark={0.5}{},
arrowmark={0.575}{},
revarrowmark={0.613}{},    revarrowmark={0.665}{},
revarrowmark={0.76}{}, revarrowmark={0.9}{}
] (axis cs: 0,-0.14)-- (axis cs:  1.05,-0.14)   node [right] {\normalsize $p$};
\path[-{>[scale=1.3]}, >=Stealth, color=red]
(axis cs:  0.611,-0.135) edge[bend left=30] (axis cs:  1,-0.135);
\path[->, >=Stealth, color=red]
(axis cs:  0.72,-0.135) edge[bend left=30] (axis cs:  1,-0.135);
\path[->, >=Stealth, color=red]
(axis cs:  0.82,-0.135) edge[bend left=30] (axis cs:  1,-0.135);
\draw [line width=0.4pt]  (axis cs:0,-0.14)  --  (axis cs:0.621,-0.14) node [pos=0, left] {0};
\draw [dotted, line width=0.8pt]  (axis cs:0,0.15)  --  (axis cs:0.605,0.15) node [pos=0, left] {1};
\draw [line width=1pt,{Circle[open]}-]  (axis cs:0.6,0.15)  --  (axis cs:1,0.15) node [right] {$\ysp (p)$};
\draw[dotted, line width=0.8]  (axis cs: 0.611, -0.14) -- (axis cs: 0.611, 0.145);
\draw[dotted, line width=0.8]  (axis cs: 0, 0.03) -- (axis cs: 0.605, 0.03) node [pos=0, left] {$z (\pm)$};
\draw[dotted, line width=0.8]  (axis cs: 0, 0.08) -- (axis cs: 0.45, 0.08) node [pos=0, left] {$z (\hat p)$};
\draw [-latex]  (axis cs:0,-0.14)  --  (axis cs:0,0.24);
\fill   (axis cs: 0.35,-0.14)  circle (1.5pt) node [below=0.15cm] {$\p_1$};
\fill[color=DarkRed]   (axis cs: 0.611,-0.14)  circle (2pt) node [below=0.1cm, xshift=0cm, text width=0.25cm] {$\pm$};
\fill   (axis cs: 0.83,-0.14)   circle (1.75pt) node [below=0.15cm] {$\p_0$};
\fill   (axis cs: 1,-0.14)  circle (1.75pt) node [below=0.15cm] {1};
\fill   (axis cs: 0.611,0.03)  circle (2.5pt);

\draw[dotted, line width=0.8]  (axis cs: 0.45, -0.14) -- (axis cs: 0.45, 0.08);
\fill [color=DarkRed]  (axis cs: 0.45, -0.14)  circle (2pt) node [below=0.1cm, xshift=0cm, text width=0.25cm] {$\hat p$};
\fill   (axis cs: 0.45,0.08)  circle (1.8pt);

\end{axis}
\end{tikzpicture}
    \caption{Monitoring Policy under MP: \ref{item:case-3}(b)}
    \label{fig:MP}
\end{figure}

\begin{prop} \label{prop:comparison-exog}  In  \ref{item:case-3}(a), both SP and MP lead to no monitoring, which is too little compared with OP. In \ref{item:case-3}(b), SP leads to full, and thus excessive, monitoring, whereas MP leads to insufficient monitoring, compared with OP.
\end{prop}

\paragraph{Payoff comparisons.}
How does MP compare with SP in total expected loss? Surprisingly, while MP prescribes distinct monitoring levels based on predictions, the two regimes entail the exact same losses in the long run:

\begin{prop}  [Payoff comparison] \label{prop:equiv} The expected loss from MP is always identical to that from SP in the long run. The expected loss is strictly lower under OP in \ref{item:case-3}.
\end{prop}
\begin{proof} It suffices to show the equivalence for \ref{item:case-3}(b), in which $\p_1 < \hat p < \pm < \p_0$. Recall from \Cref{prop:comparison-exog} that SP prescribes full monitoring in this case, so the intertemporal loss from SP is $c/r$. Meanwhile, MP implements monitoring that varies with $p_t$. To prove the equivalence, it suffices to show that the intertemporal loss from MP is $c/r$ when evaluated at our common long-run baseline $p_0 = \pi_0$, which lies within $[\hat{p}_M, 1]$, the long-run support of $p$ under MP.
 The equivalence follows since the total loss under MP is: for any $p_0 \in [\pm, 1]$, 
\begin{align*} 
  &  \mathbb{E} \left[  \left.
 \int_0^{\infty} ( p_t\l \g (1-y_t) + cy_t)e^{-rt}dt  \right| p_0=\pi_0 \right]\\
=& \int_0^{\infty}\mathbb{E} \Big[   ( p_t\l \g (1-y_t) + cy_t)\Big|
 p_0=\pi_0  \Big]e^{-rt}dt  \\    
=& \int_0^{\infty}\mathbb{E} \Big[     \mathbf{1}_{\{p_t>\pm\}}\cdot c +  \mathbf{1}_{\{p_t=\pm\}} \cdot \left(\pm\l \g (1-z(\pm)) + cz(\pm)\right) \Big|
 p_0=\pi_0  \Big]e^{-rt}dt  \\    
=& \int_0^{\infty}c e^{-rt}dt   
=  c/r,
\end{align*}
where we used the fact that $\ymp(p_t)=\mathbf{1}_{\{p_t>\pm\}} + \mathbf{1}_{\{p_t=\pm\}}\cdot z(\pm)$ and that $\pm= \frac{c}{\l \g}$. The last statement follows from the fact that OP prescribes a different action than MP in \ref{item:case-3}, making its expected loss strictly lower.
\end{proof}

The implication is striking: data-driven prediction has no value if it is used in a myopically optimal manner. MP prescribes no monitoring when $p<\pm$, whereas SP always prescribes full monitoring; and in this case, no monitoring is strictly better than full monitoring. Why doesn't MP then strictly outperform SP? Because no such $p<\pm$ is part of the support of the principal's beliefs. Prediction arises \textit{only} when there is monitoring, but monitoring never occurs when it is undesirable under MP. This makes the prediction obtained under MP redundant.

\paragraph{Passive Learning.} The equivalence between MP and SP rests crucially on the assumption that infractions are detected only through active monitoring. Although this assumption is valid for many unobservable regulatory violations, detection may also arise from passive channels, such as whistleblowing by witnesses, victim reporting, or private parties bringing suits to recover damages. \Cref{sec:passive} extends our model to allow for such a possibility; we assume that the principal detects an infraction with probability $w \in (0,1)$ \textit{passively} even when active monitoring fails to detect one.  Let $\pi_w \in (\pi_1,\pi_0)$ denote the unique belief satisfying $f (\pi_w,w)=0$, i.e.\ the long-run belief under no active monitoring when passive learning arrives at rate $w$.

\begin{prop} \label{prop:passive}
If $\pm   > \p_w$, MP achieves a strictly lower long-run loss for the principal than SP. Otherwise, MP and SP achieve the same long-run loss. 
\end{prop}

\Cref{prop:passive} demonstrates that MP benefits from passive learning, while SP does not. Passive learning enables the principal's belief to drop below the myopic cutoff $\hat{p}_M$, where MP recommends no monitoring—a strategy that is strictly optimal. Put simply, prediction occurs even when not monitoring is the best choice from a myopic perspective. However, as $w\to 0$, MP's superiority over SP vanishes. This equivalence highlights the extent to which the endogeneity of learning limits the value of data-driven policies.

\section{Endogenous Infractions and Deterrence} \label{sec:endog}

By assuming exogenous infraction behavior, the previous section focused on the principal's learning and remedial roles. However, an equally important goal of dynamic monitoring is deterring infractions. To study the deterrence implications of alternative policies, we must endogenize the agent's behavior. 

To this end, we adopt a simple model in which an agent commits an infraction only if they anticipate the monitoring rate to be less than some deterrence threshold $\bar{y}\in(0,1)$. More precisely, an agent offends with probability $x=0$ if $y>\bar{y}$, with probability $x =1$ if $y<\bar{y}$, and with any probability $x \in [0,1]$ if $y =\bar y$. This behavior can be micro-founded: suppose an agent gains a private benefit $b>0$ from an infraction but pays a penalty $\gamma\ge b$ when detected; the assumed behavior emerges with a threshold $\bar y:=b/\gamma$.\footnote{Note that agents are myopic here. This is because there is a continuum of individuals who may receive an opportunity; any effect of an infraction decision by an individual on future monitoring will be irrelevant since the probability of receiving an opportunity in the future is zero.} 

With the endogenous infraction model, the principal's ability to commit to her monitoring policy becomes relevant. If the principal could commit to her policy, she could completely deter infractions by monitoring slightly above $\bar y$. More precisely, a simple full-commitment policy is either to deter infractions fully by monitoring arbitrarily above $\bar y$ or never to monitor, whichever is cheaper. In practice, however, such formal commitment power is often lacking in regulatory contexts. Accordingly, we henceforth assume the principal cannot commit to a future monitoring level $y$, meaning she can only choose her policy in response to the likely infraction rate. Effectively, this transforms the bandit problem into a dynamic game in which the infraction rate and the monitoring policy are chosen as mutual best responses to each other.

As before, we consider three distinct regimes: SP, MP, and OP. Unlike the principal, agents do not access past detection data, meaning the infraction level $x$ is a constant (rather than a function of $p$), except now it is determined endogenously by the agents' equilibrium incentives. (In \Cref{sec:Data-aware Agents}, we will discuss the implications of agents gaining access to the same data as the principal).

Consistent with the exogenous case, we focus on a Markov perfect equilibrium in which the principal's policy depends only on her posterior, the payoff-relevant state. Since agents choose a constant probability $x\in [0,1]$, the principal's best response follows the characterization from the previous section. The only difference is that the agents' equilibrium choice of $x$ will depend on the principal's monitoring policy, and thereby on the monitoring regime. Given the endogeneity of $x$ and the lack of commitment power by the principal, the OP is not guaranteed to be optimal or even to outperform the other two regimes. We begin our analysis with the SP regime.
 
\subsection{Static Policy}
 
In this regime, neither the agents nor the principal update their beliefs based on past history. Hence, the infraction probability $x$ and the monitoring level $y$ are constants, determined solely based on the long-run prior $\p_0$. The outcome then depends on whether $\l$ is large relative to $c$ for monitoring to be worthwhile for the principal.

Suppose first $\lambda \p_0 \le c$. Then, the opportunity arises so rarely that even if the agents choose $x=1$, it is never optimal for the principal to choose any positive $y$. So $(x,y)=(1,0)$ is the only equilibrium. In that equilibrium, the principal's long-run expected loss, evaluated from the baseline belief $p_0 = \pi_0$, equals $\lambda\pi_{0}/r$.

Suppose next $\lambda \p_0> c$. In this case, a classic inspection game ensues. It is well known that there is no pure strategy equilibrium. If agents choose $\g=1$, then the principal will choose $y=1$. But then agents would deviate to $\g=0$. The principal will, in turn, deviate to $y=0$. Hence, the only equilibrium is in mixed strategies. In equilibrium, an agent chooses $\g$ so that 
\begin{align}
    \label{NP-eq} \p_0 \lambda \g=c,
\end{align}
which keeps the principal indifferent in her monitoring. In turn, the principal chooses $y=\bar y$, incentivizing agents to randomize.

The structure of this equilibrium is explained by the principal's lack of commitment power. Unable to commit to a deterring monitoring level, the principal requires infractions to occur on the path to exert monitoring effort. Indeed, the principal's expected loss equals $c/r$ regardless of $\bar y$ since the principal is indifferent to full monitoring. We summarize the results.

\begin{prop}\label{prop:endog-np} SP admits a unique equilibrium in which:
\begin{itemize}
    \item if $\lambda \le \frac{c}{\p_0}$, then $(\gnp,\ynp)=(1,0)$;
    \item if $\lambda > \frac{c}{\p_0}$, then $(\gnp,\ynp)=(\frac{c}{\p_0\l},\bar y)$.
\end{itemize}    
Under SP, the expected loss for the principal, evaluated from the baseline belief $p_0 = \pi_0$, is $\min\{\pi_{0}\lambda,c\}/r$.
\end{prop}

\subsection{Agents' Belief Formation under  Data-driven Policies}
\label{sec:criminals belief formation}

We next turn to MP and OP. Given an equilibrium infraction level $x$, our analysis from \Cref{sec:exog} shows that the principal's best response is a cutoff policy in these two regimes. Fix such a policy with any cutoff $\hat p\in (\p_1,\p_0)$:\footnote{The long-run monitoring will be degenerate otherwise: It is zero if $\hat p\ge\p_0$, and one if $\hat p\le\p_1$.} $y_{\hat p}(p):= \mathbf{1}_{\{p>\hat p\}}+ z(\hat p)\cdot \mathbf{1}_{\{p=\hat p\}}.$  

To understand agents' incentives under such a policy, we must first study their beliefs about the monitoring level $\mathbb{E}_{p}[y_{\hat p}(p)|\omega=H, x]$. The expectation is conditional on state $\omega=H$ (which the agents know upon receiving the opportunity) and the equilibrium infraction level $x$ (which influences the belief about the principal's posterior and, henceforth, the monitoring level). Since the monitoring level depends on the principal's belief $p$, the agent must form a belief about the relative likelihoods of alternative $p$'s the principal may hold. In the long run, this belief is given by the stationary distribution of the principal's belief.\footnote{The formal justification can be given by the logic of ``Poisson arrivals see time averages'' (PASTA); see \cite{wolff1982poisson}. Intuitively, for a Poisson-arriving agent, his belief that $p$ falls into some interval $(p',p'')\subset [\hat p, 1]$ equals the fraction of time $p$ lies in that interval, once proper conditioning is made on the fact that state $\omega=H$.} We thus digress briefly on how the distribution is characterized.

\subsubsection{Stationary distribution of the principal's belief.} 
 
To begin, fix a cutoff policy with an arbitrary cutoff $\hat p\in (\p_1, \p_0)$ and an arbitrary infraction level $x\in (0,1)$. Such a policy induces a Markov process on the belief $p\in [\hat p, 1]$, which admits a unique stationary distribution $\Phi:[\hat p, 1]\to [0,1]$.\footnote{\label{fn:stationary} The process is Markov since once any $p\in [\hat p, 1]$ is reached, the future process is identical and depends only on $p$, regardless of the prior history. It is also positive-recurrent, with the process returning to the same belief in finite expected time. The regenerative nature of the process means that it admits a unique limit distribution forming a unique stationary distribution (see \cite{asmussen2003applied}, p. 170, Theorem 1.2).} \Cref{appendix:criminals belief formation} characterizes the stationary distribution:  

 \begin{lem} \label{lem:stationary} The stationary distribution $\Phi$ for given $x$ and policy $y_{\hat p}$, with $\hat p\in (\p_1, \p_0)$, admits density $\phi (p)$ satisfying:
 \begin{align}
 \phi (p)  = \phi (\hat p) \exp{\left( \int_{\hat p}^{p} r(s) ds \right) }  \label{stationary-belief-1} \end{align}
 for each $p\in (\hat p,1)$, where $r (s): = - \frac{s \l \g + f' (s,1)}{f (s,1)},$ and atom $\Phi (\hat p)$ at belief $\hat p$ satisfying:
 \begin{align}
    \Phi (\hat p) z (\hat p) \hat p \lambda x =  -\phi(\hat p) f(\hat p, 1).
\label{stationary-belief-2}
\end{align}  
\end{lem}
\begin{proof}
    See \Cref{appendix:proof of lemma 2}.
\end{proof}

Intuitively, the posterior belief follows a Poisson process at each $p\in (\hat p, 1]$, leading to a well-defined density in the stationary distribution for that region. Meanwhile, the partially absorbing nature of belief $\hat p$ means that the system spends real time at that belief, explaining the atom in the stationary distribution.  

\begin{figure}[h]
\subfigure[CDF $\Phi$'s under two cutoffs $\hat p$ and $\hat p'>\hat p$]{\includegraphics[width=3in,height=2.5in]{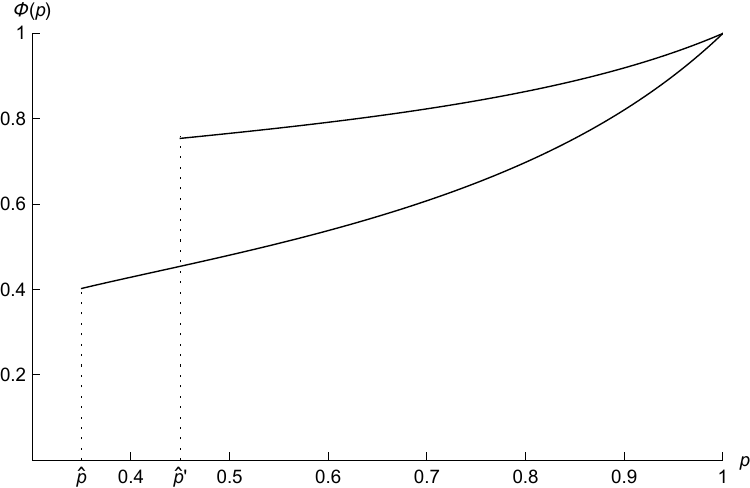}\label{fig:cdfh}}\qquad
\subfigure[CDF's $\Phi$, $\Upsilon$, $\Psi$]{\includegraphics[width=3in,height=2.5in]{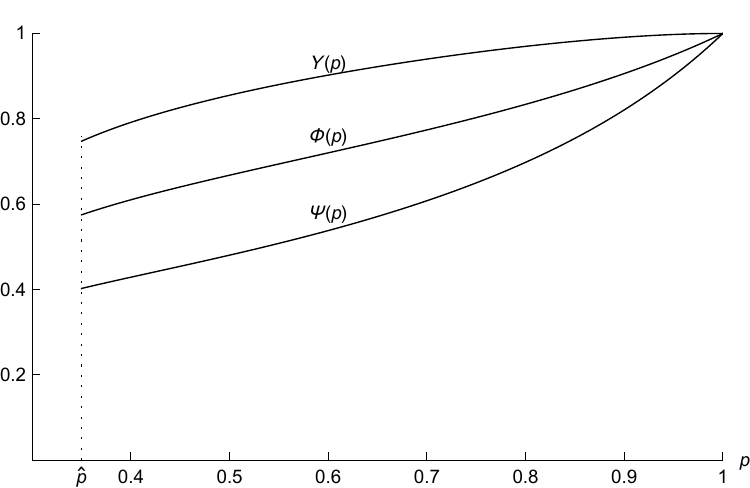}\label{fig:dist}}
\caption{Invariant distributions of posterior beliefs}\label{fig:inv-dist}
\end{figure}

\Cref{fig:cdfh} depicts $\Phi$ under two cutoffs $\hat p$ and $\hat p'>\hat p$. The policy with a lower cutoff involves more exploration. Hence, the stationary distribution corresponding to $\hat p$ is a mean-preserving spread of the one corresponding to $\hat p'$.\footnote{\label{ft:mps}This observation is proved in \Cref{lem:mps} in \Cref{appendix:criminals belief formation}.} 

Using the unconditional stationary distribution $\Phi$, we can derive the long-run distribution of the beliefs in each state. The distributions of $p$ conditional on states $H$ and $L$ are respectively:
$$\Psi(p):= \frac{ \hat p  \Phi (\hat p)  + \int_{\hat p}^p  \tilde p \phi(\tilde p)    d\tilde p}{\p_0} \mbox{ and } \Upsilon(p):=\frac{(1-\hat p)  \Phi (\hat p) + \int_{\hat p}^p (1- \tilde p)\phi(\tilde p) d\tilde p}{1-\p_0}$$
if $p \ge \hat p$, and zero otherwise. Naturally, $\Psi$ first-order stochastically dominates $\Phi$, which in turn first-order stochastically dominates $\Upsilon$. \Cref{fig:dist} depicts $\Phi, \Psi$, and $\Upsilon$, for a given cutoff policy. Observe that $\Psi$ puts a smaller mass at $\hat p$ than does $\Upsilon$. To the extent that a cutoff policy enforces strictly less at $\hat p$ than at $p>\hat p$, the policy monitors more in state $H$ than in $L$, revealing how prediction benefits monitoring (both in MP and OP). The relevant belief the agents use to evaluate the monitoring level is $\Psi(p)$, as they know the state is $H$ when making their infraction decisions.

\subsubsection{Agents' belief formation.} \label{par:belief formation}

We now characterize agents' beliefs about the monitoring level under a data-driven policy with cutoff $\hat p$. The answer is trivial if $\hat p$ lies outside $(\p_1, \p_0)$: it is zero if $\hat p\ge \p_0$, and one if $\hat p\le \p_1$. We therefore assume $\hat p \in (\p_1, \p_0)$. In this case, monitoring is binary: $y(p)=1$ for $p>\hat p$ and $y( p)=z(p)\in (0,1)$ for $p=\hat p$.

The preceding analysis suggests that agents form their beliefs about monitoring using $\Psi$ as the distribution. Let $\mu(\hat p, x):=\Psi(\hat p)$ denote the mass assigned to $\hat p$---the fraction of time the belief is at $\hat p$ given state $H$. Then, the agents face a conditional monitoring level: 
\begin{equation} \label{eq:enforcement}
    \mathbb{E}[y_{\hat p}(p)|\omega=H, x] =\mathbb{E}_{p\sim \Psi}[y_{\hat p}(p)]=\mu(\hat p, x) z(\hat p, x)+ (1-\mu(\hat p, x))\cdot 1,
\end{equation}
where we now make explicit the dependence of $z(\hat p)$ on $x$. While $z(\hat p,x)$ is decreasing in $\hat p$, we establish in \Cref{appendix:criminals belief formation} that $\mu(\hat p,x)$ is increasing in $\hat p$. It then follows from \Cref{eq:enforcement} that the anticipated monitoring level $\mathbb{E}_{p\sim \Psi}[y_{\hat p}(p)]$ is decreasing in $\hat p$. This is intuitive since a lower cutoff policy involves more monitoring. We now show that for any $x$, there exists a unique cutoff $\hat p=\overline p(\l x)$ that gives rise to the deterring level of monitoring $\overline y$.

\begin{lem} \label{lem:expected-enf}   Given any $x\in (0,1]$, there exists a cutoff $\overline p(\l x)\in (\p_1,\p_0)$ such that the associated cutoff policy $y_{\hat p}$ with $\hat p =\overline p (\l x)$ gives rise to a conditional monitoring level of $\overline y$; i.e.,
\begin{align} \label{bar-p} \mathbb{E}_{p\sim \Psi}[y_{\hat p}(p)]=\overline  y. \end{align}  
Also, the deterring cutoff $\overline p(\l x)$ is continuous in $x$ and decreases in $\overline y$, with $\lim_{\overline y\to 0}\overline p(\l x)=\p_0$ and $\lim_{\overline y\to 1}\overline p(\l x)=\p_1$ for all $x$.
\end{lem}
\begin{proof}
    See \Cref{appendix:proof of lemma 2}.
\end{proof}

This characterization is crucial for our equilibrium analysis of MP and OP. Suppose the equilibrium has an interior infraction level $x\in (0,1)$ in either the MP or OP regime. Then, agents must be indifferent, meaning they anticipate the monitoring level of $\overline y$ upon receiving an opportunity. \Cref{lem:expected-enf} means that, for such an equilibrium to obtain, the principal must find it optimal to choose a cutoff $\overline p(\l x)$. This, in turn, pins down the equilibrium infraction level $x$.

\subsection{Myopic Policy}
 
We are now ready to study the principal's behavior under MP. It is convenient to define $\lambda^M \in (\frac{c}{\p_0}, \frac{c}{\p_1})$ to be the smallest $\l$ that satisfies:
 \begin{equation} \label{eq:gp-eq0}
     \pm(\l)=\frac{c}{\l}=\overline p(\l).
\end{equation}
If $\l=\l^M$, then given $x=1$, the MP policy (with its myopic cutoff $\hat  p_M(\l)$) satisfies \Cref{bar-p} and implements the deterring monitoring level $\overline y$. Such a $\l^M$ is well-defined.

To characterize the MP equilibrium, consider first $ \lambda\le \lambda^M$. In this case, $\pm(\l x)=\frac{c}{\l x} > \overline p(\l x)$ for all $x<1$. In other words, the MP policy does not generate sufficient monitoring to deter infractions. Hence, in equilibrium, agents choose $\gsp=1$, and the principal responds with the corresponding myopic cutoff $\pm(\l)=\frac{c}{\l}$.

Consider next $ \lambda > \lambda^M$. In this case, we have a mixed strategy equilibrium in which agents commit infractions with probability $x \in (0,1)$. For them to randomize, the anticipated monitoring level must be $\overline y$. This requires the principal to find it optimal to choose $\overline p(\l x)$. Under MP, this requires that:
  \begin{equation}\label{eq:gp-eq}
  \pm(\l x)=\frac{c}{\l x} = \overline p(\l x),
  \end{equation} 
which determines the equilibrium infraction level $x$.

\begin{figure}[htb]
	\centering
	\includegraphics[scale=1]{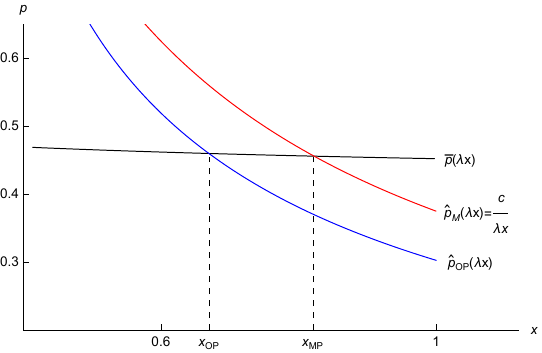}
	\caption{Equilibria under OP and MP with $\lambda=4$, $c=\frac{3}{2}$, $r=2$, $\rho_L =\rho_H =1$, and $\overline{y} =0.3$.}
 \label{fig:endo}
\end{figure}
  
\Cref{fig:endo} illustrates how the equilibrium infraction level is determined. The black curve depicts the deterring cutoff $\overline p(\l x)$ as a function of $x$.\footnote{The curve $\overline p(\cdot)$ is negatively sloped in all numerical analyses we conducted, which is intuitive. Suppose $x$ increases. First, $z(\hat p)$ must decline so that $x z(\hat p)$ remains constant, implying the principal monitors less at $\hat p$. The expected monitoring would fall—hence $\bar p$ should decrease—if the belief stays longer at $\hat p$, that is, if $\Phi(\hat p)$ increases. This tends to be the case because a higher $x$ causes the belief to drift downward at a faster rate, although a formal proof is unavailable due to the ambiguity in determining the change of $\phi(\hat p)$.} Meanwhile, the red curve depicts the myopic cutoff $\pm(\l x)=\frac{c}{\l x}$ as a function of $x$; it is downward sloping since a higher infraction rate calls for increasing monitoring, or a lower cutoff. The equilibrium condition \cref{eq:gp-eq} holds at the intersection of the two curves. 

Note from \cref{eq:gp-eq0} that $x=\l^M/\l$ solves \cref{eq:gp-eq}. We thus have the following result: 

 \begin{prop} \label{prop:endog-gp}   MP admits an equilibrium in which:
\begin{itemize}
    \item if $\lambda \le \lambda^M$, then $\gsp=1 \ge \min \{ 1, \frac{c}{\p_0 \l} \} =\gnp$, and $\ysp(\cdot)$ is a cutoff policy with $\hat p = \frac{c}{\l}$;
    \item if $\lambda > \lambda^M$, then $\gsp = \frac{\lambda^M}{\lambda} \in (0,1)$ and $\ysp(\cdot)$ is a cutoff policy with $\hat p = \frac{c}{\lambda^M}.$
\end{itemize}   
Under MP, the expected loss for the principal, evaluated from the baseline belief $p_0 = \pi_0$, is $\min\{\pi_{0}\lambda,c\}/r$.
\end{prop}

The payoff characterization follows from a simple observation. When $\p_0\lambda \le c$, the principal chooses zero monitoring in the long run, so the total loss is simply the loss from infractions, $\p_0 \l/r$. If $\p_0\lambda > c$, the principal adopts a nontrivial cutoff policy against the infraction level $\gsp=\min\{\frac{\l^M}{\l},1\}$. Since the principal is indifferent at the myopic cutoff, her payoff is the same as if she always monitors fully, making her total loss $c/r$. Note that the payoff does not depend on the deterring monitoring level $\overline y$. This irrelevance follows from the principal's lack of commitment power.

A comparison with \Cref{prop:endog-np} establishes payoff equivalence:

\begin{cor} [Payoff equivalence between SP and MP]  \label{cor:equiv-gp-np} The expected loss for the principal is the same between SP and MP.
\end{cor}
 
The intuition for this equivalence is the same as before: the information is obtained \textit{only when monitoring is myopically optimal} and hence is redundant.\footnote{If $\p_0\l>c$, the principal chooses full monitoring under SP, but posterior-dependent monitoring under MP. Yet, whenever the monitoring is partial under MP, the principal is indifferent, so she experiences the exact same expected loss as full monitoring.}

\subsection{Optimal Policy}

We now study the Markov perfect equilibrium under OP. From our analysis in \Cref{sec:exog}, for any infraction level $x$, the principal's optimal response is a cutoff policy, and the resulting outcome depends on the optimal cutoff $\pop (\lambda x)$, as characterized in \Cref{thm:main}.

To characterize the equilibrium, define $\lambda^* \in (\underline{\lambda}, \lambda^M)$ to be the smallest $\lambda$ that satisfies:
   \begin{equation}\label{eq:op-eq0}
  \pop(\l)=\overline p(\l).
\end{equation}
That is, if $\l=\l^*$, then, given $x=1$, an optimal cutoff $\hat p$ yields a deterring monitoring level $\overline y$. Such a $\l^*$ is well defined.

Consider first $\lambda \le \l^*$. In this case, we have $\pop (\l x) >\overline p(\l x)$ for all $x<1$, meaning  monitoring will be insufficient to deter infractions. Hence, in equilibrium, agents choose $x=1$, and the principal optimally responds with a cutoff $\pop (\l)$.

Consider next $\lambda > \l^*$. In this case, agents must randomize in equilibrium. This requires that the anticipated monitoring level equal $\overline y$, or:
   \begin{equation}\label{eq:op-eq}
  \pop(\l  x)=\overline p(\l  x),
  \end{equation}
which pins down the equilibrium infraction level $\xop$. \Cref{fig:endo} depicts $\xop$ as an intersection of the optimal cutoff facing $x$ (the blue curve) and the deterring-cutoff level $\overline p(x)$ (the black curve). It follows from \cref{eq:op-eq0} that $x=\l^*/\l$ solves this equation. Hence, in equilibrium, agents choose $\xop=\l^*/\l$ and the principal chooses a cutoff $\pop (\l^*)=\overline p(\l^*)$, which yields $\overline y$ in expectation, justifying the agents' randomization.

\begin{prop} \label{prop:op-endog}   The OP regime admits an equilibrium in which: 
 \begin{itemize}
     \item if $\lambda \le \l^*$, then $\xop = \gsp = 1$, and $\yop(\cdot)$ is a cutoff policy with $ \hat p = \hat p (\l)$ which is less than the MP cutoff $\pm(\l) =\frac{c}{\lambda}$;
     \item if $\lambda > \l^*$, then $\xop=\frac{\l^*}{\l} < \min \{ \frac{\l^M}{\l},1 \} =\gsp$, and $\yop (p)$ is a cutoff policy with $\hat p = \overline p(\l^*)$.
 \end{itemize}  
The principal incurs a weakly (resp. strictly) lower expected loss under OP than under MP or SP (if $\l> \ll$).
\end{prop} 
 
It is instructive to compare the equilibrium outcomes of OP and MP, as illustrated in \Cref{fig:endo}. The crucial difference between the regimes is that the optimal policy is always more aggressive than the myopic one. For any given infraction level $x$, the OP policy's informational motive leads it to choose a strictly lower monitoring cutoff, i.e., $\pop(\lambda x) < \pm(\lambda x)$. This difference in aggressiveness is what drives the infraction rate down under OP. 

To see the intuition, imagine the infraction rate is at the myopic equilibrium, $\gsp$. By definition, the myopic cutoff $\pm(\lambda \gsp)$ generates just enough monitoring to make agents indifferent. At this same infraction rate, however, the OP principal chooses the strictly lower cutoff $\pop(\lambda \gsp)$, leading to a higher expected monitoring level. This over-monitoring incentivizes agents to commit fewer infractions. Consequently, the equilibrium infraction rate must be lower under OP ($\xop< \gsp$), settling at the point where the OP principal's greater aggressiveness is perfectly balanced by the lower infraction rate to maintain the deterrence threshold.

In short, the informational motive of the principal acts as an endogenous commitment to deter infractions. This explains why the expected loss is smaller under OP than the other two regimes. Specifically, if $\l\le \ll$, all three regimes lead to no monitoring resulting in the identical expected loss of $\p_0 \l/ r$. If $\l> \ll$, the OP entails an expected loss strictly smaller than $c/r$, the loss under SP and MP.

\subsection{Data-aware Agents}
\label{sec:Data-aware Agents}

So far, we have assumed that agents cannot observe the history of detections in the way the principal can. This assumption is realistic in many regulatory contexts where agencies collect data privately and do not disclose their internal audit histories. In this section, we assume that agents are ``data-aware'': they have access to the same past detection data as the principal, using it to infer the principal's posterior belief $p$ and respond accordingly at each $p$. Considering this scenario highlights the role of the informational advantage the principal has over the agents.

In the case of SP, since the principal has no information other than $\p_0$, agents access no additional information, so the analysis remains unchanged. However, unlike the case of uninformed agents, MP and OP become strictly equivalent irrespective of $\l$.

When agents are data-aware and condition their infraction decisions on the same posterior belief as the principal, the MP and OP regimes admit the same equilibrium monitoring policy. In particular, there is a cutoff $\hat p=\frac{c}{\lambda}$ such that $y(p)=0$ for $p<\hat p$ and $y(p)=\bar y$ for $p>\hat p$.\footnote{Any $y(\hat p)\in[0,\bar y]$ is consistent with equilibrium at the cutoff $\hat p$.} In response, agents choose
$x(p)=1$ for $p<\hat p$, and
\[
x(p)=\frac{c}{p\lambda} \quad \text{for } p\ge \hat p,
\]
which keeps the principal exactly indifferent at every $p\ge \hat p$. Under MP, this strategy profile is the unique equilibrium. Moreover, because the principal is indifferent at every posterior above the cutoff, there is no residual informational gain from experimentation. Hence, the same strategy profile also constitutes an equilibrium under OP. \Cref{prop:pred-infringement} further shows that this equilibrium outcome is unique under OP within the class of \textbf{regular equilibria}, where the principal's monitoring policy is discontinuous at finitely many points. Thus, once agents observe the same posterior as the principal, they adjust their infraction decisions so as to eliminate the principal's informational motive for exploration.

\begin{prop}\label{prop:pred-infringement}
When data-aware agents condition their infraction decisions on the same posterior belief as the principal, all three regimes yield the same total expected loss in any regular equilibrium under OP.
\end{prop}
\begin{proof}
See \Cref{appendix: three way equivalence}.
\end{proof}
 
This result, together with \cref{cor:equiv-gp-np} and \cref{prop:op-endog}, establishes that the principal's informational advantage is strictly necessary for her to realize the deterrence value of dynamically optimal monitoring.

\section{Discussion} \label{sec:discussion}

In this section, we discuss two broader implications of our dynamic monitoring framework: the restoration of classic deterrence theory under dynamically optimal policies, and the extension of our model to environments with multiple independent monitoring targets.

\subsection{Monitoring Commitment and the Beckerian Prescription}

A foundational premise of the economic approach to enforcement, originating with \cite{becker1968crime}, is that increasing the penalty for an infraction should reduce its equilibrium occurrence. In our model, the principal's ability to manipulate the penalty corresponds to shifting the deterrence threshold $\overline{y}$. Specifically, if the principal increases the penalty $\gamma$ assessed upon detection, the probability of monitoring required to deter an agent, $\overline{y} = b/\gamma$, strictly decreases. A robust deterrence regime should translate this lower threshold into a reduced equilibrium infraction rate $x$.

However, our analysis reveals that when monitoring is dynamic and the principal lacks formal commitment power, the effectiveness of this Beckerian prescription depends entirely on how the principal utilizes data. Under the SP and  MP regimes, raising the penalty completely fails to reduce infractions. Recall from \Cref{prop:endog-np} and \Cref{prop:endog-gp} that the equilibrium infraction rates under these myopic regimes---$\gnp = \min\{1, \frac{c}{\pi_0 \lambda}\}$ and $\gsp = \min\{1, \frac{\lambda^M}{\lambda}\}$---are determined exclusively by the principal's indifference conditions, which are independent of $\overline{y}$. If the principal increases the penalty (lowering $\overline{y}$), agents intuitively require less monitoring to be deterred. Because the myopic principal cannot commit to maintaining a high monitoring intensity, she simply takes advantage of the higher penalty by monitoring less frequently, allowing the expected monitoring level to drop to the new, lower $\overline{y}$. The infraction rate $x$ remains exactly unchanged, keeping the principal indifferent.

In sharp contrast, the optimal policy (OP) restores the effectiveness of the Beckerian prescription. Because the OP principal explicitly values the informational option value of monitoring, her desire to explore acts as an endogenous commitment to monitor more aggressively than is myopically justified. From \Cref{lem:expected-enf}, a decrease in $\overline{y}$ implies that the required deterring cutoff $\overline{p}(\lambda x)$ must increase (the principal can afford to wait until the belief is higher before monitoring). In the equilibrium condition for OP, $\pop (\lambda \xop) = \overline{p}(\lambda \xop)$, an upward shift in the $\overline{p}$ curve intersects the optimally aggressive cutoff $\hat{p}_{OP}$ at a strictly lower infraction rate $\xop$. Consequently, under the dynamically optimal regime, an increase in the penalty successfully translates into a lower equilibrium infraction rate. This highlights a profound complementarity between the severity of punishments and the dynamic sophistication of the monitoring regime: severe penalties only deter agents if the principal is committed to gathering information.


\subsection{Multiple Arms and Whittle Dynamics}

Throughout our analysis, we have focused on a single monitoring domain facing a continuum of agents, capturing the scarcity of enforcement resources through a constant shadow cost $c>0$. In practice, the principal often allocates monitoring resources across many covariate profiles, transaction classes, agents, or network segments. For example, a cybersecurity monitor may distinguish domestic and foreign IP addresses, while a financial auditor may distinguish asset classes, business units, or counterparties. Each profile has its own evolving hidden risk state and its own posterior belief. The principal's prediction problem is then to continuously assess the unknown, changing effect of these profiles on the likelihood of an infraction.

To make the extension concrete, suppose there are several independent risky arms. Arm $i$ corresponds to one covariate profile and has posterior belief $p_i$, transition rates $\rho_{L,i}$ and $\rho_{H,i}$, opportunity rate $\lambda_i$, and infraction rate $x_i$. Write $\beta_i=\lambda_i x_i$. When monitoring intensity $y_i$ is assigned to arm $i$, the no-detection drift of its posterior is
\[
 f_i(p_i,y_i)=\rho_{L,i}(1-p_i)-\rho_{H,i}p_i-\beta_i p_i(1-p_i)y_i.
\]
The principal has a limited monitoring budget, for instance $\sum_i y_i\leq 1$. Because the structural parameters may differ across covariate profiles, raw posterior beliefs are not directly comparable. A posterior of $0.6$ for a persistent arm may represent a very different monitoring priority from a posterior of $0.6$ for a transitory arm.

A standard policy for restless multi-armed bandit problems is the Whittle-index policy.\footnote{Whittle-index policies are the canonical Lagrangian index policies for restless multi-armed bandits \citep{whittle1988restless} They generalize the logic of the Gittins index for classical rested bandits, but unlike the Gittins index, a Whittle-index policy need not be exactly optimal in a finite restless-bandit problem. The reason is that a passive arm continues to evolve, so the true opportunity cost of monitoring one arm rather than another is itself state-dependent. The Whittle construction replaces this multidimensional opportunity cost by a scalar Lagrange multiplier. This scalar-pricing approximation is what makes the policy tractable and implementable, but it is also why global finite-arm optimality requires additional conditions. Existing results nevertheless provide an important justification: Whittle argued that the policy should be nearly optimal in large systems, and subsequent work proves asymptotic optimality under global-stability or global-attractor conditions; see \citet{WeberWeiss1990} and \citet{Verloop2016}.} Like the Gittins index policy for classical rested bandits, it assigns to each arm a scalar priority index and then allocates capacity to the arms with the highest indices. The difference is that, in our environment, inactive arms are not frozen: their hidden states continue to change and their posteriors continue to drift. The Whittle policy is therefore best understood as a tractable Lagrangian way to leverage the one-arm solution rather than as a direct solution of the full multidimensional dynamic program.

The Lagrangian interpretation is simple. In the one-arm problem, the monitoring cost $c$ is the opportunity cost of using enforcement capacity on the risky arm. In a multi-arm problem, that opportunity cost is endogenous: using capacity on arm $i$ means not using it on other arms. The Whittle construction replaces this hard, state-dependent opportunity cost by a constant shadow price $c$, solves the resulting one-arm dynamic problem for each arm, and then asks which arms would still be worth monitoring at the highest shadow price. Thus the policy is dynamic within each arm but price-based across arms: each arm's index incorporates the learning-aware continuation value of monitoring that arm, while the comparison across arms is made through a scalar shadow price rather than through the full multidimensional value function.

Formally, fix an arm $i$ and interpret $c$ as the shadow price of one unit of monitoring capacity. The one-arm problem analyzed in \Cref{thm:main} yields an optimal cutoff $\widehat p_i(c)$.  The monotonicity of this cutoff  is precisely the relevant indexability property: as the shadow price $c$ rises, the set of beliefs at which arm $i$ is worth monitoring shrinks monotonically. Since $\widehat p_i(c)$ is continuously increasing in $c$, we can invert the cutoff rule and define the Whittle index of arm $i$ at posterior $p_i$ by
\[
 \mathcal I_i(p_i):=\widehat p_i^{-1}(p_i).
\]
Equivalently, $\mathcal I_i(p_i)$ is the hypothetical monitoring cost that would make the principal exactly indifferent between monitoring and not monitoring arm $i$ at belief $p_i$. A high index means that the arm remains worth monitoring even when monitoring capacity is costly; it therefore has high priority relative to other arms. This is why the index allows the principal to compare arms whose raw posterior beliefs are not directly comparable.

This interpretation also clarifies why the Whittle policy is not myopic. The index of an arm is computed from the dynamic one-arm value problem, so it incorporates the informational option value of monitoring: current mitigation, the possibility that a detection resets the posterior to one, and the way that no detections change future beliefs. The simplification is instead in the comparison across arms. The policy does not compute the full multidimensional HJB value of leaving every other arm passive; it prices that opportunity cost through the scalar index.

The resulting Whittle dynamics are simple. Let
\[
 \mathcal T(p):=\{i:\mathcal I_i(p_i)=\max_j \mathcal I_j(p_j)\}
\]
be the set of arms with the highest current index. If a single arm has the highest index, the principal monitors that arm fully. If several arms share the highest index, the continuous-time implementation must satisfy the same admissibility logic as in the single-arm cutoff problem: it must avoid instantaneous switching among tied arms. Thus the principal splits capacity across tied arms so that, absent a detection, their indices evolve together and remain tied. Formally, arms outside $\mathcal T(p)$ receive zero monitoring, while tied arms receive intensities satisfying
\[
 \sum_{i\in\mathcal T(p)} y_i=1
\]
and
\[
 \mathcal I_i'(p_i)f_i(p_i,y_i)=\mathcal I_j'(p_j)f_j(p_j,y_j)
 \qquad\text{for all } i,j\in\mathcal T(p).
\]
This equal-index-drift rule is the multi-arm analogue of the interior monitoring intensity $z(\widehat p)$ in the one-arm problem.   In the one-arm case, $z(\hat p)$ keeps the posterior at the cutoff, and equivalently keeps the arm’s index fixed at the indifference level because the outside option is constant. In the multi-arm case, there is no single fixed outside index; instead, the fractional allocation keeps the tied top indices moving together until a detection breaks the tie.

The dynamics are intuitive. As an active arm is monitored and no infraction is detected, its posterior and index tend to fall. An inactive arm's posterior continues to evolve toward its stationary benchmark, so its index may rise while the arm is not monitored. When several indices meet at the top, monitoring is shared to maintain the tie. If a detection occurs on one arm, that arm's posterior jumps to one, and its index rises discretely; that arm then receives priority until its index falls back to the common priority level.

This dynamic index policy preserves the central mechanism of the paper. It avoids the purely greedy logic that would permanently neglect profiles that currently look safe, while focusing monitoring on the covariate profiles whose beliefs, transition rates, and detection opportunities make them most valuable to inspect. In this sense, the Whittle construction provides a scalable way to extend the informational commitment logic of the single-arm optimal policy to environments with many evolving sources of risk.

\section{Conclusion} \label{sec:conclusion}

The increasing reliance on algorithmic prediction and historical data to allocate monitoring resources presents a profound dynamic mechanism design challenge. While data-driven monitoring promises enhanced efficiency, it is fundamentally constrained by the endogeneity of data generation: the principal only observes the state of the environment when actively monitoring. 

This paper provides a rigorous theoretical framework to evaluate the long-run consequences of this endogeneity. We demonstrate that the value of predictive data is not automatic. If a principal utilizes data myopically---monitoring only when the current belief exceeds a static threshold---the data-driven algorithm is rendered functionally redundant. Because the principal ceases monitoring when an entity appears safe, she creates an inferential blind spot, failing to learn when the underlying state inevitably deteriorates. Consequently, myopic monitoring performs no better in the long run than a naive policy that ignores data entirely. 

To unlock the true value of data, the principal must behave as a dynamically optimal experimenter in a changing world. Because the underlying state evolves according to a hidden Markov process, the need for learning never permanently resolves. The optimal policy explicitly internalizes the informational option value of monitoring, leading to a novel hedging behavior where the principal maintains persistent, partial vigilance even when the immediate risk appears low.

Crucially, when we embed this learning problem within a dynamic inspection game, we find that the principal's informational motive transcends mere prediction: it acts as an endogenous commitment device. The optimal principal's inherent desire to gather data organically pushes her monitoring intensity above the myopically optimal level. Knowing that the principal is eager to explore, strategic agents are forced to lower their equilibrium infraction rate to maintain the deterrence condition. Thus, dynamically optimal monitoring not only minimizes the principal's expected loss but also restores the effectiveness of classic Beckerian penalties, proving that the deterrence value of predictive data relies entirely on the principal's commitment to persistent exploration.

\bibliographystyle{economet}
\bibliography{bibckm}

\appendix

\section{Proof of \Cref{thm:main}}
\label{sec:proof of main}  
For the optimal policy, we consider a cutoff policy of the form: $y(p) = 1$  if $p >   \hat p$ and $y(p)=0$ if $p<\hat p$, for some threshold $\hat p$, and an  associated value function that  satisfies
\begin{align} \label{matching-pasting}
 V (\hat p_+)  = V (\hat p_-) \; \mbox{ and } \; V' (\hat p_+)  & = V' (\hat p_-), \end{align} known respectively as the {\it value-matching} and {\it smooth-pasting} conditions.
We refer to such a policy as a \emph{regular cutoff policy}.
Note that we do not claim the necessity of \eqref{matching-pasting}. Instead, we will show that the candidate value function and cutoff policy constructed using these conditions satisfy the HJB equation, which, by the verification theorem, implies optimality.

 
The proof of Theorem 1 then proceeds in three main steps. Our objective is to show that a regular cutoff policy—defined by a single cutoff belief $\hat{p}$—is optimal, and to characterize this cutoff.

\begin{enumerate}
    \item \textbf{Characterize Necessary Conditions}: We first establish a set of necessary conditions that any optimal regular cutoff policy and its associated value function $V(p)$ must satisfy.  We consolidate these properties in  \Cref{prop:necessary} in  \Cref{sec:necessary}.

    \item \textbf{Construct a Candidate Solution}:  In \Cref{sec:HJB-existence}, we construct a candidate solution---a \emph{unique} pair of a cutoff $\hat{p}$ and a value function $V(p)$---that satisfies all the necessary conditions from Step 1. The construction proceeds by analyzing the three distinct cases for the infraction rate outlined in Theorem 1. This step establishes the existence and uniqueness of a candidate that fits our characterization.

    \item \textbf{Verify Optimality}:  Finally,  in \Cref{sec:prop4}, we verify that the candidate solution constructed in Step 2 is indeed the optimal policy. We do this by showing that our constructed value function and policy satisfy the HJB equation for all beliefs $p \in [0,1]$. By the standard verification theorem, this confirms the policy's optimality.
\end{enumerate}

All omitted proofs in this section are provided in \Cref{sec:omitted proofs} of the Online Appendix.

\subsection{Necessary Conditions for Optimality}
\label{sec:necessary}

For a cutoff policy, the HJB equation \eqref{HJB} reduces to a two-part ordinary differential equation (ODE):
\begin{numcases}{r V(p)=}
h(p) V'(p) & for $p < \hat{p}$, \label{HJB0}\\
p\lambda x - c + p\lambda x \big[V(1) - V(p)\big] + g(p) V'(p) & for $p > \hat{p}$, \label{HJB1}
\end{numcases}
which follows from setting $y=0$ for $p<\hat p$ and $y=1$ for $p>\hat p$ in \eqref{HJB}. 

A solution to this ODE depends on two endogenous variables: the cutoff $\hat{p}$ and the boundary value $V(1)$. To characterize the optimal policy, define the marginal value of enforcement as
\begin{align}
\Delta(p) := \frac{1}{x} \frac{\partial W(p, y)}{\partial y}
= p \lambda - \frac{c}{x} + p \lambda \big[V(1) - V(p) - (1 - p) V'(p)\big].
\label{def:delta}
\end{align}

The next proposition provides conditions that determine $\hat p$ and $V(1)$ in the subsequent analysis:
\begin{prop} \label{prop:necessary}
Any regular cutoff policy  that satisfies \eqref{HJB} must satisfy:\footnote{In fact,  the proof of this proposition establishes  stronger properties than \eqref{slope-case1} and \eqref{slope-case2}: in \ref{item:case-1}, $V(p) =V' (p) =0,\forall p \le \hat p$; in \ref{item:case-2}, $V'(p) =\kappa,\forall p \ge \hat p$.}
    \begin{align}
    \label{cond-delta} \Delta (\hat p)  & =0  \end{align} 
    and 
    \begin{numcases}{    V' (\hat p)=}          0      &  if $\hat p \ge \p_0$   (\ref{item:case-1})  \label{slope-case1}\\   \kappa     &  if  $\hat p \le \p_1$  (\ref{item:case-2}) \label{slope-case2} \\   \sigma (\hat p)  &  if  $\hat p \in (\p_1,\p_0)$  (\ref{item:case-3}),  \label{slope-case3} 
\end{numcases}
 where $$\kappa:=\frac{\l x}{r + \rho_L +\rho_H} \; \mbox{ and } \; \sigma(p):=\frac{h(p) c}{p^{2}(1-p)    \lambda x(r+\rho_L+\rho_H)}.$$
\end{prop}

Condition \eqref{cond-delta} is the first-order condition ensuring that the principal is indifferent at the cutoff $\hat p$. The remaining conditions follow from the single-crossing property of $\Delta(\cdot)$ at $\hat p$: namely, $\Delta(p) \ge 0$ for $p \ge \hat p$ and $\Delta(p) \le 0$ for $p \le \hat p$. Given \eqref{cond-delta}, this requires $\Delta'(\hat p_-)\ge 0$ and $\Delta'(\hat p_+)\ge 0$. This condition binds in \ref{item:case-3}, implying $\Delta'(\hat p)=0$, which yields \eqref{slope-case3}. In contrast, it is slack in \ref{item:case-1} and \ref{item:case-2}, where \eqref{slope-case1} and \eqref{slope-case2} follow directly from the structures of the ODEs \eqref{HJB0} and \eqref{HJB1}, respectively.

\subsection{Construction of Value Function} \label{sec:HJB-existence}

We now begin the second step of our roadmap. Our goal is to construct a unique pair $(\hat{p}, V(1))$ that fulfills all the necessary conditions specified in  \Cref{prop:necessary}.

To proceed, it is convenient to first focus on the ODE for $p > \hat{p}$ and parametrize it by the unknown boundary value $K := V(1)$. This defines a family of potential value functions, $V(p;K)$, each solving the system we label $D[K]$:
\begin{flalign}
\text{$D[K]$ } && rV(p) &= p\lambda x  -c  +p\lambda x \left[ K -V(p) \right] + g(p)V'(p) \mbox{ with }  V(1) =K. & \label{HJB1-para}
\end{flalign} We will search for a unique value of $K$ and a corresponding belief $\hat{p}$ that jointly satisfy the conditions in \Cref{prop:necessary}. We let $\Delta(p;K)$ denote the function $\Delta(p)$ corresponding to the solution $V(p;K)$.   

First, we establish some essential properties of the solutions to $D[K]$ that will be used throughout the construction.   
\begin{lem} \label{lem:v-k}
For any given $\epsilon>0$, the ODE system $D[K]$ has a unique solution over $[\pi_{1}+\epsilon,1]$, denoted $V(p;K)$, which satisfies the following properties:  
\begin{itemize}
\item[(i)] $V (p;K)$ and $V' (p;K)$ are continuous  in  $(\lambda, x, c, p, K)$; 

\item[(ii)]  With $K  = \underline{K}:=\frac{(\lambda x -c)(r+\rho_L) - c \rho_H }{r(r+\rho_L+\rho_H)} $, $V' (p; K) = \kappa$ for all $p \in [0,1]$;\footnote{Note that this solution extends to the entire unit interval, not just $[\p_1+ \varepsilon, 1]$}
\item[(iii)] $V' (p;K)$ is strictly decreasing in $K$ while $V (p;K)$ is strictly increasing in $K$; 
\item[(iv)] $V''(p; K)$ is strictly increasing in $K$, and $V''(p; K)\gtreqqless 0$ whenever $K\gtreqqless \underline{K}$.
\end{itemize}
\end{lem}
Our construction of the pair $(\hat{p}, \hat{K})$ is divided into the three cases. Once we find the value function $V(p)$ on $[\hat{p}, 1]$, it can be extended to $[0, \hat{p}]$ by solving the ODE in \eqref{HJB0} with the boundary condition from value-matching (i.e., $V(\hat p_-) = V(\hat p_+)$), thus completing the construction.

We begin our analysis with \ref{item:case-2}, as it is the most straightforward.

\subsubsection{\ref{item:case-2}: $\hat p \le \p_1$}
This case corresponds to scenarios where the infraction rate $\lambda x$ is sufficiently high. The following lemmas formally characterize the parameter region for this case:
\begin{lem} \label{lem:lh-pi1-underp1}
Defining $\lh$ and $\underline{p}$ as \begin{align}
     \label{def:lh}
\lh := 
\begin{cases} \frac{c(\rho_L+\rho_H-c)}{(\rho_L-c)}  & \mbox{if }  \rho_L -c >0  \\   \infty  &  \mbox{otherwise}, \end{cases}  \; \mbox{ and } \;  \sigma (\underline{p}) = \kappa,   \end{align}  the following conditions are equivalent:
(a) $\lambda x \ge \lh$; (b) $\p_1 \ge \pm$; (c) $\p_1 \ge \underline{p}$; (d) $\underline{p} \ge \pm$. 
If any one of these inequalities holds with equality, then all of them do.
\end{lem}
\begin{lem} \label{lem:case2-char}
$\hat p  =\hat p_M \le \p_1$ if and only if $\l x \ge \lh$.     
\end{lem}
Given this parameterization, the next proposition establishes the existence and uniqueness of a candidate solution that satisfies the necessary conditions from \Cref{prop:necessary}
 for this case:
\begin{prop}
\label{prop:existence-case1}
For  $\lambda x \ge \lh $, there exists a unique pair  $(\hat{p}, \hat{K})$ satisfying the conditions of \Cref{prop:necessary} for \ref{item:case-2}. This pair is given by:
$$\hat{p} = \hat{p}_{M} \quad \text{and} \quad \hat{K} = \underline{K}$$
Furthermore, $\hat{p}=\hat{p}_{M}=\pi_{1}$ if $\lambda x=\overline{\lambda}$.  
\end{prop}
\begin{proof}  Let $K = \underline{K}$ and $\hat p = \hat p_M$.  Then, by \Cref{lem:v-k}(ii), we have $V' (\hat p) =\kappa$.  This linearity of $V$ implies that $V(1) - V(\hat p) - (1 - \hat p) V'(\hat p) =0$.  Plugging this into the definition of $\Delta(\hat{p})$ from \eqref{def:delta}, $\Delta (\hat p) = \hat p \lambda - \frac{c}{x} =0 $ since $\hat p = \hat p_M$, which proves the first statement. The second statement follows from \Cref{lem:lh-pi1-underp1}. \end{proof}

\subsubsection{\ref{item:case-3}: $\hat p \in (\p_1 ,\p_0)$}

This is the central and most involved case, where the infraction rate is neither too high nor too low. In this regime, the optimal policy features belief-freezing  at the cutoff $\hat{p}$. The following proposition establishes the existence and uniqueness of a candidate solution for this case:
\begin{prop}
\label{prop:existence-case2}  There is $\ll \in(c,\frac{c}{\p_0})$
such that for  $\lambda x \in(\ll ,\lh )$,
there is a unique pair $(\hat{p}, \hat{K})$ with $\hat{K}>\underline{K}$ and $\hat{p}\in(\pi_{1},\pi_{0})$ that satisfies the necessary conditions of  \Cref{prop:necessary} for \ref{item:case-3}. Specifically, it solves:
\begin{align} \label{two-equations}
    \Delta (\hat p ) =0 \mbox{ and } V' (\hat p) = \sigma (\hat p).
\end{align}
Furthermore, the cutoff $\hat{p}$ is continuously increasing in $c$ and decreasing in $\lambda x$, moving from $\pi_{0}$ to $\pi_{1}$ as $\lambda x$ increases from $\underline{\lambda}$ to $\overline{\lambda}$.
\end{prop}
 \begin{proof}

\label{sec:prop2}

We first  establish a couple of lemmas:   
\begin{lem}
\label{lem:v-under-k}
With $K=\underline{K}$,    $V'(p; K)=\kappa=\sigma(\underline{p}),\forall p\in[\underline{p},1]$. Moreover,  for  $\lambda x < \lh$, we have
 $\underline{p}\in(\p_1,\min\{\pm,\p_0\})$ that is decreasing in $\lambda x$ and increasing in $c$. 
\end{lem}

\begin{lem}\label{lem:v-lam}
Let $(\lambda_i,x_i,c_i)$ for $i=1,2$ satisfy
$\lambda_1 x_1 \le \lambda_2 x_2 < \lh$ and $c_1 \ge c_2$, with at least one strict.
Let $(\underline p_i,\underline K_i,V_i(\cdot;K))$  denote $\underline{p}$, $\underline K$, and the solution to  $D[K]$ associated with $(\lambda_i,x_i,c_i)$.
Then for all $K \ge \underline K_1$ and $p \ge \underline p_1$,
$V_1'(p;K) < V_2'(p;K) $ and $
V_1(p;K) > V_2(p;K).$\end{lem}

The rest of the proof proceeds in several steps. To begin, let us plug $p=1$  into \eqref{HJB1-para} to  obtain 
\begin{equation}
rK = \lambda x -c-\rho_HV  '(1; K)\;\mbox{ or }\;V '(1; K)=\frac{\lambda x-c-rK}{\rho_H}. \label{eq:4}
\end{equation}  Define $\overline{K}$ to be such that $V' (1; \overline{K}) =0$ in \eqref{eq:4}, that is,  \begin{align}
    \label{K-bar} \overline{K}:=\frac{\lambda x -c}{r} >\frac{(\lambda x -c)(r+\rho_L) - c \rho_H }{r(r+\rho_L+\rho_H)} =\underline{K}.
\end{align} 

\noindent  \textsc{\textbf{Step 1}:} \textbf{Define  $p(K)$.} \textit{Fix $(\lambda, x, c)$ with $\lambda x < \lh$. Then, there exists $K_0 \in (\underline{K}, \overline{K})$ and a continuous, strictly increasing function $p : [\underline{K}, K_0] \to [\underline{p}, \p_0]$ such that for each $K \in [\underline{K}, K_0]$, $p(K)$ is the unique solution to $V'(p; K) = \sigma(p)$. Moreover, $p(\underline{K}) = \underline{p}$ and $p(K_0) = \p_0$, and hence $V'(\p_0; K_0) = \sigma(\p_0) = 0$ and $V'(\underline{p}; \underline{K}) = \sigma(\underline{p})$. }
\vspace{0.2cm}

First, $V'(1;\overline{K})=0$  implies $V'(p;\overline{K})<0$ for all $ p < 1$ since $V'(\cdot;\overline{K})$
is increasing by  \cref{lem:v-k}(iv). Thus, $V'(\p_0;\overline{K})<\sigma(\p_0)=0$. 
Observe next that we have 
\[
V'(\p_0;\underline{K})=V'(\underline{p};\underline{K})=\sigma(\underline{p})>\sigma(\p_0),
\]
where the equalities hold due to \cref{lem:v-under-k} and the
inequality due to the fact that $\p_0>\underline{p}$ and $\sigma$
is decreasing. Since $V'(\p_0;\underline{K})>\sigma(\p_0)>V'(\p_0;\overline{K}  )$
and since $V'$ is continuously and strictly decreasing in $K$, there
exists a unique $K_0\in(\underline{K},\overline{K})$
such that $V'(\p_0;K_0)=\sigma(\p_0) =0$, meaning $p (K_0) =\p_0$.

For any $K\in(\underline{K},K_0)$,
we have 
\begin{align} \label{def:p(k)}
{\textstyle \sigma(\underline{p})=V'(\underline{p};\underline{K})>V'(\underline{p};K)\mbox{ and }\sigma(\p_0)=V'(\p_0;K_0)<V'(\p_0;K),}
\end{align}
where the first equality follows from \cref{lem:v-under-k} while the inequalities hold due to  \cref{lem:v-k}(iii). 
These inequalities, together with the monotonicity of $\sigma$ and $V'$ with respect to $p$, imply that for each $K\in(\underline K,K_0)$, there exists a unique $p(K)\in(\underline p,\pi_0)$ such that $V'(p(K);K)=\sigma(p(K))$. As established above, $p(\underline K)=\underline p$ and $p(K_0)=\pi_0$.

That $p(K)$ is continuously increasing in $K$ is straightforward
from the fact that $V'$ continuously increases in $p$ and decreases in $K$ while $\sigma$ continuously decreases in $p$.

\vspace{0.2cm}

\noindent \textsc{\textbf{Step 2}:} \textbf{No solution with $K \le \underline{K}$.} \textit{$\Delta(p(K);K)<0$ for any $K \le \underline{K}$.}
\vspace{0.2cm}

By   \cref{lem:v-k}(ii), \cref{lem:v-k}(iii), and Step 1, we have for any $K \le \underline{K}$ and $p > p (\underline{K}) =\underline{p}$,
\[
V'(p; K)\ge V'(p;\underline{K})=V'(\underline{p};\underline{K})=\sigma(\underline{p})>\sigma(p), 
\]
which  implies that $p(K)\le\underline{p}$ since $V'( p(K);K)=\sigma(p (K))$.
Thus, $p(K)\le\underline{p}<\frac{c}{\lambda x}$. Then,
\begin{align*}
\Delta(p(K); K) & =p (K) \lambda- \textstyle{\frac{c}{x}}+  p (K) \lambda\left[V(1; K)-V(p(K); K)-(1-p(K))V'(p (K); K)\right]\\
 & <\lambda p (K)\left[V(1;K)-V(p (K); K)-(1-p (K))V'(p (K);K)\right]\le 0,
\end{align*}
where the second inequality follows from  \cref{lem:v-k}(iv).
\vspace{0.2cm}

\noindent \textsc{\textbf{Step 3}:} \textbf{Monotonicity of  $\Delta(p(K);K)$ in $K$.} \textit{$\Delta(p(K);K)$ is strictly increasing
in $K\ge\underline{K}$.}
\vspace{0.2cm}

Note first that using $g(p)=h(p)-\lambda  x p(1-p)$, we can rewrite
\eqref{HJB1} as 
\begin{align}
rV(p)=h(p)V'(p)+x\Delta(p)\mbox{ for }p\ge\hat{p}.\label{eq:HJB-Delta}
\end{align} By  \eqref{eq:HJB-Delta} and the definition of $p(K)$ from \textsc{Step 1}, we have 
\begin{align}
x\Delta(p(K);K) & =rV(p(K);K)-h(p(K))V'(p(K);K)\nonumber \\
 & =rV(p(K);K)-h(p(K))\sigma(p(K)).\label{eq:delta-pk-k}
\end{align}

To prove the desired result, consider any $K_{2}>K_{1}\ge\underline{K}$
so that $p(K_{2})>p(K_{1})$. Observe that $h(p(K_{1}))\sigma(p(K_{1}))>h(p(K_{2}))\sigma(p(K_{2}))$
since both $h$ and $\sigma$ are decreasing. Observe also that, by
 \cref{lem:v-k}(iv), $V'(p;K_{1})\ge V'(p(K_{1});K_{1})=\sigma(p(K_{1}))\ge0,\forall p\in[p(K_{1}),p(K_{2})]$,
which implies that $V(p(K_{2});K_{1})\ge V(p(K_{1});K_{1})$. By
these observations and \eqref{eq:delta-pk-k}, 
\begin{align*}
x\Delta(p(K_{1});K_{1}) & =rV(p(K_{1});K_{1})-h(p(K_{1}))\sigma(p(K_{1}))\\
 & <rV(p(K_{2});K_{1})-h(p(K_{2}))\sigma(p(K_{2}))\\
 & <rV(p(K_{2});K_{2})-h(p(K_{2}))\sigma(p(K_{2}))=x\Delta(p(K_{2});K_{2}),
\end{align*}
where the second inequality follows from  \cref{lem:v-k}(iii).

\vspace{0.2cm}

\noindent \textsc{\textbf{Step 4}:} \textbf{Existence of unique solution.} \textit{For $\lambda x \in  [\frac{c}{\p_0}, \lh )$,  there exists  a unique
pair $(\hat{p},\hat K)$ with $K > \underline{K}$ and $\hat{p}\in(\p_1,\p_0)$ satisfying \eqref{two-equations}.}
\vspace{0.2cm}

 Observe first that $p(K_0)=\p_0\ge\frac{c}{\lambda x}$
or $\lambda p(K_0)\ge \frac{c}{x}$, so 
\begin{align*}
\Delta(p(K_0);K_0)\ge p(K_0) \lambda\left[V(1;K_0)-V(\p_0;K_0)-(1-\p_0)V'(\p_0;K_0)\right]>0,
\end{align*}
where the inequality is due to the strict convexity of $V$ as shown in  \cref{lem:v-k}(iv)  with  $K_0 > \underline{K} $. By this inequality and \textsc{Step 2}, we have
$\Delta(p(\underline{K});\underline{K})<0<\Delta(p(K_0);K_0)$.
Given this, the continuity and strict monotonicity of $\Delta(p(K),K)$
with respect to $K$ imply that there exists a unique $\hat K\in(\underline{K},K_0)$
such that $\Delta(p(\hat K),\hat K)=0$ and $p(\hat K)\in(\underline{p},\p_0)$. Note
also that $V'(p(\hat K))=\sigma(p(\hat K))$ by definition of $p(K)$.
Given \textsc{Step 2}, the strict monotonicity of $\Delta(p(K);K)$
for $K\ge\underline{K}$ implies  that a pair $(\hat{p}, \hat K)$ satisfying \eqref{two-equations} is unique. That $\hat p > \p_1$ follows  from the fact that $\hat p = p(\hat K) > \underline{p}$ and $\underline p > \p_1$ if $\lambda x < \lh$ by  \cref{lem:lh-pi1-underp1}.

\vspace{0.2cm}

\noindent \textsc{\textbf{Step 5}:} \textbf{Comparative statics of  $\hat{p}$.} \textit{Let $(\lambda_i, x_i, c_i)$ for $i=1,2$ satisfy $\lambda_i x_i < \lh_i$, where $\lh_i$ is $\lh$ with $c=c_i$.
Suppose $\lambda_1 x_1 \le \lambda_2 x_2$ and $c_1 \ge c_2$, with at least one strict inequality.
If $(\lambda_1, x_1, c_1)$ admits a pair $(\hat p_1, K_1)$ with $\hat p_1 \in (\pi_1, \pi_0)$ satisfying \eqref{two-equations}, then so does $(\lambda_2, x_2, c_2)$, with $\hat p_2 < \hat p_1$.}\footnote{Note that $\pi_1$   depends on $(\l_i, \g_i, c_i)$.}
\vspace{0.2cm}

Let us use the subscript $i$ to denote functions or variables  associated with $(\lambda_i,x_i, c_i) $.

Consider a pair $(\hat{p}_{1},\hat K_{1})$ solving \eqref{two-equations} under $(\lambda_1,x_1, c_1)$. Observe first
that by  \cref{lem:v-lam}, $V_{2}'(\hat{p}_{1}; \hat K_{1})>V_{1}'(\hat{p}_{1}; \hat K_{1})=\sigma_{1}(\hat{p}_{1})>\sigma_{2}(\hat{p}_{1})$ (where the second inequality holds since $\sigma$ is decreasing in $\lambda x$ and increasing in $c$).
Also, $V_{2}'(\hat{p}_{1};\overline{K}_2)<0$ (recall the
argument from Step 1). Thus, one can find $ K \in(\hat K_{1},\overline{K}_2 )$
such that $V_{2}'(\hat{p}_{1};K)=\sigma_{2}(\hat{p}_{1})$, which
means $\hat{p}_{1}=p_{2}(K)$. Note that $\hat{p}_{1}>\underline{p}_{1}>\underline{p}_{2}$,
which implies $K>\underline{K}_2$ since $p_{2}(\underline{K}_2)=\underline{p}_{2}$
and $p_{2}(\cdot)$ is increasing. Next, we show that $\Delta_{2}(p_{2}(K); K) =\Delta_{2}(\hat{p}_{1}; K) >0$.
Consider first the case in which $V_{2}(\hat{p}_{1};K)>V_{1}(\hat{p}_{1};\hat K_{1})$.
Using \eqref{eq:delta-pk-k} and the fact that $\sigma_{1}(\hat{p}_{1}) >  \sigma_{2}(\hat{p}_{1})$,
we obtain 
\begin{align*} 
0=x_1 \Delta_{1}(\hat{p}_{1}; \hat K_{1}) & =rV_{1}(\hat{p}_{1};\hat K_{1})-h(\hat{p}_{1})\sigma_{1}(\hat{p}_{1})
 <rV_{2}(\hat{p}_{1};K)-h(\hat{p}_{1})\sigma_{2}(\hat{p}_{1})=x_2\Delta_{2}(\hat{p}_{1};K).
\end{align*}
Consider next the case in which $V_{2}(\hat{p}_{1};K)\le V_{1}(\hat{p}_{1}; \hat K_{1})$.
Using \eqref{def:delta}, we again obtain 
\begin{align*}  
0  =x_1 \Delta_{1}(\hat{p}_{1};\hat K_{1})& =\hat{p}_{1} \lambda_{1} x_1 - c_1+\hat{p}_{1} \lambda_{1} x_1\left[V_{1}(1; \hat K_{1})-V_{1}(\hat{p}_{1};\hat K_{1})-(1-\hat{p}_{1})V_{1}'(\hat{p}_{1};\hat K_{1})\right]\\ 
 & <\hat{p}_{1}  \lambda_{2} x_2 -c_2+\hat{p}_{1}\lambda_{1}x_1\left[V_{2}(1;K)-V_{2}(\hat{p}_{1};K)-(1-\hat{p}_{1})V_{2}'(\hat{p}_{1};K)\right]\\ \nonumber
 & <\hat{p}_{1}\lambda_{2}x_2-c_2+\hat{p}_{1}\lambda_{2}x_2 \left[V_{2}(1;K)-V_{2}(\hat{p}_{1};K)-(1-\hat{p}_{1})V_{2}'(\hat{p}_{1};K)\right] =x_2 \Delta_{2}(\hat{p}_{1};K),
\end{align*}
where the first inequality holds due to the fact that $V_{1}(1;K_{1})=K_{1}<K=V_{2}(1;K)$
and $V_{1}'(\hat{p}_{1};\hat K_{1})=\sigma_{1}(\hat{p}_{1})>\sigma_{2}(\hat{p}_{1})=V_{2}'(\hat{p}_{1};K)$.
The second inequality holds since the expression in the square brackets is positive due to the convexity
of $V_{2}(\cdot;K)$. Recalling $\hat p_1 = p_2 (K)$, we have thus proven 
$\Delta_{2}(\hat{p}_{1};K)=\Delta_{2}(p_{2}(K);K) >0 > \Delta_{2}(p_{2}(\underline{K}_2);\underline{K}_2)$, which implies by  \textsc{Step 3} that
 there exists a unique $\hat K_{2}\in(\underline{K}_2,K)$
such that $\Delta(p_{2}(\hat K_{2});\hat K_{2})=0$. Moreover,   $\hat p_2 =p_{2}(\hat K_{2})<p_{2}(K)=\hat{p}_{1}$
by the monotonicity of $p_{2}(\cdot)$.

\vspace{0.2cm}

\noindent \textsc{\textbf{Step 6}:} \textbf{No solution at $\l x  =c$.} \textit{With $\lambda x =c$, there exists
no pair $(\hat{p},\hat K)$ satisfying \eqref{two-equations}.}
\vspace{0.2cm}

Given \textsc{Step 2} and \textsc{Step 3}, it suffices to show that with $\lambda x=c$,
$\Delta(p(K_0);K_0)<0$, since it will imply $\Delta(p(K);K)<0,\forall K\in[\underline{K},K_0]$.
Recall that $p(K_0)=\p_0$. Then, we have $\sigma(p(K_0))=0=V'(p(K_0);K_0)$.
Using this and \eqref{HJB1-para}, we also have $V(p(K_0);K_0)=\frac{ \p_0 \lambda x (1+K_0)-c}{r+p(K_0)\lambda x }.$
Plugging these into the definition of $\Delta$ and rearranging, we
obtain 
\begin{align*}
\Delta(p(K_0);K_0)=\frac{r \left( \p_0 \lambda  x  (1+K_0) -c\right)}{x  \left(r  + \p_0 \lambda x  \right)} & < \frac{r \left(  \lambda  x  \left( 1+  \frac{\lambda x -c}{r}\right) -c\right)}{x  \left(r  + \p_0 \lambda  x  \right)}  = \frac{(\lambda  x   -c) (r+  \lambda x )}{x  \left(r  + \p_0 \lambda  x  \right)}   =0,
\end{align*}
where the first inequality holds since $\p_0 < 1$ and  $K_0< \overline{K}  = \frac{\lambda x -c}{r}$.
\vspace{0.2cm}

\noindent \textsc{\textbf{Step 7}:} \textbf{Identify $\ll$.} \textit{There is $\ll \in( c,\frac{c}{\p_0})$
such that any $\lambda x \in(\ll,\lh )$  admits a unique pair $(\hat{p},\hat K)$ with $\hat p \in (\p_1, \p_0)$ satisfying  \eqref{two-equations}.}
\vspace{0.2cm}

The preceding steps establish that a unique solution for Case 3 exists when $\lambda x$ is in the range $[c/\pi_0, \overline{\lambda})$ but not when $\lambda x=c$. By continuity, there must exist a unique threshold $\underline{\lambda} \in (c, c/\pi_0)$ such that a solution only exists for $\lambda x > \underline{\lambda}$. 
\vspace{0.2cm}

\noindent \textsc{\textbf{Step 8}:} \textbf{Limit behavior of $\hat p$ at $\l x = \ll$ or $\lh$.} \textit{As $\lambda x \searrow \underline{\lambda}$,  $\hat{p} \nearrow \pi_0$;  as $\lambda x \nearrow \overline{\lambda}$,  $\hat{p} \searrow \p_1$.} \vspace{0.2cm}

The proof for  this step is  in \Cref{sec:step 8} of  the Online Appendix.   \end{proof}

\subsubsection{\ref{item:case-1}: $\hat p \ge \p_0$}

This final case corresponds to scenarios where the infraction rate $\lambda x$ is low, specifically when $\lambda x \le \underline{\lambda}$, where $\underline{\lambda}$ was defined above. 

\begin{prop} \label{prop:existence-case3}
For  $\lambda x\in(c,\underline{\lambda}]$, there exists a unique pair $(\hat{p}, \hat{K})$ with $\hat{K}> \underline{K}$ and $\hat{p}\in[\pi_{0},1)$ that satisfies the necessary conditions of \Cref{prop:necessary} for \ref{item:case-1}. Specifically, it solves
\begin{equation} \label{two-equations-1}
V^{\prime}(\hat{p};K)=0 \quad \text{and} \quad \Delta(\hat{p};K)=0.
\end{equation}
Furthermore, the cutoff $\hat{p}$ is continuously decreasing in $\lambda x$ and increasing in $c$.
\end{prop}
\begin{proof}[Proof Sketch] The proof follows a logic analogous to  that of \Cref{prop:existence-case2}. It involves defining a function $p(K)$ for $K \in [K_0, \overline{K})$ where $V'(p(K);K)=0$, and then showing that the function $\Delta(p(K);K)$ is monotonic in $K$ and crosses zero at a unique value $\hat{K}$, which defines the solution $(\hat{p}, \hat{K})$. Refer to the Online Appendix for details. \end{proof}

\subsection{Comparative Statics and Verification}
\label{sec:prop4} 
Using the value functions we have constructed so far, we prove  the  comparative statics and verification results to complete the proof of  \Cref{thm:main}.

\begin{prop}[Comparative Statics]
\label{prop:monotone}    $\hat{p}\le \pm = \frac{c}{\lambda x}$, with  the strict inequality if and only if   $\hat{p}>\p_1$.
 Also, $\hat p$ decreases continuously in $\lambda x$ and  increases continuously in $c$.
\end{prop}
\begin{proof}
    Refer to the Online Appendix.
\end{proof}

\begin{prop}[Verification]  \label{thm:existence}
The  cutoff policy identified  in  \Cref{sec:HJB-existence}    satisfies   \eqref{HJB}. 
\end{prop}

\begin{proof}  It suffices to prove  $\Delta (p) \ge (\le) 0 $ for $p  >  (<) \hat p$.  The proof of this result proceeds
in a few steps. To begin, one can derive:\footnote{For the derivation, see \eqref{eq:Delta-prime2} and
\eqref{eq:Delta-prime1} in the Online Appendix.} 
\begin{numcases}{\Delta'(p)=}  
\frac{\Delta (p)}{p} +  \frac{p (1- p)\lambda (r+\rho_L +\rho_H)}{h (p)}\left[\sigma (p)-V'(p)\right] &  for $p < \hat{p}$  \label{Delta-Prime-I}\\
\frac{h(p)}{g(p)}\left(\frac{\Delta(p)}{p}\right)  +  \frac{p (1-p)\lambda (r+\rho_L +\rho_H)}{g (p)} \left[\sigma (p)- V'(p)\right] &  for $p > \hat{p}$. \label{Delta-Prime-II}
\end{numcases}

We first prove the following  claim: 
\begin{claim} \label{claim:verification}
    If $\Delta(p)\ge0$ at $p=\max\{\hat{p},\p_0 \}$,
then $\Delta(p)\ge0,\forall p > \max\{\hat p,\p_0\}$.
\end{claim}
\begin{proof}
Note that $g(p)<0$ and $h(p)\le0$ for $p\ge\max\{\hat{p},\p_0\}$.
Hence, since $V'(p)\ge0$,
\eqref{Delta-Prime-II} implies that $\Delta'(p)>0$ whenever $\Delta(p)\ge0$.
In particular, if $\Delta(p)\ge0$ at $p=\max\{\p_0,\hat{p}\}$,  it follows that $\Delta(p)\ge0$ for all $p>\max\{\p_0,\hat{p}\}$.    
\end{proof}

\paragraph{Verification for \ref{item:case-1} ($\hat p \ge \p_0$)}

\begin{itemize}
    \item \textbf{For $p >  \hat p$}: In this case, we have $\max\{\hat{p},\p_0\}  =\hat p$, so $\Delta(p)=0$
at $p=\max\{\hat{p},\p_0\}$, which implies by \Cref{claim:verification} that $\Delta(p)\ge0,\forall p > \hat{p}$.

\item \textbf{For $p < \hat p$}: To show that $\Delta(p)\le0$, note that $V'(p)=0$ for all $p\le \hat p$,
and that $h(\hat{p})< 0 $ and $g(\hat{p})<0$. Consequently,
by \eqref{Delta-Prime-I},  \eqref{Delta-Prime-II}, and $\Delta (\hat p)=0$,
\[
\Delta'(\hat{p}_{-})=\frac{\hat p (1- \hat p)\lambda (r+\rho_L +\rho_H)}{h (\hat p)} \sigma (\hat p)>0,
\] and \[
\Delta'(\hat{p}_{+})= \frac{\hat p (1-\hat p)\lambda (r+\rho_L +\rho_H)}{g (\hat p)} \sigma (\hat p) >0. 
\]
This means that there exists $\varepsilon>0$ such that for $p\in(\hat{p}-\varepsilon,\hat{p})$,
$\Delta(p)<0$, and for $p\in(\hat{p},\hat{p}+\varepsilon)$, $\Delta(p)>0$.
If $\Delta(p)>0$ for some $p<\hat{p}$, we must have $\check{p}\in[p,\hat{p}]$
such that $\Delta'(\check{p})<0$ and $\Delta(\check{p})=0$. But
we have 
\[
\Delta'(\check{p})=\frac{\check p (1- \check p)\lambda (r+\rho_L +\rho_H)}{h (\check p)} \sigma (\check p)>0,
\]
a contradiction.

\end{itemize}

\paragraph{Verification for \ref{item:case-3} ($\hat{p}\in(\p_1,\p_0)$)}

\begin{itemize}
    \item \textbf{For $p > \hat p$}: The result will follow from \Cref{claim:verification} once we prove $\Delta(p)\ge0$
for $p\in[\hat{p},\p_0]$. Let us thus fix any $p\in[\hat{p},\p_0]$.  Rewrite (\ref{Delta-Prime-II}) for the part of $p\ge\hat{p}$ as
\begin{align}
\Delta'(p)=\frac{h(p)}{g(p)}\left(\frac{\Delta(p)}{p}\right)-\frac{\delta(p)}{g(p)p},\label{delta-prime-1}
\end{align} where
\begin{align*}
     \delta(p):=p^{2}(1-p)\lambda(r+\rho_L +\rho_H)V'(p)-\frac{c}{x} h(p)  = p^{2}(1-p)\lambda(r+\rho_L +\rho_H) \left[ V'(p)  - \sigma (p)\right].
\end{align*}

\begin{claim} \label{claim:delta-1}
If $\Delta (p) \ge 0$ and $\delta (p) =0$ for $p \in  [\hat p, \p_0)$, then $\delta ' (p) >0$.\footnote{With $p =\hat p$,  the conclusion should be taken as $\delta'(\hat p_+) >0$. The proof of this result is provided in the Online Appendix.}
\end{claim}

By \cref{prop:existence-case2}, we have  $\Delta(\hat{p})=0$ and  
$V'(\hat{p})=\sigma(\hat{p})\ge0$---and hence  $\delta(\hat{p})=0$.  Then, by \Cref{claim:delta-1},  $\delta'(\hat{p}_{+})>0$.  Thus, using \eqref{delta-prime-1}  and  $\Delta (\hat p_+) = \Delta'(\hat{p}_+)= \delta (\hat p_+) =0$,
we obtain 
\[
\Delta''(\hat{p}_+)=-\frac{\delta'(\hat{p}_+)}{g(\hat{p})\hat{p}}>0,
\]
from which it follows that $\Delta(p')>0$ for all $p'\in(\hat{p},\hat{p}+\varepsilon]$,
for some $\varepsilon>0$.\footnote{Note that in \ref{item:case-3}, $V'(\hat p) = \sigma (\hat p)$ is equivalent to $\Delta' (\hat p) =0$, as shown in \Cref{lem:necessary-case-2} in the Online Appendix.} We now prove the following claim: 
\begin{claim}  \label{claim:delta-3}
If $\Delta(p'')<0$ for some $p''\in[\hat{p},\p_0]$, then there exists
$\check{p}\in(\hat{p},p'')$ such that $\Delta(\check{p})\ge0$, $\delta(\check{p})=0$
and $\delta'(\check{p})\le 0$. \end{claim} 
\begin{proof}
Let $p_{0}=\sup\{p'\in(\hat{p}+\varepsilon,p''):\Delta(p)>0,\forall p<p'\}$.
Then, we must have $\Delta(p_{0})\le0$. Given this, it cannot be
the case that $\delta(p)>0,\forall p\in[\hat{p},p_{0}]$, since it
would imply that whenever $\Delta(p)=0$, we have $\Delta'(p)>0$
by \eqref{delta-prime-1} and $g(p)<0$, which in turn implies $\Delta(p_{0})>0$.
Thus, we must have some $p_{1}\in(\hat{p},p_{0}]$ with $\delta(p_{1})\le0$.
Since $\delta(\hat{p}_{+})>0$, this implies that there must be some
$\check{p}\in(\hat{p},p_{1}]$ such that $\delta(\check{p})=0$ and
$\delta'(\check{p})\le0$. 
\end{proof}

Suppose for contradiction that $\Delta(p'')<0$ for some $p''\in[\hat{p},\p_0]$.
By \Cref{claim:delta-3}, one can find $\check{p}\in(\hat{p},p'')$ that
$\Delta(\check{p})\ge0$, $\delta(\check{p})=0$ and $\delta'(\check{p})\le0$.
However, this is a contradiction since, by \Cref{claim:delta-1}, $\Delta(\check{p})\ge0$ and  $\delta(\check{p})=0$ imply $\delta'(\check{p})>0$.

\item  \textbf{For $p < \hat{p}$}: Rewriting (\ref{Delta-Prime-I})
 as 
\[
\Delta'(p)=\frac{\Delta(p)}{p}-\frac{\delta(p)}{ph(p)},
\]  we prove:
\begin{claim} \label{claim:delta-2}
If $\delta (p) =0$  for $p \in  (\p_1, \hat p]$, then   $\delta' (p) >0$.\footnote{With $p =\hat p$, the conclusion should be taken as $\delta'(\hat p_-) >0$.}
\end{claim}

Then, since $\delta(\hat{p})=0$, we obtain
$\delta'(\hat{p}_{-})>0$. 
Using this, one can follow an
argument analogous to that in \textsc{Step 3} to show that $\Delta(p)\le0,\forall p < \hat{p}$.
\end{itemize}

\paragraph{Verification for \ref{item:case-2} ($\hat{p} \le \p_1$)} 

\begin{itemize}
\item \textbf{For $p > \hat p$}: In this case,  we have $V'(p) =\kappa $ for all $p\in [\hat p,1]$, as seen from the proof of \Cref{prop:existence-case2}. Thus, $\Delta (p) = p \l -\frac{c}{x} > 0$ for $p > \hat p  =\hat p_M = \frac{c}{\l x}$.

    \item \textbf{For $p < \hat p$}: In this case, $\hat p =\pm \le \underline p$ by \cref{prop:existence-case1} and \cref{lem:lh-pi1-underp1},  so  we
have   for any $p<\hat{p}$, $$\sigma (p)  - V' (p)   > \sigma (\underline p) -V' (\hat p)  =  \sigma (\underline p)- \kappa  = 0, $$
where the  inequality follows from the fact that $V$ is strictly convex while  $\sigma$ is decreasing.\footnote{For the strict convexity of $V$ on $[0,\hat p]$, we observe that $$ V(p) = \left( \frac{h(p)}{h(\hat p)}\right)^{-\frac{r}{\rho_L +\rho_H}} V(\hat p) $$ solves the ODE \eqref{HJB0}. It is straightforward to check that $V''(p) >0$, using $h(\hat p) >0$.}  Thus,  by  \eqref{Delta-Prime-I}, we conclude that  for any $p< \hat p$, $\Delta' (p) >0$ whenever $\Delta (p) =0$.
   Given that 
$\Delta(\hat{p})=0$, this observation implies that $\Delta(p)\le0$
for all $p < \hat{p}$.  
\end{itemize} This completes the verification. \end{proof}

\section{Proof of \Cref{prop:pred-infringement}}
\label{appendix: three way equivalence}

Under SP, the principal's equilibrium loss is
\[
\frac{\min\{c,\pi_0\lambda\}}{r}.
\]
We show that any regular equilibrium under OP yields the same loss. Given an agent response
$x(\cdot)$, the principal's value satisfies
\begin{align}
rV(p)=h(p)V'(p)+\max_{y\in[0,1]}y\,\Gamma(p),
\label{eq:OP_HJB_reg2}
\end{align}
where
\[
\Gamma(p):=(p\lambda x(p)-c)
+p\lambda x(p)\bigl[V(1)-V(p)-(1-p)V'(p)\bigr],
\]
and \(h(p)=\rho_L(1-p)-\rho_Hp\).

First, \(y(p)>\bar y\) cannot occur in equilibrium. If \(y(p)>\bar y\), then the agents'
best response gives \(x(p)=0\), in which case the coefficient of \(y\) in the HJB is \(-c<0\);
hence \(y=0\) is strictly optimal, a contradiction.

We next rule out interior monitoring on a nondegenerate interval. Suppose that
\(y(p)\in(0,\bar y)\) for all \(p\) in some interval \(I\). Then \(x(p)=1\) on \(I\), and the
principal's interior optimality condition implies
\begin{equation}
(p\lambda-c)+p\lambda\bigl[V(1)-V(p)-(1-p)V'(p)\bigr]=0.
\label{A.3.short}
\end{equation}
The HJB then reduces to
\begin{equation}
rV(p)=h(p)V'(p).
\label{A.4.short}
\end{equation}
Differentiating \eqref{A.3.short} and using \eqref{A.3.short} again gives
\[
V''(p)=\frac{c}{\lambda p^2(1-p)}.
\]
Differentiating \eqref{A.4.short} gives
\[
(r+\rho_L+\rho_H)V'(p)=h(p)V''(p),
\]
and hence
\[
V'(p)=\frac{c\,h(p)}
{\lambda p^2(1-p)(r+\rho_L+\rho_H)}.
\]
Differentiating this expression and comparing with the previous expression for
\(V''(p)\) yields
\[
(r-\rho_L-\rho_H)p^2+(3\rho_L-r)p-2\rho_L=0
\quad\text{for all }p\in I,
\]
which is impossible on a nondegenerate interval. Thus, on every continuity interval,
\(y\) is either \(0\) or \(\bar y\).

On any such interval, \(V\) satisfies
\begin{equation}
rV(p)=h(p)V'(p).
\label{eq:global_ode_piece_short}
\end{equation}
Indeed, if \(y=0\), this follows directly from the HJB. If \(y=\bar y\), then \(\bar y\) is an
interior action for the principal, so \(\Gamma(p)=0\), and the same equation follows. Therefore
\[
V(p)=K|h(p)|^{-r/(\rho_L+\rho_H)}
\]
on each continuity interval. Since \(V\) is continuous, the constant \(K\) must be the same
across adjacent intervals on the same side of \(\pi_0\). Hence there are constants \(K_-\) and
\(K_+\) such that this expression holds on \((0,\pi_0)\) and \((\pi_0,1)\), respectively. Since
\(|h(p)|^{-r/(\rho_L+\rho_H)}\to\infty\) as \(p\to\pi_0\) whereas \(V\) is bounded, we must
have \(K_-=K_+=0\). Thus \(V\equiv0\).

With \(V\equiv0\), the HJB coefficient reduces to
\[
\Gamma(p)=p\lambda x(p)-c.
\]
If \(y(p)=0\), then the agents' best response gives \(x(p)=1\), while optimality of
\(y=0\) requires \(p\lambda x(p)-c\le0\). Hence such points must satisfy \(p\le c/\lambda\).
If \(y(p)=\bar y\), then interior optimality requires
\[
p\lambda x(p)-c=0,
\]
so \(x(p)=c/(p\lambda)\), which is feasible only if \(p\ge c/\lambda\). Therefore the only
regular equilibrium policy has cutoff \(\hat p=c/\lambda\):
\[
y(p)=
\begin{cases}
0 & \text{for } p<\hat p,\\
a & \text{for } p=\hat p,\\
\bar y & \text{for } p>\hat p,
\end{cases}
\qquad
x(p)=
\begin{cases}
1 & \text{for } p\le \hat p,\\[3pt]
\dfrac{c}{p\lambda} & \text{for } p\ge \hat p.
\end{cases}
\]

It remains to compute the loss. If \(\hat p>\pi_0\), the belief converges to \(\pi_0\), where
the principal does not monitor, so the loss is \(\pi_0\lambda/r\). If \(\hat p\le \pi_0\), the
long-run belief lies in \([\max\{\hat p,\pi_1\},1]\), where the principal is indifferent and
incurs flow loss \(c\). Hence the total loss is
\[
\frac{\min\{c,\pi_0\lambda\}}{r},
\]
as under SP.

Finally, the same cutoff profile is the unique equilibrium under MP. If \(p<\hat p\), then
\(p\lambda x(p)\le p\lambda<c\), so the myopic principal chooses \(y(p)=0\), and agents choose
\(x(p)=1\). If \(p>\hat p\), then \(y(p)>\bar y\) is impossible because it induces \(x(p)=0\)
and hence \(y=0\) would be optimal; while \(y(p)<\bar y\) would induce \(x(p)=1\), making
\(p\lambda>c\) and contradicting the optimality of \(y(p)<\bar y\). Thus \(y(p)=\bar y\) and
\(x(p)=c/(p\lambda)\). At \(p=\hat p\), any \(x(\hat p)<1\) would imply
\(\hat p\lambda x(\hat p)<c\), so the principal would choose \(y(\hat p)=0\), inducing
\(x(\hat p)=1\), a contradiction. Hence \(x(\hat p)=1\).

\clearpage

\setcounter{equation}{0}
\pagenumbering{arabic}
\renewcommand*{\thepage}{SA.\arabic{page}}
\renewcommand{\theequation}{SA.\arabic{equation}}
\begin{center}
    \huge{Online Appendix for: \vskip 2pt Predictive Enforcement}
\end{center}

\section{Equivalence between $V$ Formulation and $L$ Formulation}
\label{sec:omitted proofs}
\begin{lem} \label{lem:equivalence}
 Any policy  $y(\cdot)$   solves \cref{HJB} if and only if  it minimizes the loss function in \cref{eq:loss}. 
\end{lem}

\begin{proof}
Fix any admissible Markov policy $y:[0,1]\to[0,1]$. Let $L_y(p)$ denote the principal's total discounted loss under policy $y$ when the current belief is $p$. By considering a short interval $dt$, we can write
\begin{align*}
     L_y (p) = \big[ p \lambda x (1-y  (p)  ) +c y (p)  \big] dt  &  + p\lambda x y (p) dt   (1-r dt ) L_y (1)   \\  & + (1-r dt) (1-p \lambda x  y (p)  dt ) L_y (p_{t +dt}) +o(dt).
 \end{align*}
Taking the limit as $dt\to 0$ and using the belief drift in \cref{eq:drift}, we obtain
\begin{align}
rL_y(p)
=
p\lambda x(1-y(p))+cy(p)
+
p\lambda x y(p)\bigl(L_y(1)-L_y(p)\bigr)
+
f(p,y(p))L_y'(p).
\label{A1}    
\end{align} Thus, any optimal policy that minimizes \eqref{eq:loss} must minimize the RHS of \eqref{A1}.  

Next, let $C(p)$ denote the total discounted expected amount of infractions, whether detected or undetected, when the current belief is $p$. By the same  argument,
\[ C(p)
=
p\lambda x\,dt
+
p\lambda x y(p)\,dt\,(1-rdt)C(1)
+
(1-rdt)\bigl(1-p\lambda x y(p)\,dt\bigr)C(p_{t+dt})
+
o(dt),
\]
which yields
\begin{align}
rC(p)
=
p\lambda x
+
p\lambda x y(p)\bigl(C(1)-C(p)\bigr)
+
f(p,y(p))C'(p).
\label{A2}    
\end{align}

It follows immediately from the definition that
\begin{align} \label{total-crime}
C(p) = L_y(p) + V_y(p).    
\end{align}
Subtracting \eqref{A1} from \eqref{A2}, we obtain
\begin{align}
rV_y(p)
=
(p\lambda x-c)y(p)
+
p\lambda x y(p)\bigl(V_y(1)-V_y(p)\bigr)
+
f(p,y(p))V_y'(p).
\label{A3}
\end{align}
Since $C(p)$ is independent of the policy, the relationship in  \eqref{total-crime}  implies that any  policy $y$ which minimizes the RHS of   \eqref{A1} maximizes the RHS of \eqref{A3}, meaning exactly \eqref{HJB}.\footnote{The function $C(p)$ does not depend on the monitoring policy. The reason is that, when choosing a monitoring policy, the principal treats the infraction rate $x$ as fixed, while infraction opportunities arrive at rate $\lambda$ only in state $H$. Hence, the total flow rate of infractions is $\lambda x \mathbf{1}\{\omega_t=H\}$, which does not depend on the monitoring policy. The policy affects only whether an infraction is detected and prevented from causing harm. Hence the total flow rate of infractions is simply $\lambda x\,\mathbf{1}\{\omega_t=H\}$, which is independent of $y$. }
\end{proof}

\section{Omitted Proofs for \Cref{thm:main}}
\label{sec:omitted proofs}
\subsection{Proof of \Cref{lem:v-k}}
The differential equation  $D [K]$ can be rewritten as \[
V'(p)=\frac{r+p\lambda x}{g(p)} V(p) +  \frac{c- p\lambda x(1+K)}{g(p)},
\]  which clearly has a unique solution for $p  \in  [\p_1 +\varepsilon,1]$ that satisfies the desired continuity, proving Part (i).  


Part (ii) is established by finding a linear solution to  $D[K]$ constructively. 
To do so, let  $V(p)= K +a(p-1)$ for $p\in[\hat{p},1]$. Then,
the RHS of \eqref{HJB1-para} becomes 
\begin{align*}
 &    p \lambda x -c   +p\lambda x  a (1-p)  +  [\rho_L(1-p)-\rho_Hp-p(1-p){\lambda} x]a\\
= & -c+ \rho_La+[\lambda x-(\rho_L+\rho_H)a]p,
\end{align*}
which must equal the LHS of \eqref{HJB1-para}, $rV(p)=r(K+a(p-1))=r(K-a)+rap.$
We must thus have $\rho_La=r(K-a)$ and $\lambda x-(\rho_L+\rho_H)a=ra$,
which can be solved to yield $a=\frac{\lambda x}{r+\rho_L+\rho_H}$
and $K=\frac{(\lambda x -c)(r+\rho_L) - c \rho_H }{r(r+\rho_L+\rho_H)}$, as desired.

To prove Part (iii),  let us consider any $K_{2}>K_{1}$. For simplicity, let $V_{i}(\cdot)$
denote  $V (\cdot; K_i)$ for $i=1,2$.
To begin, observe that we have $V_{1}'(1)>V_{2}'(1)$ by \eqref{eq:4}. Suppose
now for contradiction that there is some $p <1$ such that
$V_{1}'(p)\le V_{2}'(p)$. One
can then find some $p'<1$ such that $V_{1}'(p')=V_{2}'(p')$
and $V_{1}'(p)>V_{2}'(p),\forall p>p'$. Note first that 
\begin{align}
V_{1}(p')=V_{1}(1)-\int_{p'}^{1}V_{1}'(p)dp<V_{2}(1)-\int_{p'}^{1}V_{2}'(p)dp=V_{2}(p')\label{ineq:v-k}
\end{align}
since $V_{1}(1)=K_{1}<K_{2}=V_{2}(1)$ and $V_{1}'(p)<V_{2}'(p),\forall p\in[p',1]$.
We then rewrite  \eqref{HJB1-para} with $p=p'$, $V=V_{i}$, and
$K=V_{i}(1)$ as 
\[
g(p')V_{i}'(p')+p'\lambda x -c=rV_{i}(p')-p'\lambda x[V_{i}(1)-V_{i}(p')].
\]
With $V_{1}'(p')=V_{2}'(p')$, we have the LHS, and thus RHS, of this equation unchanged across $i=1,2$. However, 
\begin{align*}
rV_{1}(p')-p'\lambda x[V_{1}(1)-V_{1}(p')] & =rV_{1}(p')-p'\lambda x\int_{p'}^{1}V_{1}'(p)dp\\
 & <rV_{2} (p')-p'\lambda x\int_{p'}^{1}V_{2}'(p)dp\\
 & =rV_{2}(p')-p'\lambda x[V_{2}(1)-V_{2}(p')],
\end{align*}
where the inequality holds since $V_{2}(p')>V_{1}(p')$
from \eqref{ineq:v-k} and $V_{2}'(p)<V_{1}'(p),\forall p>p'$.
We have thus obtained a contradiction. That $V (p;K)$ is strictly increasing in $K$ can be established  by an analogous  argument, so its proof is omitted.

For Part (iv), let us  differentiate both sides of \eqref{HJB1-para} to obtain 
\begin{align*}
rV'(p;K )= \lambda x(1+K) -\lambda x V(p;K) + \left( g' (p) -p\lambda x   \right)  V'(p;K)+g(p)V''(p;K),
\end{align*} which 
can be combined with  \eqref{HJB1-para} to  remove
$V(p)$ and obtain \begin{align}
V''(p;K)=\frac{\left[r^{2}+r(\lambda x+\rho_L+\rho_H)+\lambda x\rho_L\right]V'(p;K)-\lambda x(c+r+r K)}{(r+\lambda  p x)g(p)}. \label{Vdoubleprime}
\end{align}
Since  the numerator is strictly  decreasing
in $K$ and the denominator is negative (recall $g(p)<0$
for $p>\p_1$), $V'' (p; K)$ is  strictly increasing in $K$. Combining this with Part (ii) proves the rest of Part (iv).

\subsection{Proof of \Cref{lem:lh-pi1-underp1}}
We first prove a couple of claims:   \begin{claim}
\label{claim:sigma} $\sigma(p)$  strictly decreases from $\infty$ to $0$ as $p$ increases from 0 to $\p_0$. 
\end{claim}
\begin{proof}
Note first that $\frac{h(p)}{p^{2}(1-p)}$ is strictly decreasing on $[0,\p_0]$:  
\begin{align*}
\frac{d}{dp}\left(\frac{h(p)}{p^{2}(1-p)}\right) & =\frac{-2p^{2}(\rho_H+\rho_L)+p(\rho_H+4\rho_L)-2\rho_L}{(1-p)^{2}p^{3}}\\
 & =\frac{-2(\rho_H+\rho_L)(p-\p_0)^{2}+\frac{2(\rho_L)^{2}}{\rho_H+\rho_L}+p\rho_H-2\rho_L}{(1-p)^{2}p^{3}}<0,
\end{align*}
since the numerator is maximized at $p=\p_0$ within the range $[0,\p_0]$  and the maximized value
is equal to $\frac{-\rho_L\rho_H}{\rho_H+\rho_L}<0$.  Thus, $\sigma (p)$ is strictly decreasing.  Then, the result follows from observing  that $\sigma (0)=  \infty$ and $\sigma (\p_0) =0$ since $h(0) > 0 = h(\p_0).$\end{proof}

\begin{claim}
\label{claim:equiv-lam-gam}
$\lambda x \ge \lh$ is equivalent to  \begin{align}
    \label{equiv1} \lambda x (\rho_L-c)+ c(c-\rho_L-\rho_H)\ge 0.
\end{align}
\end{claim}
\begin{proof}
Notice that   if  $\rho_L -c > 0$, then  \cref{equiv1} is just  a rearrangement of $\lambda x \ge \lh $. It thus suffices to show that \cref{equiv1} implies $\rho_L >  c$. Otherwise,  we would have 
\[
\mbox{LHS of \eqref{equiv1}}\le  c (\rho_L-c)+c(c-\rho_L-\rho_H)= -c \rho_H < 0,
\]
where the first inequality holds since $\lambda x > c$.\end{proof}

To prove the equivalence between (a) and (b),    observe first that    solving  $g (p)  =0 $ yields $\p_1=\frac{\lambda x+\rho_L+\rho_H-\sqrt{(\lambda x+\rho_L+\rho_H)^{2}-4\lambda x\rho_L}}{2\lambda x}.$ Thus,   
\begin{align}
 \frac{c}{\lambda x} \le \p_1   &  \Leftrightarrow     \lambda x  +\rho_L +\rho_H - 2c\ge\sqrt{(\lambda x+\rho_L+\rho_H)^{2}-4\lambda x\rho_L}.\label{ineq3}
  \\  &  \Leftrightarrow \left( \lambda x+\rho_L+\rho_H -2c\right)^{2}\ge(\lambda x+\rho_L+\rho_H)^{2}-4\lambda x\rho_L. \nonumber \\
 & \Leftrightarrow   \lambda x (\rho_L-c)+ c(c-\rho_L-\rho_H)\ge 0, \label{equiv4}
\end{align} if  the LHS of   \eqref{ineq3} is nonnegative.   Given this and  \cref{claim:equiv-lam-gam}, the equivalence between $\p_1 \ge \frac{c}{\lambda x}$ and   $\lambda x \ge \lh $ will follow if  \eqref{equiv1} implies that the LHS of \eqref{ineq3} is nonnegative.  
To see this, recall that  \eqref{equiv1} implies  $\rho_L  > c$, so   $\lambda + \rho_L +\rho_H  -2c   >  \lambda +\rho_H -c > 0$ since $\lambda\ge \lambda x >c$.

To prove the equivalence between (b) and (c), observe that  since  
$0 = g (\p_1 ) = h (\p_1 ) - \p_1 (1-\p_1) \lambda x $, 
we have   \begin{align} \nonumber 
\sigma (\underline{p})  -\sigma (\p_1)    =  \frac{\lambda x}{r+\rho_L+\rho_H}-\sigma(\p_1) & =\frac{\lambda x}{r+\rho_L+\rho_H}-\frac{h(\p_1) c}{\p_1^{2}(1- \p_1)  \lambda x (r+\rho_L+\rho_H)}\\ \label{ineq4}
   & =\frac{c}{r+\rho_L+\rho_H}\left(\frac{\lambda x }{c}-\frac{1}{\p_1}\right)\ge 0.
\end{align} Since   $\sigma$ is decreasing, we  have  $\p_1 \ge \underline{p} $ if and only if $\p_1 \ge \pm$.

For the equivalence between (d)  and (a), observe 
$$ \sigma (\pm)  - \sigma (\underline p) = \frac{ \lambda x (\rho_L-c)+ c(c-\rho_L-\rho_H)}{c (\lambda  x -c) \left(r+\rho_L+\rho_H\right)}.$$ Thus, given that $\sigma$ is decreasing,   $\underline p \ge \pm $ is equivalent to  $ \lambda x (\rho_L-c)+ c(c-\rho_L-\rho_H) \ge 0$, which is equivalent to (a) due to \cref{claim:equiv-lam-gam}.

The second statement follows since all inequalities in the above derivations bind simultaneously whenever one of them does.

\subsection{Proof of \Cref{prop:necessary}}

To first derive the condition  \eqref{cond-delta}, let us use  \eqref{HJB0},  \eqref{HJB1}, \eqref{def:delta},  and the value matching to write 
\begin{align*}
h (\hat p) V' (\hat p_+) +  x \Delta (\hat p_+) = r V (\hat p_+) = r V (\hat p_-) = h (\hat p) V' (\hat p_-),
\end{align*} which implies  $\Delta (\hat p_+) =0$  by the smooth pasting.  Also, by value matching and smooth pasting, $\Delta (\hat p_+) = \Delta (\hat p_-)$, which yields \eqref{cond-delta}.

The conditions \eqref{slope-case1} to \eqref{slope-case3}  are derived by proving  the next three  lemmas. 

\begin{lem}
\label{lem:necessary-case}  If  $\hat p \ge \p_0$  (\ref{item:case-1}), then $V(p) = V'(p) =0$ for all $p \in [0,\hat p]$. 
\end{lem}
\begin{proof}
Observe first  that  with $\hat p \ge \p_0$, \eqref{HJB0} and $h(\p_0)=0$ together imply $V(\p_0)=0$.  Suppose  for  contradiction that there is some  $p<\p_0$ such that
$V(p)<0$. Then, we must have some $\check{p}\in(p,\p_0)$
such that $V(\check{p})<0$ and $V'(\check{p})>0$.
Since $h(\check{p})>0$ and thus $h(\check{p})V'(\check{p})>0$, \eqref{HJB0}
implies $V(\check{p})>0$, a contradiction. Suppose next
that there is $p<\p_0$ such that $V(p)>0$.
Then, we must have some $\check{p}\in(p,\p_0)$ such that $V(\check{p})> 0$
and $V'(\check{p})<0$. Since $h(\check{p})>0$ and thus $h(\check{p})V'(\check{p})<0$,
\eqref{HJB0} implies $V(\check{p})< 0$, a contradiction.
The argument so far establishes $V(p)=0$ for
all $p\le\p_0$. An analogous argument can be used to establish the
same result for the range $[\p_0,\hat{p}]$.    \end{proof}

 \begin{lem} \label{lem:necessary-case-1} If  $\hat p \le \pi_1$  (\ref{item:case-2}), then   $V' (p) =\kappa $ for all  $p \in [\hat p, 1]$. \end{lem}
 \begin{proof}
    Let us differentiate both sides of \eqref{HJB1} to obtain 
\begin{align*}
rV'(p)=g'(p)V'(p)+g(p)V''(p)-\lambda x V(p)-p\lambda x V'(p)+\lambda x(1+V(1)).
\end{align*}
We can combine this equation and \eqref{HJB1} to  remove
$V(p)$ and obtain 
\begin{align}
V'(p)= & \frac{\lambda x(c + r +r V(1))+(r+ p \lambda x)g(p)V''(p)}{r^{2}+r(\lambda x+\rho_L+\rho_H)+\lambda x\rho_L}.\label{eq:vf-slope}
\end{align}
Let us denote $a=\frac{\lambda x(c+r + r V(1))}{r^{2}+r(\lambda x+\rho_L+\rho_H)+\lambda x\rho_L}$.
We show that $V'(p)=a,\forall p\ge\hat{p}.$ Note first that this is true
for $p=\p_1$ since $g(\p_1)=0$ and \eqref{eq:vf-slope}
imply $V'(\p_1)=a$.  To first prove the linearity of $V$ over $(\p_1, 1 ]$, suppose not, i.e.,   $V'(p)\ne a$
for some $p\in(\p_1,1]$.  Consider first the case    $V'(p)>a$. We must then
have some $\check{p}\in(\p_1,p)$ such that $V''(\check{p})>0$
and $V'(\check{p})>a$.  Since $g(\check{p})<0$
and thus $g(\check{p})V''(\check{p})<0$, the equation \eqref{eq:vf-slope}
implies $V'(\check{p})<\frac{\lambda x(c+r +r V(1))}{r^{2}+r(\lambda x+\rho_L+\rho_H)+\lambda x\rho_L}=a$, a contradiction. Consider next  the case   $V'(p)<a$. Then, we must have
some $\check{p}\in(\p_1,p)$ such that $V''(\check{p})<0$ and
$V'(\check{p})<a$. Since $g(\check{p})<0$ and thus $g(\check{p})V''(\check{p})>0$,
the equation \eqref{eq:vf-slope} implies $V'(\check{p})>\frac{\lambda x(c+ r+ r V(1))}{r^{2}+r(\lambda x+\rho_L+\rho_H)+\lambda x\rho_L}=a$,
a contradiction. The argument so far establishes that $V'(p)=a,\forall p\in[\p_1,1]$.
An analogous argument can be used to show that $V (p)$ is linear on  $[\hat{p},\p_1]$ as well. 

According to \Cref{lem:v-k}(iv), $V$ can be linear only if $K = \underline{K}$, in which case $V' (p) = \kappa,\forall p$ according to \Cref{lem:v-k}(ii), as desired.  
 \end{proof}

\begin{lem}
\label{lem:necessary-case-2} If $\hat{p}\in(\p_1,\p_0)$  (\ref{item:case-3}), then  
$V'(\hat{p})=\sigma(\hat{p})$---equivalently $\Delta' (\hat p) =0 $.
\end{lem}
\begin{proof} Given the cutoff class, the HJB equation \eqref{HJB} requires $y =0 (1)$ to be optimal for $p < (>) \hat p$. Thus,  $\frac{\partial W(p,y)}{\partial y} \leq 0$ if $p <  \hat p$ and $\frac{\partial W(p,y)}{\partial y} \geq 0$ if $p > \hat p$. Since $\Delta(p) = \frac{1}{x}\frac{\partial W(p,y)}{\partial y}=0$ at $p=\hat p$, this condition implies  $\Delta'(\hat p_-)\geq 0$ and $\Delta'(\hat p_+)\geq 0$.

    Let us differentiate both sides of \eqref{HJB0} to obtain $rV'(p)=h'(p)V'(p)+h(p)V''(p)$ or 
    \begin{align}
        \label{V-dprime2}
        V'' (p) = \frac{(r+\rho_L +\rho_H) V' (p)}{h (p)}.
    \end{align}
 Substituting this into the differentiation of $\Delta (p)$ in \eqref{def:delta}, we obtain for $p < \hat{p}$, 
\begin{align} \nonumber
\Delta'(p) & =\frac{\Delta(p)}{p}+\frac{c}{ px}-p(1-p)\lambda V''(p)\\
 & =\frac{\Delta(p)}{p}+\frac{c}{ p x}-\frac{p(1-p)\lambda(r+\rho_L+\rho_H)}{h(p)}V'(p), \nonumber 
 \\  & = \frac{\Delta (p)}{p} +  \frac{p (1- p)\lambda (r+\rho_L +\rho_H)}{h (p)}\left[\sigma (p)-V'(p)\right],\label{eq:Delta-prime2}  
\end{align}  Since $\Delta (\hat p) =0$, this equation and $\Delta'(\hat p_-)\geq 0$ imply 
\begin{align}
\Delta'(\hat{p}_{-})  =\frac{\hat p (1-\hat p)\lambda (r+\rho_L +\rho_H)}{h (\hat{p})}\left[\sigma (\hat p)-V'(\hat{p})\right] \ge  0.  \label{case2-unimprove2}
\end{align}

Let us next differentiate both sides of \eqref{HJB1} to obtain 
\begin{align*}
rV'(p)=\lambda \g+\lambda \g [V(1) - V(p)] -p\lambda \g V'(p)
+  g'(p)V'(p)+g(p)V''(p), 
\end{align*} which can be rewritten as 
\begin{align}
g(p)V''(p) & =(r-g'(p)+p\lambda x)V'(p)+\lambda x[V(p)-(1+V(1))]\nonumber \\
 & =(r+\rho_L+\rho_H)V'(p)+(1-p)\lambda x V'(p)+\lambda x[V(p)-(1+V(1))]\nonumber \\
 & =(r+\rho_L+\rho_H) V'(p)-\frac{c}{p}-\frac{\Delta(p)x}{p}.\label{V-dprime1} 
\end{align}
Substituting this into the differentiation of $\Delta (p)$, we obtain for $p\ge\hat{p}$ 
\begin{align} \nonumber
\Delta'(p) & =\frac{\Delta(p)}{p}+\frac{c}{px}-p(1-p)\lambda V''(p)\\  \nonumber
 & =\frac{\Delta(p)}{p}+\frac{c}{px}-p(1-p)\lambda\frac{(r+\rho_L+\rho_H)V'(p)-\frac{c}{p}-\frac{\Delta(p)x}{p}}{g(p)}\\ \nonumber
 &  =\left(1+\frac{p(1-p)\lambda x}{g(p)}\right)\left(\frac{\Delta(p)}{p} + \frac{c}{px} \right)-\frac{p(1-p)\lambda (r+\rho_L+\rho_H)}{g(p)}V'(p)\\ 
 & =\frac{h(p)}{g(p)}\left(\frac{\Delta(p)}{p}+\frac{c}{px}\right)-\frac{p(1-p)\lambda (r+\rho_L+\rho_H)}{g(p)}V'(p) \nonumber  \\ & =\frac{h(p)}{g(p)}\left(\frac{\Delta(p)}{p}\right)  +  \frac{p (1-p)\lambda (r+\rho_L +\rho_H)}{g (p)} \left[\sigma (p)- V'(p)\right].  \label{eq:Delta-prime1}
\end{align}   Since $\Delta (\hat p) =0$, this equation and $\Delta'(\hat p_+)\geq 0$ imply 
\begin{align}
\Delta'(\hat{p}_{+}) = \frac{\hat p (1-\hat p)\lambda (r+\rho_L +\rho_H)}{g (\hat{p})} \left[\sigma (\hat p)- V'(\hat{p})\right] \ge  0.\label{case2-unimprove1}
\end{align}

For the interior cutoff case where $\hat{p} \in (\pi_1, \pi_0)$, the natural drift of the hidden state implies $h(\hat{p}) > 0$ (the belief drifts up without monitoring), while $g(\hat{p}) < 0$ (the belief drifts down under full monitoring).

Applying these opposing signs to our single-crossing requirements yields a binding condition. First,  since $h(\hat{p}) > 0$, \eqref{case2-unimprove2} requires that $\sigma(\hat{p}) \ge V'(\hat{p})$ for $\Delta'(\hat{p}_{-}) \ge 0$ to hold.  Second, since $g(\hat{p}) < 0$, equation \eqref{case2-unimprove1}  requires that $\sigma(\hat{p}) \le V'(\hat{p})$ for $\Delta'(\hat{p}_{+}) \ge 0$ to hold.
 These two conditions can hold simultaneously if and only if $V'(\hat{p}) = \sigma(\hat{p})$. Thus, the single-crossing property must strictly bind, equivalently yielding $\Delta'(\hat{p}) = 0$.  
\end{proof}

\subsection{Proof of \Cref{lem:case2-char}}
The \emph{only if} direction directly follows from \cref{lem:lh-pi1-underp1}  which shows that  $\pm \le \p_1$ is equivalent to $\l x \ge \lh$.

To prove the \emph{if} direction, suppose that $\l x \ge \lh$ and hence $\p_1 \ge \max \{ \hat p_M, \underline{p} \}$ by  \cref{lem:lh-pi1-underp1}.  Suppose for  contradiction that $\hat p > \p_1$ and hence $\hat p > \max \{ \pm, \underline{p} \}$.   We consider two cases depending on $\hat p \in (\pi_1, \pi_0)$ or $\hat p \ge \pi_0$. 

Consider first the case  $\hat p  \in (\pi_1, \pi_0)$. Recall from \cref{lem:v-k}(ii) that for any $\varepsilon >0$,  $V' (p ; \underline{K}) = \frac{\lambda x}{r+\rho_L+\rho_H}  =\sigma (\underline{p}),\forall p \in [\p_1 +\varepsilon,1]$. Using this,  we argue that $V(1) \ge \underline{K} $. Else if $V (1) < \underline{K} $, then we would have for all $p \ge \hat p$ 
$$  V' (\hat p)  = V' (\hat p; V (1))  > V' (\hat p ;  \underline{K}  )   = \sigma (\underline{p})  > \sigma (\hat p), $$
 where the first inequality follows from \cref{lem:v-k}(iii) while the second inequality  holds since $\sigma$ is decreasing.  This inequality contradicts the condition in \eqref{case2-unimprove2} that requires $V' (\hat p) \le \sigma (\hat p)$ since $h (\hat p) >0$.
 
 That $V(1) \ge \underline{K} $ implies  $V$ is (weakly) convex over $[\hat p,1]$ due to \cref{lem:v-k}(iv). Then, we have a contradiction since, by \cref{def:delta}, $$\Delta (\hat p) = {\hat p}\lambda- \frac{c}{x }  + \hat p \lambda\left[V(1)-V(\hat{p})-(1-\hat{p})V'(\hat{p})\right]\ge {\hat p}\lambda- \frac{c}{x } >0, $$ where the first inequality follows from the convexity of $V$ while the second from $\hat p > \pm$.

Consider next the case $\hat p \ge \pi_0$. By \cref{lem:necessary-case} (together with  value-matching and smooth pasting), we have $V(\hat p) = V' (\hat p) =0$. Then, by \eqref{def:delta}, 
$$\Delta (\hat p) = {\hat p}\lambda- \frac{c}{x }  + \hat p \lambda V(1) \ge {\hat p}\lambda- \frac{c}{x } >0, $$ a contradiction, which establishes $\hat p \le \p_1$.

To lastly prove $\hat p  =\pm$, observe that if $\hat p \le \p_1$, then, by \Cref{lem:necessary-case-1},  $V' (p) =\kappa,\forall p \in [\hat p,1]$.  
This linearity of $V$ over $[\hat{p},1]$ implies  $V(1)-V(p)-V'(p)(1-p)=0$ for all $p\ge\hat{p}$, which implies  by \eqref{def:delta} that  
$\Delta(p)= p \lambda-\frac{c}{x}$ and hence  $\hat{p}=\frac{c}{\lambda x} =\pm$ due to \eqref{cond-delta}.

\subsection{Proof of \Cref{prop:existence-case1}}

\subsubsection{Proof of \Cref{lem:v-under-k}}
The first statement follows directly from \Cref{lem:v-k}(ii) and the definition of $\underline{p}$.
 
To prove the second statement, note that  by \cref{lem:lh-pi1-underp1}, $\lambda x < \lh $ is equivalent  to  $\p_1  < \underline p < \pm$.   
Thus,  $\underline{p}\in(\p_1,\min\{\pm,\p_0\})$.
The monotonicity of $\underline{p}$ with respect to $\lambda$
follows from the fact that $\sigma$ is decreasing in $p$ and $\lambda x$ and increasing in $c$
while $\kappa =\frac{\lambda x}{r+\rho_L+\rho_H}$ is increasing
in $\lambda x$.

\subsubsection{Proof of \Cref{lem:v-lam}}
Observe first that $\underline{p}_2 < \underline{p}_1$   since $\underline{p}$ is decreasing in $\lambda x$ and increasing in $c$ as shown in  \cref{lem:v-under-k}.

To simplify notations, let $V_i (\cdot)$ denote $V_i (\cdot;K)$. 
Let us first establish  $V_1' (p) < V_2' (p),\forall p \ge \underline{p}_1 $. To do so, observe
first that $V_{1}'(1)<V_{2}'(1)$ due to   \eqref{eq:4}, $\lambda_{1} x_1 \le \lambda_{2}x_2$, and $c_1 \ge c_2$ (with at least one inequality being strict).
Suppose for contradiction that there is some $p\in[\underline{p}_{1},1]$
such that $V_{1}'(p)\ge V_{2}'(p)$. Then, there must exist $\check{p}\in[p,1]$
such that $V_{1}'(\check{p})=V_{2}'(\check{p})$ and $V_{1}'(p)<V_{2}'(p),\forall p>\check{p}$.
Given this and $V_{1}(1)=V_{2}(1)=K$, 
\begin{align}
V_{1}(\check{p})=V_{1}(1)-\int_{\check{p}}^{1}V_{1}'(p)dp>V_{2}(1)-\int_{\check{p}}^{1}V_{2}'(p)dp=V_{2}(\check{p}).\label{ineq:v-lam}
\end{align}
Using \eqref{HJB1-para} and the fact that $V_{1}'(\check{p})=V_{2}'(\check{p})$,
we can write 
\begin{align*}
 & \phantom{<} r(V_{1}(\check{p})-V_{2}(\check{p}))\\
 & =(\lambda_{2} x_2-\lambda_{1} x_1)\check{p}(1-\check{p})V_{1}'(\check{p})- (c_{1}-c_{2})+ \check{p}(\lambda_1 x_1 - \lambda_2 x_2) + \check{p}\lambda_{1}x_1\left[K-V_{1}(\check{p})\right]-\check{p}\lambda_{2}x_2\left[K-V_{2}(\check{p})\right]\\
 & \le (\lambda_{2}x_2 -\lambda_{1} x_1)\check{p}(1-\check{p})V_{1}'(\check{p}) + \check{p}(\lambda_{1}x_1-\lambda_{2}x_2)\left[V_{1}(1)-V_{1}(\check{p})\right]\\
 & =\check{p}(\lambda_{1} x_1-\lambda_{2}x_2)\left[V_{1}(1)-V_{1}(\check{p})-(1-\check{p})V_{1}'(\check{p})\right]\le 0,
\end{align*}
where the first inequality follows from the fact that $K=V_{1}(1)$
and $V_{1}(\check{p})>V_{2}(\check{p})$ while the second inequality
from  the convexity of $V_{1}$ for $K \ge \underline{K}_1$ as established in \cref{lem:v-k}(iv). This inequality contradicts the inequality in \eqref{ineq:v-lam}.

Then, $V_{1}(p)>V_{2}(p),\forall p\in[\underline{p}_{1},1)$ follows
from the same argument as in \eqref{ineq:v-lam}, using the fact that
$V_{1}'(p)<V_{2}'(p),\forall p\in[\underline{p}_{1},1]$.

\subsubsection{Proof of \textsc{Step 8}}
\label{sec:step 8}
 Conside fist the case $\l x  \searrow \ll$.   Let $(p^-, K^-)$ denote the limit of  $(\hat p, K)$  satisfying \eqref{two-equations} as   $\lambda x$ decreases to $\ll $. By  the continuity of $V$ and $V'$ with respect to $p$, $K$, and $(\l,\g,c)$,  the pair  $(p^-, K^-)$   is a unique solution    at $\lambda x = \ll $.\footnote{The uniqueness follows from Step 1 and 4.}

To show that   $p^- = \p_0$, suppose for a contradiction that $p^-  < \p_0$.\footnote{Note that we cannot have $p^- > \p_0$ since $p^-$ is the limit of $\hat p < \p_0$.} Then,  we must
have $K^- < K_0$ since $p^- =p(K^-)$ and $\p_0=p(K_0)$ while
$p(\cdot)$ is increasing. This implies $\Delta(p(K_0);K_0)>0=\Delta(p(K^-);K^-)$
since $\Delta(p(\cdot);\cdot)$ is increasing. Now let us choose some $(\lambda_1,x_1,c)$  such that 
$\lambda_1  x_1=\ll -\varepsilon$ with small
$\varepsilon>0$. Let $V_1 (\cdot;\cdot)$, $\Delta_1 (\cdot;\cdot)$, $\sigma_1 (\cdot)$, and  $p_1 (\cdot)$
denote the corresponding functions. By the continuity of these functions with respect to $(\l,\g,c)$, we can make $\varepsilon>0$
sufficiently small that $\Delta_1 (p_{1}(K_0);K_0)>0$. Let $K'$ be such that $\sigma_1 (p^-) = V_1' (p^- ; K') $, i.e., $p_1 (K') = p^-$. Then, we have $K' < K^-$ since  $\sigma_1 (p^-) > \sigma (p^-) = V' (p^-; K^-)  > V_1' (p^-; K^-) $ (where the second inequality holds due to \cref{lem:v-lam}) and since $V_1' (p^-; \cdot)$ is decreasing.  Also,
using $\lambda_{1}x_1<\ll$ and  derivations analogous to those in Step 5, we can show that $\Delta_{1}(p^- ;K')<\Delta(p^- ;K^-)=0$. Given that $\Delta_{1}(p_{1}(K_0);K_0)>0>\Delta_{1}(p_{1}(K');K')$,
we can find some $K_{1}\in(K',K_0)$ such that $p_{1}(K_{1})<\p_0$
and $\Delta_{1}(p_{1}(K_{1});K_{1})=0$ while $\sigma_1 (p_1 (K_1) ) = V_1' (p_1 (K_1) ; K_1) $, which contradicts the definition
of $\ll $ (since $\lambda_{1} x_1<\ll$).

Consider next the case $\l x \nearrow \lh$. To prove the convergence, let $(p^+,K^+)$ denote the limit of $(\hat p, \hat  K)$ satisfying \eqref{two-equations} as $\lambda x$ increases to $\lh $. Clearly, $p^+ \ge  \p_1$. Suppose $p^+ > \p_1$ for a contradiction.  Since, at $\lambda x  =\lh $, we have $\p_1  = \underline{p} =\frac{c}{\lambda x}$ by  \Cref{lem:lh-pi1-underp1},  we can choose $\lambda x$ sufficiently close  to $\lh $  (or sufficiently large if $\lh  =\infty$)  that   $\p_1$  is close to  $ \frac{c}{\lambda x}$. Letting $(\hat p, K)$ denote a pair satisfying \eqref{two-equations}  for such $\lambda x$, we have $\hat p > \p_1$, which implies that $\hat p > \frac{c}{\lambda x} $ and $\hat p > \underline{p}$ since $\hat p$ is close to $p^+ > \p_1$.  Then, $\Delta (\hat p;K) =0$ requires $V(1; K) - V(\hat p;K) - (1-\hat p) V' (\hat p; K)  < 0$, which implies $K  <   \underline{K}  $ by \cref{lem:v-k}(iv). Then, by \cref{lem:v-k}(iii) and \cref{claim:sigma}, $V' (\hat p ; K ) >  V' (\hat p; \underline{K}) = \kappa = \sigma (\underline{p})  > \sigma (\hat p)$, a contradiction.

\subsection{Proof of \Cref{prop:existence-case3}}

\label{sec:prop3}
The proof proceeds in a few steps and will employ the notations and
results from the proof of \cref{prop:existence-case2}.
For instance, we continue to use the notations $K_0$
and $\overline{K}$. 

\vspace{0.3cm}
\noindent \textsc{\textbf{Step 1}:} \textit{No pair $(\hat{p}, \hat K)$ with $\hat K<K_0$ or $\hat K\ge\overline{K}$
can satisfy \eqref{two-equations-1}.}
\vspace{0.2cm}

Recall first that $V'(\p_0;K_0)=0$ by the definition of
$K_0$.  We can then observe that if $\hat K<K_0$, then,
 for all $p \ge \p_0$,  $$V'(p; \hat K)\ge V'(\p_0; \hat K)>V'(\p_0;K_0)=0$$
since $V'$ is decreasing in $K$ and increasing in $p$. Observe
also that if $\hat K\ge\overline{K}$, then $V'(p;\hat K)<V'(1;\hat K)\le V'(1;\overline{K})=0,\forall p<1.$

\vspace{0.3cm}

\noindent \textsc{\textbf{Step 2}:} \textit{For any $K\in[K_0,\overline{K})$, there
is a unique $p(K)\in[\p_0,1)$ that satisfies $V'(p(K);K)=0$ and is
continuously increasing in $K$.}
\vspace{0.2cm}

For any $K\in[K_0,\overline{K})$, 
\begin{align*}
V'(\p_0;K)\le V'(\p_0;K_0)=0<V'(1;K),
\end{align*}
where the second inequality holds since $V'(1;K)=\frac{\lambda x-c-rK}{\rho_H}>0$
for $K<\overline{K} =\frac{\lambda x -c}{r}$. Thus, there exists a unique $p(K)\in(\p_0,1)$
that satisfies the desired property, since $V'(\cdot;K)$ is strictly
increasing.

The monotonicity of $p(K)$ follows easily from the fact that $V'$
is increasing in $p$ and decreasing in $K$.

\vspace{0.3cm}

\noindent \textsc{\textbf{Step 3}:} \textit{For each $\lambda x \in(c,\ll]$,
there exists a unique pair $(\hat{p},\hat K)$ satisfying \eqref{two-equations-1} with $\hat{p} \in [\p_0, 1)$ and $\hat K \in [ K_0 , \overline{K}  )$.}
\vspace{0.2cm}

Consider any $\lambda_{1} x_1 \in(c,\ll]$ and  $\lambda_{2} x_2=\ll $. 
 Let $V_{i}(p;K)$, $\sigma_{i} (p)$, $\Delta_{i}(p;K)$,
and $p_{i}(K)$ denote the functions associated with $(\l_i,\g_i,c)$.  Let $K_{i}$ and $\overline{K}_i$ denote $K_0$ and $\overline{K}$ corresponding to $(\l_i,\g_i,c)$.   Let us first observe that  $K_{2} >  K_{1}$. By \cref{lem:v-lam}, $V_{2}'(\p_0;K_1)>V_{1}'(\p_0;K_1)=0=V_{2}'(\p_0;K_2)$,
which implies $K_1<K_2$
since $V_{2}'$ is decreasing in $K$ by  \cref{lem:v-k}(iii).

By the definition of $K_0$ and $p (\cdot)$, $p_2 (K_2) = \p_0$. Then, by \textsc{Step 8} in  the proof of  \cref{prop:existence-case2} and the continuity of $\Delta$, we have  $\Delta_{2}(p_{2}(K_{2});K_{2})=0$.
Thus, using \eqref{eq:HJB-Delta}, we obtain 
\begin{align}  \nonumber 
0=\Delta_{2}(p_{2}(K_{2});K_{2}) &  =\frac{r V_{2}(\p_0;K_{2})}{x}  \\ &    \ge  \frac{r V_{1}(\p_0;K_{2})}{x}  >  \frac{r V_{1}(\p_0;K_{1})}{x}  =   \Delta_{1}(p_{1}(K_{1});K_{1}), \label{ineq}
\end{align}
where the first inequality follows from    \cref{lem:v-lam} while
the second inequality from  \cref{lem:v-k}(iv) and $K_{2} >  K_{1}$.

Observe next that $p_{1}(\overline{K}_1)=1$ since
$V_{1}'(1;\overline{K}_1)=0$. Thus, using \eqref{def:delta},
we obtain 
\[
\Delta_{1}(p_{1}(\overline{K}_1);\overline{K}_1)=\Delta_{1}(1;\overline{K}_1)=\lambda_{1}-\frac{c}{x}>0.
\]
Combining this with \eqref{ineq} and using the strict
monotonicity of $\Delta_{1}(p_{1}(\cdot);\cdot)$, we can conclude
that there is a unique $\hat K\in[K_{1},\overline{K}_1)$
such that $\Delta_{1}(p_{1}(\hat K); \hat K)=0$ and $p_{1}(\hat K)\in[\p_0,1)$.

\vspace{0.3cm}

\noindent \textsc{\textbf{Step 4}:} \textit{For the pair $(\hat{p},\hat K)$ satisfying \eqref{two-equations-1}, $\hat{p}$ is  decreasing in $\lambda x$ and increasing in $c$.}
\vspace{0.2cm}

The proof of this step  is analogous to that of \textsc{Step 5} in the proof of  \cref{prop:existence-case2}, and hence omitted.

\subsection{Proofs of \Cref{claim:delta-1} and \Cref{claim:delta-2}}

Differentiating $\delta (\cdot)$ and letting $\alpha := \rho_L +\rho_H$, we  obtain 
\begin{align}
\delta'(p)=(2p-3p^{2})\lambda(r+\alpha)V'(p)+\frac{c \alpha}{x}  +p^{2}(1-p)\lambda(r+\alpha)V''(p).\label{Delta-Prime-1}
\end{align}

To prove \Cref{claim:delta-1},   we have  $V''(p)=\frac{(r+\alpha)V'(p)-\frac{c}{p}-\frac{\Delta(p)x}{p}}{g(p)}$ from \eqref{V-dprime1}
and substitute this into \eqref{Delta-Prime-1} to obtain 
\begin{align}
\delta'(p)= & (2p-3p^{2})\lambda(r+\alpha)V'(p)+\frac{c \alpha}{x} +p^{2}(1-p)\lambda(r+\alpha)\frac{(r+\alpha)V'(p)-\frac{c}{p}-\frac{\Delta(p)x}{p}}{g(p)}\nonumber \\
= & (2p-3p^{2})\lambda(r+\alpha)V'(p)+\frac{c \alpha}{x}-p(1-p)\lambda(r+\alpha)\frac{\Delta(p)x}{g(p)}-p(1-p)\lambda(r+\alpha)\frac{c}{g(p)}\nonumber \\
 & \qquad+\frac{p^{2}(1-p)\lambda(r+\alpha)^{2}V'(p)}{g(p)}\nonumber \\
= & (2p-3p^{2})\lambda(r+\alpha)V'(p)+\frac{c \alpha}{x}-p(1-p)\lambda(r+\alpha)\frac{\Delta(p)x}{g(p)}-p(1-p)\lambda(r+\alpha)\frac{c}{g(p)}\nonumber \\
 & \qquad+(r+\alpha)\frac{p^{2}(1-p)\lambda(r+\alpha)V'(p)- \frac{c}{x} h(p)}{g(p)}+(r+\alpha)\frac{c h(p)}{x g(p)}\nonumber \\
= & (2p-3p^{2})\lambda(r+\alpha)V'(p)+\frac{c \alpha}{x}-p(1-p)\lambda(r+\alpha)\frac{\Delta(p)x}{g(p)}+(r+\alpha)\frac{\delta(p)}{g(p)}+\frac{c(r+\alpha)}{x}\nonumber \\
= & 2p(1-p)\lambda(r+\alpha)V'(p)-p^{2}\lambda(r+\alpha)V'(p)\nonumber \\
 & \qquad+\frac{c \alpha}{x}-p(1-p)\lambda(r+\alpha)\frac{\Delta(p)x}{g(p)}+(r+\alpha)\frac{\delta(p)}{g(p)}+\frac{c(r+\alpha)}{x}\nonumber \\
= & 2p(1-p)\lambda(r+\alpha)V'(p)-\frac{p^{2}(1-p)\lambda(r+\alpha)V'(p)-h(p)\frac{c}{x}}{1-p}-\frac{h(p)c}{(1-p)x}\nonumber \\
 & \qquad+\frac{c \alpha}{x}-p(1-p)\lambda(r+\alpha)\frac{\Delta(p)x}{g(p)}+(r+\alpha)\frac{\delta(p)}{g(p)}+\frac{c(r+\alpha)}{x}\nonumber \\
= & 2p(1-p)\lambda(r+\alpha)V'(p)-\frac{\delta(p)}{1-p}-\frac{h(p)c}{(1-p)x}\nonumber \\
 & \qquad+\frac{c \alpha}{x}-p(1-p)\lambda(r+\alpha)\frac{\Delta(p)x}{g(p)}+(r+\alpha)\frac{\delta(p)}{g(p)}+\frac{c(r+\alpha)}{x}\nonumber \\
= & 2p(1-p)\lambda(r+\alpha)V'(p)-\frac{\delta(p)}{1-p}-\frac{c \alpha}{x}+\frac{\rho_H c}{(1-p)x}\nonumber \\
 & \qquad+\frac{c \alpha}{x}-p(1-p)\lambda(r+\alpha)\frac{\Delta(p)x}{g(p)}+(r+\alpha)\frac{\delta(p)}{g(p)}+\frac{c(r+\alpha)}{x}\nonumber \\
= & 2p(1-p)\lambda(r+\alpha)V'(p)-\delta(p)\left(\frac{1}{1-p}-\frac{r+\alpha}{g(p)}\right)+\frac{\rho_Hc}{(1-p)x}\nonumber \\
 & \qquad-p(1-p)\lambda(r+\alpha)\frac{\Delta(p)x}{g(p)}+\frac{c(r+\alpha)}{x}.\nonumber
\end{align}
Since $V' (p) \ge 0$ and $g (p) <0$,   we have $\delta'(p) >0$ if $\Delta (p) \ge 0 $ and $\delta (p) =0$.

To prove \Cref{claim:delta-2}, let  us  substitute  \eqref{V-dprime2} into \eqref{Delta-Prime-1} to obtain 
\begin{align}
\delta'(p)= & (2p-3p^{2})\lambda(r+\alpha)V'(p)+\frac{c \alpha}{x}+p^{2}(1-p) \lambda (r+\alpha)^{2}\frac{V'(p)}{h(p)}\nonumber \\
= & (2p-3p^{2})\lambda(r+\alpha)V'(p)+\frac{(r+\alpha)}{h(p)}\left(p^{2}(1-p)(r+\alpha)\lambda V'(p)-\frac{c}{x}  h (p) \right)+\frac{c(r+2\alpha)}{x}\nonumber \\
= & 2p(1-p)\lambda(r+\alpha)V'(p)-\frac{p^{2}(1-p)\lambda(r+\alpha)V'(p)-\frac{c}{x} h(p)}{1-p} -\frac{c}{(1-p)x}h (p)\nonumber \\
 & \qquad+\frac{(r+\alpha)}{h(p)}\delta(p)+\frac{c(r+2\alpha)}{x}\nonumber \\
= & 2p(1-p)\lambda(r+\alpha)V'(p)-\frac{\delta(p)}{1-p}-\frac{c\alpha}{x}+  \frac{\rho_H c}{(1-p)x} +\frac{(r+\alpha)}{h(p)}\delta(p)+\frac{c(r+2\alpha)}{x}\nonumber \\
= & 2p(1-p)\lambda(r+\alpha)V'(p)-\frac{\delta(p)}{1-p}+\frac{\rho_Hc}{(1-p)x}+\frac{(r+\alpha)}{h(p)}\delta(p)+\frac{c (r+\alpha)}{x}.\label{delta-prime-2}
\end{align}  Since $V' (p) \ge 0$, we have $\delta' (p) >0$ if $\delta (p) =0$.

\subsection{Proof of \Cref{prop:monotone}}

The first statement  follows     from \cref{prop:existence-case1}  in the case $\hat p \le \p_1$, where  $\hat p =\frac{c}{\lambda x}$. To deal with the case  $\hat p > \p_1$, rewrite  $\Delta (\hat p) =0$ as 
\begin{align}
    \label{eq:p-c} \hat{p}\lambda-\textstyle{\frac{c}{x}} = -\lambda\hat{p}\left[V(1)-V(\hat{p})-(1-\hat{p})V'(\hat{p})\right].
\end{align}
From the proofs of \cref{prop:existence-case2} and \cref{prop:existence-case3}, we know that for $p \ge \hat p$,  $V (p) = V (p; K) $ for some $K > \underline{K}$.  Thus, by \cref{lem:v-k}(iv), $V$ is strictly convex, which implies that the expression inside the square bracket in \eqref{eq:p-c} is strictly positive, so $\hat p < \frac{c}{\lambda x}$.

To prove the second statement, let us focus on establishing the comparative statics with $\lambda x$ since doing so  with $c$ is analogous. To do so, consider any $\lambda_{2} x_2>\lambda_{1} x_1$.
Let  $\p_{0i}$, $\p_{1i}$ and $\hat{p}_{i}$ denote the values of $\p_0$, $\p_1$
and $\hat{p}$, respectively, under  $\lambda_{i} x_i,i=1,2$. Note that $\p_{01} = \p_{02}$ and $\p_{11}  >   \p_{12}$.

The desired result, $\hat p_1 > \hat p_2$, follows  immediately from \cref{prop:existence-case1} to
\cref{prop:existence-case3} if $\lambda_{2}x_2>\lambda_{1} x_1\ge\lh $
or if $\lh  > \lambda_{2} x_2 > \lambda_{1} x_1 > \ll $
or if $\lambda_{1}<\lambda_{2}\le\ll $.
In the case $\lambda_{1} x_1<\lh\le\lambda_{2} x_2$, we have  $\hat{p}_{1}>\p_{11}>\p_{12}\ge\hat{p}_{2}$
as desired.  In the case $\lambda_{1} x_1\le\ll <\lambda_{2} x_2<\lh $, we have $\hat{p}_{1}\ge\p_{01} = \p_{02} >\hat{p}_{2}$.

\section{Passive Learning} \label{sec:passive}

In this section, we consider  a situation in which the detection can occur without monitoring, which we call \emph{passive learning}: for instance, there is a community reporting by the victim of an infraction.    Let $w \in (0,1)$ denote the probability of such detection (upon an infraction being committed). Assuming that the passive learning occurs independently of  the principle's monitoring, an infraction is detected with the probability $y+w -wy$, so the rate at which the principal learns $\omega =H$ is $p_t \lambda x (y +w -yw) $. We assume that passive learning, which occurs without any monitoring,  has no effect on reducing harm. Thus, by choosing  $y$ units of enforcement, the principal with posterior belief $p_t$  mitigates the harm by the same rate as before, that is, $p_t \lambda x y $.
As a result, the principal's flow loss and threshold belief for the myopically optimal policy are unchanged. Therefore, passive learning affects the principal's discounted total loss only through her belief updating.  

The belief updating rule \eqref{eq:drift} changes to 
\begin{align}  
 \dot{p} = \rho_L(1-p)-\rho_Hp - p(1-p)\lambda  x (y +w - yw)=: f_w (p,y).
      \label{eq:drift-passive} 
\end{align} 
With passive learning, learning takes place even when the principal does not enforce. This means that the posterior belief may drift below $\p_0$ even when $y=0$.  Indeed, one can  find a  unique  $\p_w \in (\p_1,\p_0) $  such that  $f_w (\p_w,0) =0$.\footnote{This is because   $f_w (\p_0, 0) = f (\p_0, w) <  f (\p_0, 0) = 0  = f (\p_1, 1) < f (\p_1, w)  =f_w (\p_1, 0)$. Note also that the benchmark $\p_1$ remains relevant: Since $f_w (p,1) = f (p,1)$, we have $f_w (\p_1,1) = f (\p_1,1) =0$.}  

Recall that, without passive learning, the long-term belief under  MP either stays at $\pi_0$ or cycles over $[\pm, 1]$. In either case, the principal obtains the same payoff under MP as under SP, for information is never collected when monitoring is strictly suboptimal. This is no longer the case when there is passive learning.  Suppose $\pi_w < \pm$. Then, learning occurs with positive probability even when the posterior falls below $\pm$ (where enforcement is strictly suboptimal)  until it reaches $\pi_w$.   In this case, MP prescribes the  (myopically) correct action, i.e., no monitoring, unlike SP, so the former performs strictly better:

\begin{proprestated}{4}
If $\hat p_M > \pi_w$, MP achieves a strictly lower long-run loss for the principal than SP. Otherwise, MP and SP achieve the same long-run loss.
\end{proprestated}

\begin{proof} We first establish the following result: 

\begin{claim} \label{claim:stationary} $  \mathbb{E} [p_t  | p_0 = \pi_0] = \pi_0$.   
\end{claim} 
\begin{proof}
The posterior belief conditional on  the initial belief $p_0 =\pi_0$ and any history of monitoring and detection through time $t$, denoted by $h^t$, is equal to   $$\mbox{Pr} [\omega_t = H  | p_0 = \pi_0, h^t ] =  \mbox{Pr}  [\omega_t = H | \omega_0 = H, h^t] \pi_0  + \mbox{Pr}  [\omega_t = H | \omega_0 = L, h^t] (1-\pi_0)  . $$ Thus, \begin{align*}
    \mathbb{E} [p_t  | p_0 = \pi_0]  &  =  \mathbb{E} [\mbox{Pr} [\omega_t = H  | p_0 = \pi_0, h^t ]] \\ 
 & = \mathbb{E}_{h^t} \Big[ \mbox{Pr}  [\omega_t = H | \omega_0 = H, h^t] \pi_0  + \mbox{Pr}  [\omega_t = H | \omega_0 = L, h^t] (1-\pi_0)  \Big]   \\  & = \mbox{Pr}  [\omega_t = H | \omega_0 = H] \pi_0  + \mbox{Pr}  [\omega_t = H | \omega_0 = L] (1-\pi_0)     = \pi_0,
  \end{align*} where the last equality follows from the fact that $\pi_0$ is the stationary belief that the Markov chain $\omega_t$ is at state $H$.  
\end{proof}

The expected total loss under MP with the initial belief $p_0= \pi_0$ is given by   \begin{align*} 
L_{\tiny MP}(\p_0)=  &\mathbb{E} \left[  \left.  \int_0^{\infty} ( p_t\l \g (1-\ysp(p_t)) + c\ysp(p_t))e^{-rt}dt  \right| p_0  =\pi_0 \right]  \\   =  &\mathbb{E}    \left[ \left. \int_0^{\infty}  
\min \{p_t \l \g,  c\} e^{-rt}dt  \right| p_0 =\pi_0  \right] \\
=  &     \int_0^{\infty}  
\mathbb{E}\Big[\min \{p_t \l \g,  c\} \Big| p_0 =\pi_0  \Big] e^{-rt}dt  \\
\le &     \int_0^{\infty}  
\min \Big\{ \mathbb{E}  \big[p_t \l \g  \big| p_0 =\pi_0   \big], c\Big\} e^{-rt}dt \\
= &     \int_0^{\infty}  
\min \big\{ \p_0 \l \g  , c\big\} e^{-rt}dt \\
= &  \min\{\p_0\l\g, c\}/r= L_{\tiny SP}(\p_0),
\end{align*}  
where the inequality follows from Jensen's inequality and the equality right after that inequality follows from \Cref{claim:stationary}. 

The weak inequality becomes an equality if  $\pm\le \p_w$.  This is because $\min \{p_t \l \g,  c\}=c$ for all $p_t\in [\p_w, 1]$.  However, if  $\pm> \p_w$, then both $p_t \l \g< c$ and $p_t \l \g> c$ occurs with positive  probability.  Hence, in this case, the inequality becomes strict.  \end{proof}

\section{Analysis for \Cref{sec:endog}} \label{appendix:criminals belief formation}

\begin{proof}[Proof of \Cref{lem:stationary}]
To identify the stationary distribution of the posterior belief, we invoke a {\it balance equation}: for any subset $S \subset [0,1]$, the mass of beliefs flowing out of $S$ must equal the mass of beliefs flowing into $S$. It suffices to consider two cases: $S = \{\hat p\}$ and $S = (p,1]$ for each $p > \hat p$.

For $S = (p,1]$ for $p > \hat p$, the balance condition requires
\begin{align} \label{belief-change}
    [\Phi(p+dp)-\Phi(p)](1- p \lambda x dt)
= \left( \int_{\hat p}^p \phi(s)\, s \lambda x \, ds + \Phi(\hat p)\hat p \lambda x z(\hat p) \right) dt  + o(dt),
\end{align}
where the LHS represents the mass flowing out of the set $(p,1]$ while the RHS represents the mass flowing into it.  Since the instantaneous rate of belief change is $f(p,1)<0$ by  \cref{eq:drift},  we have  $
dp = -f(p,1)\, dt + o(dt).$
Using this,  \eqref{belief-change} simplifies to
\[
-\phi(p)f(p,1) = \int_{\hat p}^p \phi(s)\, s \lambda x \, ds + \Phi(\hat p)\hat p \lambda x z(\hat p).
\]
Differentiating both sides with respect to $p$ yields
\[
\phi(p)\, p \lambda x = -\phi'(p) f(p,1) - \phi(p) f'(p,1),
\]
where $f'(p,1) = \frac{\partial f(p,1)}{\partial p}$.  
Thus, we obtain the ODE
\begin{align}
\label{ode-phi} \phi'(p) f(p,1) = - \phi(p)\big(p \lambda x + f'(p,1)\big),    
\end{align}
whose solution is given in \eqref{stationary-belief-1}.

Turning to $S = \{\hat p\}$,   the balance equation requires
\begin{align}
\label{invariance: mass point-0}
\Phi (\hat p)\, \hat p \lambda x z(\hat p)\, dt
= [\Phi(\hat p + dp)-\Phi(\hat p)](1-\hat p \lambda x dt) + o(dt).
\end{align}
The LHS represents the outflow of mass, which  occurs when the belief jumps from 
$\hat p$ to 1 with probability $\hat p z(\hat p)\lambda x\, dt$ during $dt$. The RHS represents the inflow of mass into $\hat p$.
Substituting this  and $
dp = -f(\hat p,1)\, dt + o(dt)$ into the RHS of \eqref{invariance: mass point-0}  yields  \eqref{stationary-belief-2}.

Given \eqref{stationary-belief-1}, the mass $\Phi (\hat p)$ must satisfy
\begin{align}
\label{invariance-1}
1 - \Phi (\hat p)  
= \int_{\hat p}^1 \phi(p)\, dp
= \phi(\hat p)\int_{\hat p}^1 \exp\!\Big(\int_{\hat p}^p r(s)\, ds\Big)\, dp.
\end{align}
Using \eqref{stationary-belief-2}, we obtain $
\Phi (\hat p) = -\frac{\phi(\hat p) f(\hat p,1)}{\hat p \lambda x z(\hat p)} $,
which can be substituted into \eqref{invariance-1} to yield
\begin{align}
\label{phi-hat-p}
\phi(\hat p)\left[
\int_{\hat p}^1 \exp\!\Big(\int_{\hat p}^p r(s)\, ds\Big)\, dp
- \frac{f(\hat p,1)}{\hat p \lambda x z(\hat p)}
\right] = 1.
\end{align}
Since the expression in brackets is positive, there exists $\phi(\hat p)>0$ satisfying this equation.    \end{proof}

 \begin{lem} \label{lem:mps}
    As $\hat p$ becomes lower, $\Phi (p)$ falls  for all $p \ge \hat p$. Hence, $\Phi$ exhibits a mean-preserving spread as $\hat p$ becomes lower. 
\end{lem}
\begin{proof}
  For $p \ge \hat p$, we have  $\Phi (p) = \Phi (\hat p) + \phi (\hat p) \int_{\hat p}^p \exp \left( \int_{\hat p}^t r (s) ds\right) dt  $ and thus  \begin{align*}
    \frac{\partial \Phi (p)}{\partial \hat p}  & = \Phi' (\hat p)+ \phi' (\hat p) \int_{\hat p}^p \exp \left( \int_{\hat p}^t r (s) ds\right) dt  - \phi (\hat p) \left[  1 + r (\hat p) \int_{\hat p}^p \exp \left( \int_{\hat p}^t r (s) ds\right) dt \right] \\ & = \Phi' (\hat p) - \phi (\hat p) +  \left[  \phi' (\hat p) - \phi (\hat p) r (\hat p) \right] \int_{\hat p}^p \exp \left( \int_{\hat p}^t r (s) ds\right) dt = \Phi' (\hat p) - \phi (\hat p) >0,
\end{align*} where the second equality follows from the  definition of $\phi$. To prove the inequality, observe that $$\Phi (\hat p) =   -\frac{\phi (\hat p) f (\hat p, 1)}{\hat p \lambda x z (\hat p) }   =  -\frac{\phi (\hat p) f (\hat p, 1) (1- \hat p)}{h (\hat p)}  \; \mbox{ or } \;  \Phi (\hat p) h(\hat p) = - \phi (\hat p) f (\hat p, 1) (1- \hat p),$$ where the second equality follows from  the definitioin of $z(\cdot)$. By differentiating both sides, we obtain 
\begin{align*}
    \Phi' (\hat p) h(\hat p) + \Phi (\hat p) h' (\hat p) & = -\phi' (\hat p) f (\hat p,1) (1-\hat p) - \phi (\hat p) ( f' (\hat p,1) (1-\hat p) - f(\hat p,1)) \\ & = \phi (\hat p) (\l x \hat p + f' (\hat p,1))  (1-\hat p) - \phi (\hat p) [ f' (\hat p,1) (1-\hat p) - f(\hat p,1)] \\  & =\phi (\hat p) (\l x \hat p (1-\hat p) +f (\hat p,1)) = \phi (\hat p) h (\hat p),
\end{align*} where the second equality follows from \eqref{ode-phi}. Rearranging the resulting equation yields $$  \Phi' (\hat p) - \phi (\hat p)  = - \frac{\Phi (\hat p) h' (\hat p)}{h (\hat p)} = \frac{\Phi(\hat p) (\rho_L +\rho_H)}{(1-\hat p)\rho_L -\hat p \rho_H } >0 ,$$
 where the inequality holds since $\hat p < \p_0 =\frac{\rho_L}{\rho_L +\rho_H}$.
\end{proof}

\begin{proof}[Proof of  \Cref{lem:expected-enf}]
\label{appendix:proof of lemma 2}

First, the long-run distribution of posterior belief conditional on $\omega=H$ can be obtained as
$$\Psi(p):=\Pr\{p'\le p| \omega=H\} = \frac{\Pr\{p'\le p \mbox{ and } \omega=H\} }{\Pr\{\omega=H\}}= \begin{cases}
   0    & \mbox{ if }  p < \hat p \\   \frac{ \Phi (\hat p)\hat p + \int_{\hat p}^p \phi(s)s ds }{\pi_0}  & \mbox{ if } p \ge \hat p. 
\end{cases}
$$

The expected enforcement conditional on $\omega =H$ is  thus
\begin{align}  
\mathbb{E}_{p \sim \Psi}[y_{\hat p}(p)  | \omega =H, x ]= \frac{\Phi (\hat p) \hat p }{\p_0}  z (\hat p,x)  +\left(1- \frac{\Phi (\hat p) \hat p }{\p_0} \right)         =  & 1- \frac{1}{\pi_0} \Phi (\hat p) \hat p (1-z (\hat p,x)).\label{expected-enfo}
\end{align}   We first show that this expression is decreasing in $\hat p$. To do so,  it suffices to prove that $\Phi (\hat p)$ is increasing since $\hat p (1- z(\hat p,x))$ is increasing. 

To this end,  let $$\eta(\hat p) : =\int_{\hat p}^{1}   \exp{\left( \int_{\hat p}^{p} r(s) ds \right)} dp. $$ Using this and \cref{phi-hat-p}, we obtain $$\phi (\hat p) =\frac{1}{\eta (\hat p) - \frac{f(\hat p,1)}{ \hat p\lambda x z(\hat p,x)}},$$ which can be  plugged   into \cref{invariance-1} to yield
\begin{align*}
    1- \Phi (\hat p) = \frac{\eta (\hat p)}{\eta (\hat p)  -\frac{f(\hat p,1)}{\hat p\lambda x z(\hat p,x)}} = \frac{1}{1 + \left(\frac{1}{\hat p\lambda x z(\hat p,x)}\right) \left( \frac{-f(\hat p,1)}{\eta (\hat p)}\right)
    },
\end{align*} which is continuous in $\hat p \in (\p_1, \p_0)$.

Observe first that $  0= f (\hat p, z( \hat p,x)) = h (\hat p)  - \l \g \hat p (1- \hat p) z(\hat p,x)$ and thus  $$  \hat p \l \g z(\hat p,x)= \frac{h(\hat p)}{1-\hat p} = \rho_L - \frac{\hat p}{1-\hat p} \rho_H, $$  which is decreasing in $\hat p$. So, it suffices to show that $\left( \frac{-f(\hat p,1)}{\eta (\hat p)}\right)$ is  increasing in $\hat p$.  For this, observe  
\begin{align*}
    \frac{d}{d \hat p} \left( \frac{-f(\hat p,1)}{\eta (\hat p)}\right) &  = \frac{- f' (\hat p,1) \eta (\hat p) + f (\hat p,1) \eta' (\hat p) }{\eta (\hat p)^2} \\  & =  \frac{- f'
 (\hat p,1) \eta (\hat p) + f (\hat p,1) (- r (\hat p) \eta (\hat p)  -1 ) }{\eta (\hat p)^2}  \\ & = \frac{ \hat p \lambda x \eta (\hat p) - f (\hat p,1)}{\eta (\hat p)^2} >0,
\end{align*}  where the second equality follows from  the fact that  $\eta' (\hat p) = - r(\hat p)  \eta (\hat p)-1$ (by the definition of $\eta$), the third equality from  the definition of $r(\hat p)$, and the inequality from  $f(\hat p, 1) <0$.

As $\hat p \to \p_1$,  we have  $z (\hat p,x) \to 1$. Then, by \eqref{stationary-belief-2}, we must have $\Phi (\hat p) \to 0$ since $f (\hat p,1)\to 0$, which, by \eqref{expected-enfo}, implies   $\mathbb{E}_{p \sim \Psi}[y_{\hat p}(p)| \omega =H ]\to 1$ as $\hat p\to \p_1$.

As $\hat p \to \p_0$,  we  have  $z (\hat p,x) \to 0$ by definition of $z (\hat p,x)$.  Then, by \eqref{stationary-belief-2}, we have $ \phi  (\hat p)  \to  0 $, which implies     $\Phi (\hat p) \to 1$ by \eqref{invariance-1}. Thus, by \eqref{expected-enfo},  $\mathbb{E}_{p \sim \Psi}[y_{\hat p}(p) | \omega =H]\to 0$  as $\hat p\to \p_0$. 

Thus, there exists $\bar p (\lambda x) \in (\p_1,\p_0)$ that satisfies \eqref{bar-p}. The continuity of $\bar p(\lambda x)$ in $\lambda x$ follows from the continuity of  $z(\cdot)$ and $\Phi (\cdot)$ in $x$. 

Given that  $\mathbb{E}_{p \sim \Psi}[y_{\hat p}(p) | \omega =H, x ]$ is decreasing from 1 to 0 as $\hat p$ increases from $\pi_1$ to $\pi_0$, it is clear that $\bar p (\lambda x)$ is decreasing from $\pi_0$  to $\pi_1$ as $\bar y$ increases from 0 to 1. 
\end{proof}

\end{document}